\documentclass[twocol]{ametsocV6.1}

\usepackage{booktabs}
\usepackage{graphicx}%
\usepackage{adjustbox,lipsum}

\DeclareMathOperator{\EX}{\mathbb{E}}
\DeclareMathOperator{\PX}{\mathbb{P}}

\title{Algorithmic Hallucinations of Near-Surface Winds:\\Statistical Downscaling with Generative Adversarial Networks to Convection-Permitting Scales}

\authors{Nicolaas J. Annau\aff{a, b}\correspondingauthor{Nicolaas J. Annau, nannau@uvic.ca}, Alex J. Cannon\aff{b}, Adam H. Monahan\aff{a}}

\affiliation{
    \aff{a}{School of Earth and Ocean Sciences, University of Victoria, Victoria, BC, Canada}\\
    \aff{b}{Climate Research Division, Environment and Climate Change Canada, Victoria, BC, Canada}
}

\abstract{
This paper explores the application of emerging machine learning methods from image super-resolution (SR)  to the task of statistical downscaling. We specifically focus on convolutional neural network-based Generative Adversarial Networks (GANs). Our GANs are conditioned on low-resolution (LR) inputs to generate high-resolution (HR) surface winds emulating Weather Research and Forecasting (WRF) model simulations over North America. Unlike traditional SR models, where LR inputs are idealized coarsened versions of the HR images, WRF emulation involves using non-idealized LR and HR pairs resulting in shared-scale mismatches due to internal variability. Our study builds upon current SR-based statistical downscaling by experimenting with a novel frequency-separation (FS) approach from the computer vision field. To assess the skill of SR models, we carefully select evaluation metrics, and focus on performance measures based on spatial power spectra. Our analyses reveal how GAN configurations influence spatial structures in the generated fields, particularly biases in spatial variability spectra. Using power spectra to evaluate the FS experiments reveals that successful applications of FS in computer vision do not translate to climate fields. However, the FS experiments demonstrate the sensitivity of power spectra to a commonly used GAN-based SR objective function, which helps interpret and understand its role in determining spatial structures. This result motivates the development of a novel partial frequency-separation scheme as a promising configuration option. We also quantify the influence on GAN performance of non-idealized LR fields resulting from internal variability. Furthermore, we conduct a spectra-based feature-importance experiment allowing us to explore the dependence of the spatial structure of generated fields on different physically relevant LR covariates.
}

\begin{document}

\maketitle

\statement
We use artificial intelligence algorithms to mimic wind patterns from high-resolution climate models, offering a faster alternative to running these models directly. Unlike many similar approaches, we use datasets that acknowledge the essentially stochastic nature of the downscaling problem. Drawing inspiration from computer vision studies, we design several experiments to explore how different configurations impact our results. We find evaluation methods based on spatial frequencies in the climate fields to be quite effective at understanding how algorithms behave. Our results provide valuable insights and interpretations of the methods for future research in this field.

\section{Introduction}
\subsection{Motivation}
Atmospheric flow structures exist on spatial scales ranging from centimetres to thousands of kilometres. Accurately representing these scales in computational simulations of the atmosphere is a great challenge, especially since processes at differing scales are not generally independent of each other \citep{judt_insights_2018}. 

Readily available low-resolution (LR) climate models (ranging from 10 - 100 km horizontal grid spacing) cannot resolve important small-scale processes, particularly for variables strongly influenced by surface heterogeneities (e.g., wind and precipitation) \citep{whiteman_mountain_2000,  frei_daily_2003, kharin_changes_2007, stephens_dreary_2010, sillmann_climate_2013, ban_heavy_2015, torma_added_2015, schlager_spatial_2019, song_evaluating_2020}. At horizontal grid-spacings of approximately 4 km or finer, high-resolution (HR) convection-permitting models offer improved representations of small-scale variability over LR models owing to the representation of convection and finer orographic resolution \citep{kopparla_improved_2013, prein_precipitation_2016, innocenti_observed_2019}. However, due to their computational cost, convection-permitting models are currently limited in scope either spatially or temporally, and large initial condition or multi-model ensemble experiments -- which are highly desirable for climate impacts and adaptation studies -- are unavailable. 

Since fast assessments of meteorological conditions are required for such studies, additional methods that can effectively model small-scale processes have been developed. As one option, statistical downscaling seeks to exploit empirical links between large-scale and small-scale processes using statistical models \citep{wilby_downscaling_1997, cannon_probabilistic_2008, sobie_high-resolution_2017, li_indices_2018}. However, standard statistical downscaling approaches are often limited in their ability to model the range of spatiotemporal variability required in many climate impacts studies \citep{maraun_precipitation_2010}.

Machine learning is the branch of artificial intelligence concerned with having computers learn how to perform certain tasks. Deep learning, which is an approach to machine learning based on artificial neural networks \citep{gardner_artificial_1998}, can be used to implement highly non-linear, high-dimensional, and flexible statistical models. As one example, Convolutional Neural Networks (CNNs) are a class of deep learning models constructed for image analysis with spatial awareness \citep{krizhevsky_imagenet_2012, karpathy_cs231n_2022}. In the field of image processing, super-resolution (SR) aims to develop deep learning models that produce plausible HR details from LR inputs. Owing to their ability to represent spatially-organized structures in images,  CNNs have led to substantial improvements in SR quality \citep{dong_learning_2014, dong_image_2015, zhang_residual_2018, zhu_gan-based_2020}. Even further improvements were found by adopting generative adversarial networks (GANs)  \citep{goodfellow_generative_2014, mirza_conditional_2014} using CNNs for SR tasks \citep{ledig_photo-realistic_2017, zhu_gan-based_2020}. GANs are a machine learning architecture that consists of duelling functions (often CNNs) trained simultaneously with opposing objectives. These objectives shape two networks that communicate with each other, namely, the \textit{Generator}, which aims to generate realistic information, and the \textit{Discriminator} (or \textit{Critic}), which aims to judge or critique this generated information and provide feedback to the Generator. 

Given the natural parallels between SR and statistical downscaling tasks, researchers have started to apply CNNs to the field of climate downscaling. Several applications have focused on temperature and precipitation. For instance, \citet{sha_deep-learning-based_2020} employed CNNs to downscale temperature over the continental United States, while \cite{wang_deep_2021} utilized a deep residual network for downscaling daily precipitation and temperature. In another study, \cite{kumar_deep_2021} used the \textit{Super-Resolution Convolutional Neural Network} to downscale rainfall data for regional climate forecasting. GANs configured for SR have also shown promise in downscaling for precipitation, wind and solar irradiance fields \citep{singh_downscaling_2019, stengel_adversarial_2020, leinonen_stochastic_2020, harris_generative_2022, price_increasing_2022}.

\subsection{Problem formulation}
The focus of this study is the evaluation of GAN and CNN-based SR methods for downscaling from LR climate model scales to HR scales. Specifically, we assess SR models adapted directly from computer vision for the multivariate statistical downscaling of near-surface winds, encompassing both the $u$ and $v$ wind components simultaneously. HR wind fields are crucial for numerous weather and climate applications, such as fire weather, pollutant dispersal, infrastructure design, and wind turbine siting. In contrast to most applications, we adopt an \emph{emulation} approach, training the machine learning models on existing pairs of HR fields and covariates from the LR models used to drive them rather than deriving them from the convection-permitting model fields themselves (i.e. through coarsening). The LR and HR fields may therefore contain mismatches on shared scales that result from the internal variability of the convection-permitting model. One benefit of this approach is that it allows us to sample from the distribution of internal variability that is physically consistent with the conditioning fields.  The generation of fine-scale features using SR has been compared to the concept of ``hallucinating,'' where plausible details are generated that may not precisely match the ``true'' HR features present in the training data \citep{zhang_texture_2020}. This ability to hallucinate these details is considered desirable for climate applications  \citep{bessac_stochastic_2019}.

\subsection{Research Questions}
To build on existing literature, we narrow in on three core questions in this paper: (i) How do the generated outputs change when we manipulate the objective functions taken directly from the computer vision literature? (ii) What capacity do the networks have to deal with non-idealized LR/HR pairs? and (iii) What role do select LR covariates play in super-resolved near-surface wind fields?

Existing climate applications of SR often overlook the importance of objective functions borrowed from the computer vision field. In this manuscript, we address this issue by focusing on the intersection of SR and statistical downscaling. We conduct experiments to explore various configurations of SR models, including objective functions, data sources, and hyperparameters. Our goal is not to develop highly optimized models for specific configurations but rather to provide insights into the effectiveness and sensitivity of the SR objective function for statistical downscaling. We propose ways to improve the configuration through the assessment of multiple skill metrics and evaluation techniques. Additionally, we investigate the influence of low-resolution (LR) inputs and the emulation approach (i.e. with the presence of internal variability) on the generated fields.

We first introduce the existing configurations of SR methods and explain how our chosen configuration and methods relate to them (Section \ref{section:previous_work}). Subsequently, we provide details on the training methods and data in Section \ref{section:training}. To organize our work, we introduce the methodology required to conduct two experiments we refer to as ``Experiment 1: Frequency Separation'' (Section 3\ref{section:exp1}) and ``Experiment 2: Partial Frequency Separation'' (Section 3\ref{section:exp2}) that we use to address research question (i). In Section \ref{section:results}, we present the results from Experiment 1 and 2. Additionally, we conduct further analysis in ``Experiment 3: Low-resolution Covariates''  in Section 4\ref{section:exp3} that address research questions (ii) and (iii). We provide a discussion in Section \ref{section:discussion}. Additionally, given the novelty and unique challenges of SR methods for statistical downscaling, we review SR and statistical downscaling in the supplemental material. We provide an acronym definition list in the supplemental material.

\section{Previous Work}
\label{section:previous_work}
\subsection{Stochastic vs. Deterministic GANs}
GAN-based SR methods have been successfully applied to statistical downscaling of climate fields using two main approaches: deterministic and stochastic. In deterministic SR, a unique realization is generated for unique LR inputs (e.g. wind and solar irradiance fields in \citealt{singh_downscaling_2019, stengel_adversarial_2020}). Alternatively, stochastic SR allows for the sampling of multiple realizations given single LR conditioning fields by providing noise to the generator network \citep{leinonen_stochastic_2020, harris_generative_2022}.

We focus on purely deterministic SR models (i.e. single HR fields for given LR input fields) to simplify the analysis and reduce the number of free parameters (such as how to configure the generator to accept noise). While stochastic SR is an important avenue of research, it introduces complexity, design choices, and additional currently unresolved issues, such as under-dispersion in the generated ensembles \citep{goodfellow_generative_2014, arjovsky_wasserstein_2017, harris_generative_2022}. Furthermore, stochastic and deterministic SR are typically optimized using similar (if not identical) objective functions and so we believe that our findings can inform design choices for stochastic approaches. There are, however, some key differences between the two approaches that we will discuss further in Section \ref{section:training}.

\subsection{Low-resolution Covariates}
In addition to developing stochastic and deterministic SR models, existing SR studies have configured the LR input data in numerous ways. For example, some studies (in both computer vision and climate/weather) use LR covariates that are coarsened versions of their HR targets forming a perfect and idealized LR/HR pair at shared scales \citep{dong_image_2015, ledig_photo-realistic_2017, wang_esrgan_2018, singh_downscaling_2019, sha_deep-learning-based_2020, cheng_generating_2020, stengel_adversarial_2020, wang_deep_2021, kumar_deep_2021, adewoyin_tru-net_2021}. More recent studies (e.g. \citealt{adewoyin_tru-net_2021}, \citealt{harris_generative_2022} and \citealt{price_increasing_2022}) have considered LR inputs that are synchronous with the HR target, but are not perfectly matched because they come from different sources -- i.e observations as HR targets with LR reanalyses or forecasts as LR inputs. Systematic biases can exist between the HR and LR fields because of their different sources.  Such biases can in principle be addressed by bias corrections.  However, since convection-permitting models will develop internal variability that differs from that of the LR driving model, mismatches may occur on shared scales which cannot be remediated by bias correction techniques \citep{lucas-picher_investigation_2008}. One of the goals in SR for statistical downscaling is to mimic the internal variability of the convection-permitting model, rather than match HR features exactly with the GAN approach.  Because of the desire to develop a tool to sample realizations of HR fields conditional on LR fields, the ability to model this internal variability is a strength of our approach. 

In computer vision, the process of obtaining non-idealized LR/HR pairs for training poses significant challenges, which in turn hinders the generalization capabilities of state-of-the-art SR methods when applied to real-world images. These methods are trained using idealized inputs that do not exhibit issues commonly faced in photography such as aliasing effects, sensor noise, and compression artifacts. Consequently, when applied to real-world scenarios, these SR models tend to produce high-frequency artifacts and distortions, as documented in studies like \citet{shocher_zero-shot_2018, fritsche_frequency_2019}. To combat this, \cite{fritsche_frequency_2019} design a training method that synthesizes non-idealized LR images from HR input images -- essentially developing a training set that contains non-idealized LR and HR pairs. This approach improved the ability of the networks to generalize to real-world (imperfect) data. Interestingly, while non-idealized training image pairs are difficult to come by in computer vision, the analogous configuration for climate fields is more readily available because of the role of internal variability.

For forecasting and observational datasets, \cite{price_increasing_2022} recognizes that systematic error and biases may contribute prominently to shared-scale mismatches for their data. To deal with this challenge, they train the networks to correct the LR input fields to match the HR domain beforehand by including explicit ``correction'' layers. In our work, we show how idealized vs. non-idealized LR/HR pairs influence the generated fields but do not include additional correction techniques; as demonstrated through analyses of model biases, our mismatches are predominantly caused by random internal variability rather than systematic errors.

Additionally, studies have either considered mapping between  LR and HR fields of the same physical quantity (e.g. \citealt{singh_downscaling_2019, stengel_adversarial_2020, leinonen_stochastic_2020}) or have provided additional input information in the form of LR climate variables or information about the model surface (e.g. \citealt{price_increasing_2022, harris_generative_2022}). However, to our knowledge, the value of including additional covariates has not yet been explicitly addressed. We include additional LR covariate fields and design experiments to measure their influence over the generated fields.

\section{Methods and Data}
\label{section:training}
\subsection{Objective functions}
 We adopt the Wasserstein GAN with Gradient Penalty (WGAN-GP) from \cite{arjovsky_wasserstein_2017} for our SR models that use a Critic, $C$, network to estimate the Wasserstein distance between the generated and target distributions ($\PX_g$ and $\PX_r$ respectively). Using WGAN-GP, we implement super-resolution GAN (SRGAN) networks from \cite{ledig_photo-realistic_2017}. More details on the networks we use are provided in Section 3\ref{section:model_training}. 
 
 For the Critic's objective function, we adopt that of \cite{gulrajani_improved_2017}. For the Generator, $G$, we use the following:

\begin{equation}\label{eq:generator_sr}
\mathcal{L}_G =  \underbrace{ - \EX_{G(\textbf{x}) \sim \PX_g} [C(G(\textbf{x}))]}_{\textnormal{Adversarial component/loss}} + \underbrace{\alpha\EX_{\textbf{y} \sim \PX_r} l_c(\textbf{y}, G(\textbf{x}))}_{\textnormal{Content loss}}
\end{equation}

\noindent where \textbf{x} are LR covariates, \textbf{y} are ``true'' HR fields, and $\alpha$ is a hyperparameter that weights the relative importance of the content loss, $l_c$, and the adversarial loss. The content loss is meant to guide the generated fields towards $\textbf{y}$ -- it is desirable that generated and true realizations agree on larger common scales -- and is necessary for stability while training the SR models. This present work uses the grid-based mean absolute error (MAE) for the content loss.

GANs enable the SR methods to minimize both distributional distances (i.e. convergence of distributions) and grid-point-based metrics (i.e. convergence of realizations) in the training process. When using grid-point-based (or pixel-wise) metrics, such as mean squared error (MSE) or MAE, enforcing strict adherence to pixel-wise errors penalizes physically realizable -- but non-congruent (i.e. mismatched) -- high-frequency patterns. This problem is similar to the limitations of grid-point-based error measures for precipitation fields, known as the double-penalty problem \citep{rossa_overview_2008, michaelides_precipitation_2008, harris_generative_2022}. Using a distributional distance in the objective function helps to mitigate the double-penalty problem. Further details on the GAN implementation can be found in the supplemental material.

\subsection{Meteorological Datasets}

We develop deterministic WGAN-GP SR models that generate 10 m wind component fields (respectively $u10$ and $v10$ for the zonal and meridional components) using simulations by the Weather Research and Forecasting (WRF) model over subregions in the High-Resolution Contiguous United States (HRCONUS) domain \citep{rasmussen_high_2017, liu_continental-scale_2017} as training data. The WRF HRCONUS simulations are at a convection-permitting resolution (4 km grid spacing) and are driven using 6-hourly ERA-Interim (80 km grid spacing) reanalysis output \citep{dee_era-interim_2011} during the historical period from October 2000 to September 2013 ($18991$ 6-hourly fields). Using GANs for SR, we aim to generate HR fields -- conditioned on LR reanalysis fields -- that are consistent with WRF HRCONUS, effectively emulating the simulated HR WRF wind fields.
 
Through its boundary forcing, and spectral nudging at large scales above the boundary layer, WRF HRCONUS is synchronous in time with ERA-Interim \citep{rasmussen_high_2017}. This synchronization creates reasonable agreement at large scales between ERA-Interim and WRF HRCONUS. However, due to upscale energy transfers, they may not match exactly since smaller scales can evolve freely as a result of WRF's internal variability. This non-idealized pairing places more responsibility on the GAN to correctly produce details consistent with the convection-permitting model, thereby testing the extent to which the Critic captures this internal variability and enables the Generator to produce them.

WRF HRCONUS outputs are not provided on a regular latitude/longitude grid. So, to begin, WRF HRCONUS is re-gridded to a regular grid through nearest neighbour interpolation. While the native WRF HRCONUS grid spacing is $\sim$4 km, WRF HRCONUS is re-gridded to $\sim$10 km, resulting in a scale factor of eight with respect to ERA-Interim's 80 km grid spacing. Nearest neighbour interpolation is intentionally used to limit unintended smoothing by other methods, such as bilinear interpolation. 

In addition to the LR $u10$ and $v10$ ERA-Interim fields, five additional LR fields from reanalysis products are used as covariates. The additional covariates and motivations for their use are:
\begin{itemize}
    \item  Convective Available Potential Energy (CAPE) is selected for its influence on wind conditions in convective systems;
    \item topography is a coarse digital elevation map, selected for its role in influencing wind speed and direction;
    \item land-sea-fraction indicates the ocean-to-land fraction of a coarse grid and influences wind patterns around coastlines;
    \item surface roughness length determines the (generally heterogeneous) strength of surface drag on the flow; and
    \item surface pressure plays a role in the surface wind momentum budget through the pressure gradient.
\end{itemize}

These are provided to the modified Generator as extra channels. Due to the unavailability of CAPE in ERA-Interim at six-hourly time steps, CAPE from ERA5 \citep{hersbach_era5_2020}, interpolated from the native 30 km grid spacing to 80 km, is used instead. ERA5 and ERA-Interim represent the same historical atmospheric conditions and so it is assumed that any mismatches introduced are small between both WRF and ERA5 as well as between ERA-Interim and ERA5.

Within the WRF HRCONUS domain, SR models are developed for three subregions with different climatological conditions:
\begin{itemize}
  \item the Western region, which covers southern British Columbia, Washington State, and Oregon, is characterized by complex topography that includes mountainous terrain and complex shorelines;
  \item the Central region, which covers North and South Dakota, as well as Minnesota and northern Iowa, southern Manitoba and the southwestern part of Ontario, has a continental climate with large lakes and relatively frequent mesoscale convective features; and
  \item the Southeast region, which includes Florida, Cuba, and adjacent waters, is subject to tropical cyclones and frequent mesoscale convective features.
\end{itemize}

\begin{figure*}
\noindent\includegraphics[width=\textwidth]{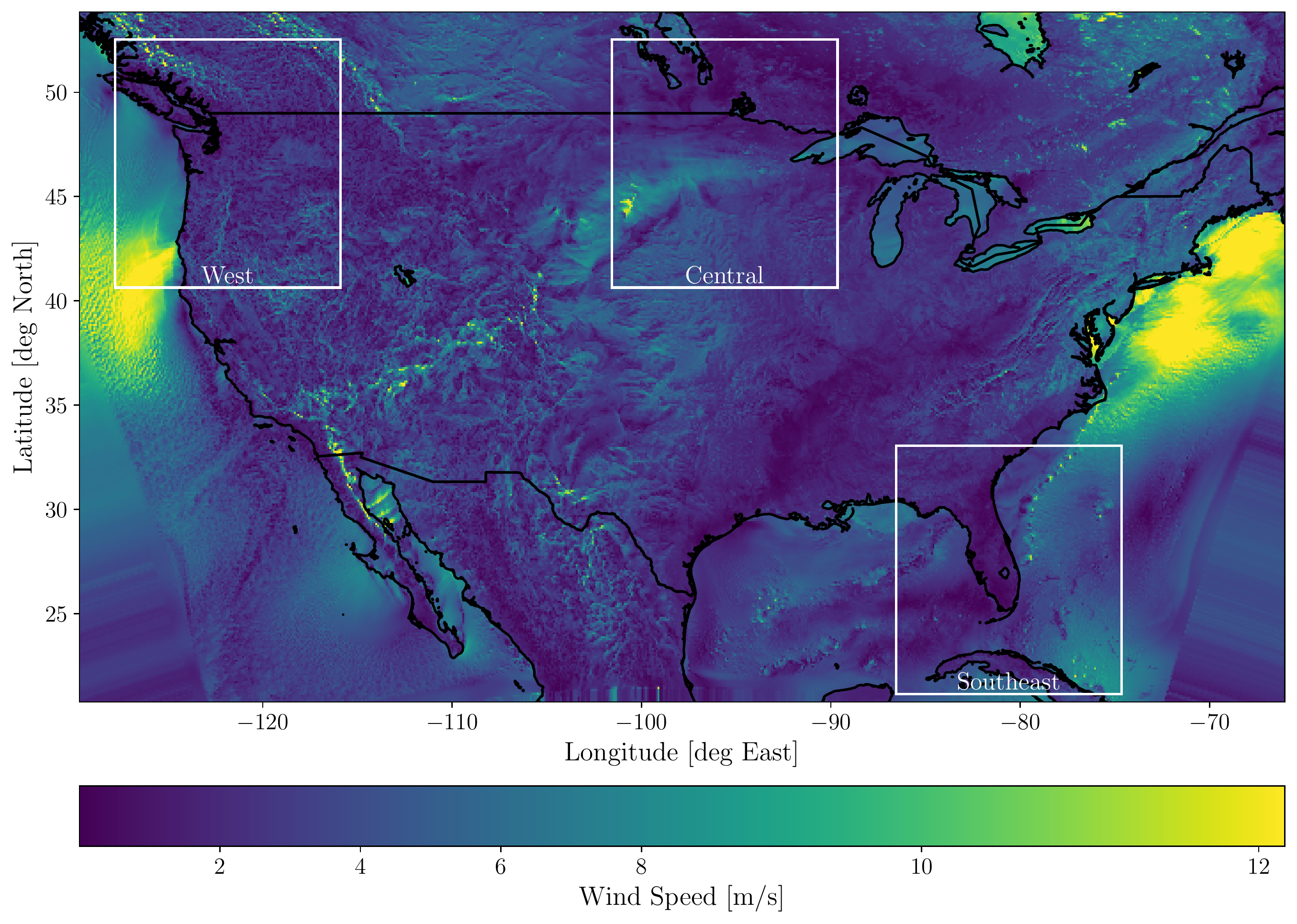}
\caption{Sample wind speed map of WRF HRCONUS for September 28, 2002, at 12:00:00 UTC. Regions selected for this work are in white boxes with their respective names. The southwest and southeast corners of this figure show edge effects caused by the nearest-neighbour interpolation to a regular latitude/longitude grid.}
\label{fig:fig1}
\end{figure*}

\noindent Figure \ref{fig:fig1} shows a representative instance of the wind speed field over the WRF HRCONUS domain. Each of the three subregions contains 16$\times$16 LR grid points and 128$\times$128 HR grid cells.

\subsection{Experiment 1: Frequency Separation}
\label{section:exp1}
In deterministic GAN SR, the objective of the MAE content loss in the Generator's objective function is to produce single realizations that look like the conditional median, which drives the outputs to appear smooth.  Simultaneously, the goal of the adversarial loss is to ensure that single realizations are drawn from the entire distribution of possible realizations instead of just the conditional median, and so it encourages generating possible arrangements of fine-scale features. A challenge with deterministic GAN SR is that the content loss compares fine-scale features from different realizations of the generated and ``true'' fields while the adversarial loss compares the distributions of the generated and ``true'' fields using the Critic. It follows that the content/adversarial loss can be viewed as implicitly oppositional (not adversarial!) because differences in physically realizable fine-scale features (which are made possible by the adversarial loss) are penalized by the content loss. While training GAN SR models, one can view the two terms as existing in ``tension'' with one another.

In their work, \cite{fritsche_frequency_2019} recognized that this tension in SR tasks can be addressed by delegating spatial frequencies in the images to select terms in the objective functions. The resulting approach, called frequency separation (FS), separates the spatial frequencies of the HR fields into high and low-frequency pairs, applying adversarial loss and content loss to each frequency range, respectively.

The concept behind FS is to use the Generator's MAE content loss to encourage realization convergence at low frequencies in the fields, rather than across the entire range of image frequencies as in typical SR configurations. As we have discussed, encouraging high-frequency realization convergence is not always appropriate for images or weather and climate fields as high-resolution features are not determined uniquely by low-resolution ones. In FS, high frequencies are isolated and provided to the Generator's adversarial loss (i.e. the Critic) which strives for distributional convergence between training and generated data, rather than individual realizations. This approach was found by \cite{fritsche_frequency_2019} to yield perceptually improved results with images and is considered in this study to evaluate its effectiveness for wind fields.

\begin{figure*}
    \noindent\includegraphics[width=\textwidth]{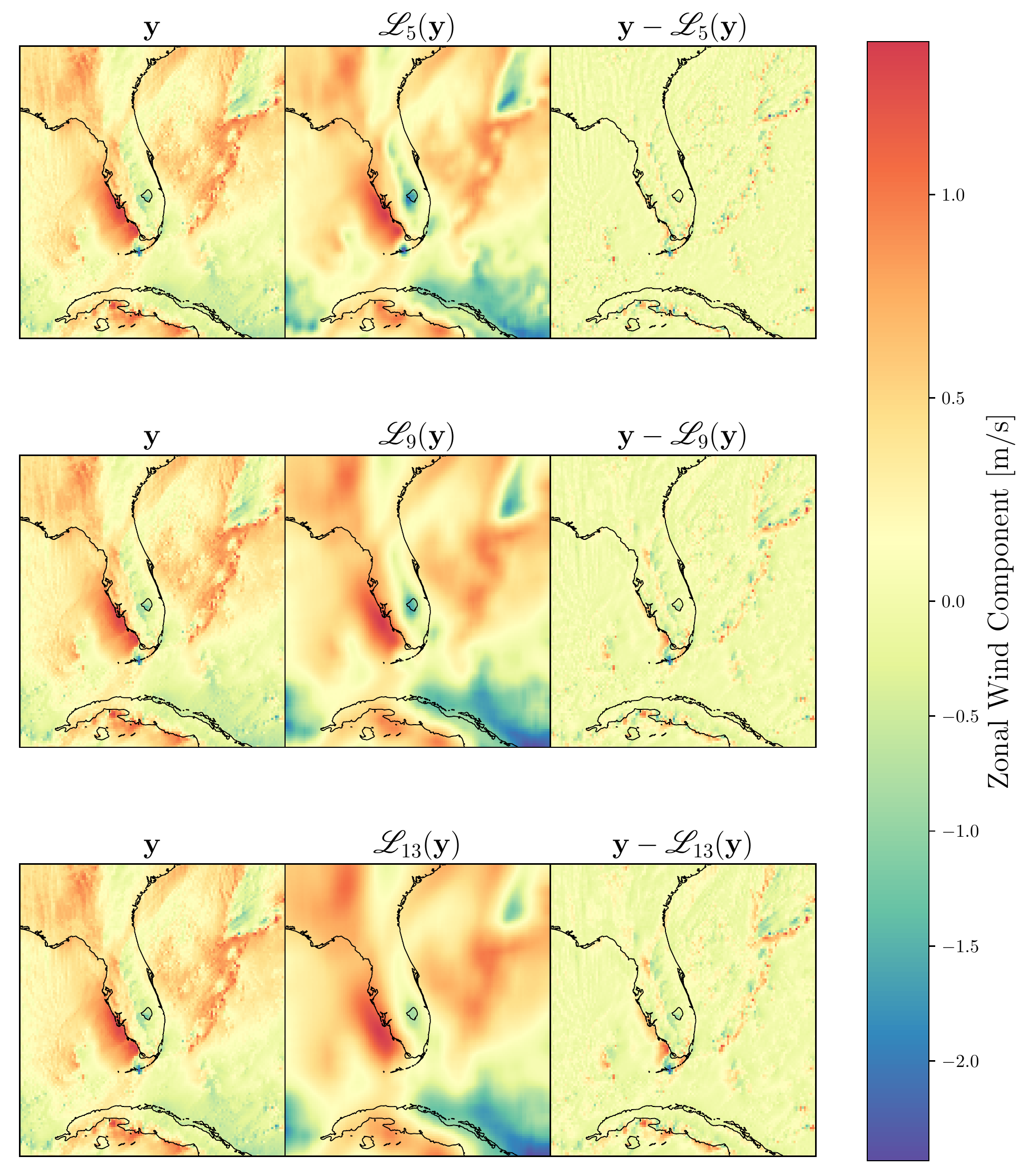}
    \caption{Various examples of frequency separation applied to a $u10$ field in the Southeast region. Each row corresponds to  different values of the kernel size $N$ for the low pass filter $\mathscr{L}_N$. The left panel is the raw input, the center panel has an average 2D pool filter applied to the input, and the right panel shows the high-frequency field determined by the difference between the left and center panels.}
    \label{fig:fig2}
\end{figure*}

Following the methodology proposed by \cite{fritsche_frequency_2019}, we apply a low-pass filter to the HR fields using a 2D spatial averaging kernel denoted as $\mathscr{L}_N$. This kernel functions as a convolution operation with a square filter of size $N \times N$, where each weight is set as $1/N^2$. We explore various values of $N$, as summarized in Table \ref{table:table1}, where smaller values allow higher frequency information to pass through the filter as shown in Figure \ref{fig:fig2}. The high frequencies can be found using the following equations:

\begin{equation}\label{eq:freq_sep_high}
\begin{split}
    \textbf{y}_h & = \textbf{y} - \mathscr{L}(\textbf{y}) \\
    G(\textbf{x})_h & = G(\textbf{x}) - \mathscr{L}(G(\textbf{x}))
\end{split}
\end{equation}

\noindent It is important to note that by separating these frequencies, the Wasserstein distance is estimated on a different probability distribution containing only high frequencies. Therefore, the Wasserstein distance cannot be directly compared between FS models for different values of $N$. We collectively refer to GANs trained using FS as FS GANs.
For a more detailed implementation description of frequency separation, refer to the supplemental material.

\subsection{Experiment 2: Partial Frequency Separation}
\label{section:exp2}

There is an interesting -- yet potentially subtle -- difference in how SR objective functions can be used for stochastic vs. deterministic SR. This difference motivates a second experiment we conduct on our deterministic GANs related to FS, which we call ``Partial'' Frequency Separation. If multiple realizations are generated -- as is done in stochastic SR -- the fine scales of the ``true'' fields can be compared to the ensemble median of the generated realizations using the content loss. As the resulting ensemble median tends to suppress fine-scale features, differences on common scales would be more strongly penalized than differences in fine-scale features between the individual realizations. Furthermore, generated individual realizations would not be encouraged to look like the conditional median. Just like in regular FS, penalizing differences on common scales is a desirable outcome of applying the content loss. Such an approach is taken by \cite{harris_array_2020} for stochastic SR precipitation fields.

For deterministic GAN SR, while we sample from the distribution of HR fields conditioned on the LR fields, for a given network the same sample is always drawn for the same conditioning fields.  To mimic the approach of \cite{harris_array_2020} in a deterministic setting, low-frequency filters can be applied to suppress fine-scale features in the HR fields instead of computing the ensemble median from several realizations. As done in the FS GANs, the content loss from stochastic SR can be mimicked in deterministic GAN SR by delegating low frequencies to the MAE so that common scales are more strongly penalized. However, unlike the FS GANs, in partial FS the adversarial loss is applied to all frequencies instead of just the high frequencies. We emphasize that we are mimicking stochastic SR because simply applying a low-frequency filter to an HR field is not the same as estimating the actual ensemble median. The practical benefits of partial FS are that it can significantly save GPU memory requirements (and training time) by not requiring the Critic network to evaluate several ensemble members.

\subsection{Experiment 3: Low-resolution Covariates}
As a separate analysis to the frequency separation experiments, we narrow our focus on the LR covariates to investigate their impact on the performance of the GANs considered in this study. We particularly focus on how the LR covariates influence spatial frequency structure. This analysis is organized into two streams that explore (1) how differences between ERA-Interim and WRF HRCONUS may influence GAN performance, and (2) what role the additional physically relevant covariates play in generating spatial structures. The details are described below with results presented in Section \ref{section:results}.

\subsubsection{Idealized Covariates}
Two additional non-FS GANs are trained using only $u10$ and $v10$ fields as LR covariates. One GAN is conditioned with ERA-Interim, without the additional covariates, and the other uses artificially coarsened (by a scale factor of eight) WRF HRCONUS HR wind components. The idealized pairing of  original HR and coarsened WRF  fields emulates approaches common to both the computer vision and climate literature \citep{ledig_photo-realistic_2017, singh_downscaling_2019, sha_deep-learning-based_2020,  leinonen_stochastic_2020, stengel_adversarial_2020, kumar_deep_2021, wang_deep_2021}. 

\subsubsection{Additional Covariates}

As a further analysis, we compare the spectra of fields produced by the non-FS GANs with all seven covariate fields  to the spectra produced by the GAN with only ERA $u10$ and $v10$. This is a simple experiment intended to evaluate the collective effect of including these additional covariates, however, does not illuminate the importance of individual covariates.

To explore how sensitive SR models are to individual covariates, an experiment is devised to randomly shuffle individual covariate fields of the already-trained non-FS GAN, and measure across wavenumbers the resulting changes in the spectra. The relative difference, $RD$, between the power spectra of the modified, $P_s(\mathbf{k})$, and unmodified baseline, $P_b(\mathbf{k})$ is quantified, and the resulting variance at each wavenumber for each perturbed covariate is computed. The above approach is known as singe-pass \textit{permutation importance} and is a common feature importance experiment \citep{MakingtheBlackBoxMoreTransparentUnderstandingthePhysicalImplicationsofMachineLearning}. Further details about our implementation are provided in the supplemental material.

\subsection{Model Training}
\label{section:model_training}
\begin{figure*}
    \noindent\includegraphics[width=\textwidth]{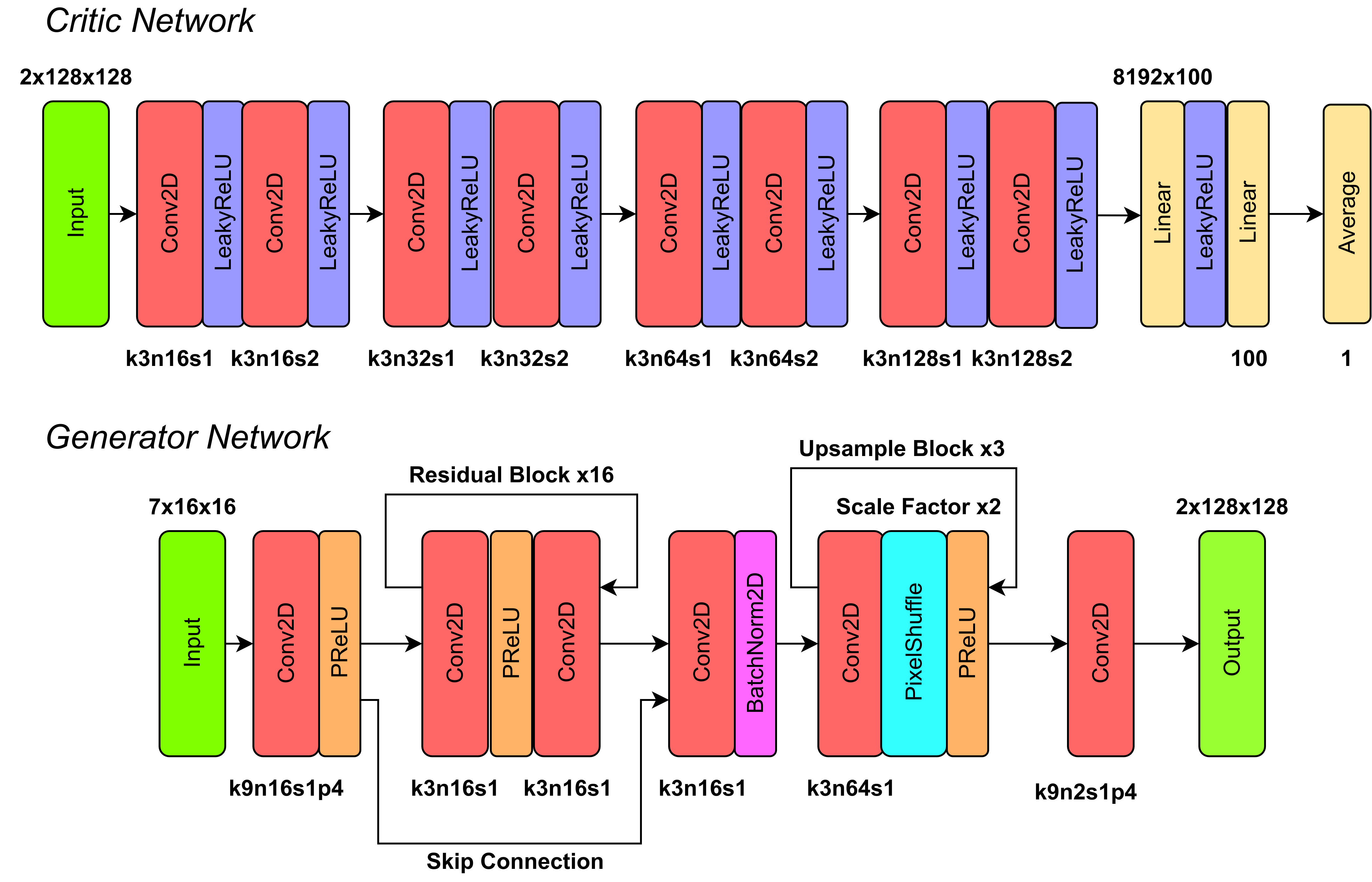}
    \caption{Schematic of the networks used with WRF HRCONUS (Critic) and ERA (Generator) with WGAN-GP. The inputs to each network are two-dimensional fields. The label for each Conv2D layer describes its function. $k$ indicates the size of the square kernel/filter used, e.g. a kernel size of 3 has 9 trainable parameters (plus one if a bias parameter is fit for that particular filter). $n$ represents the total number of filters in the layer, and $s$ is the stride at which it is applied to the input. Where relevant, the shape of matrices/vectors is included on top of the input and linear layers to indicate the dimensions of the data representations at these stages. Due to GPU memory constraints, in order to maintain a batch size of 64, the number of linear layer weights is reduced from 1024 in the original SRGAN implementation to 100 in this implementation. We also note that the linear layer in our Critic requires fixed input dimensions $2 \times 128 \times 128$ as it is not fully convolutional. The Generator is very similar to the SRGAN Generator but is adjusted to accept additional covariates and an additional upsampling block that downscales the fields by a scale factor of eight. The supporting software and data used in this work can be found in the Data availability statement.}
    \label{fig:networks}
\end{figure*}

The Critic network is adopted from the SRGAN Discriminator of \cite{ledig_photo-realistic_2017}, but without batch normalization layers \citep{gulrajani_improved_2017} for compatibility with WGAN-GP. We use a Generator network similar to SRGAN, but with additional LR inputs and one additional upsampling block. Network details are shown in Figure \ref{fig:networks}.

The models are trained using a single NVIDIA GTX 1060 GPU with 6 GB of VRAM. The Adam optimizer, a form of stochastic gradient descent \citep{kingma_adam_2017}, is used to train the models. Of the $18991$ fields in the 2000-2013 WRF HRCONUS simulation, 80\% are used for training (15704 fields), and 20\% (3287 fields comprising years 2000, 2006, and 2010) are used for testing and evaluation. The network parameters are not updated using any data from the years 2000, 2006, or 2010 test set. We would like to emphasize that most modelling decisions have been made prior to training by adopting hyperparameter and model choices from existing work. As such, we do not specify a separate validation set since we do not perform hyperparameter tuning. Instead, we focus on performing sensitivity analyses with our experimental configurations.

Each GAN takes approximately 48 hours to complete 1000 passes (i.e., epochs) through the entire training set. Hyperparameter values, which are summarized in Table \ref{table:table1}, are mostly taken directly from those recommended in the existing literature (e.g. \cite{gulrajani_improved_2017}). All results are produced by models after reaching the full 1000 training epochs.

While rescaling $\mathcal{L}_G$ by a constant does not affect its optimization, the relative magnitudes of the adversarial component and the content loss are important. For the FS GANs the values of $\alpha$, which controls this weighting, are selected such that the content loss and adversarial loss are of roughly the same magnitude. Although we do not seek optimal values of $\alpha$, the values used here (summarized in Table \ref{table:table1}) result in stable training. The partial FS GANs were trained using two different values of $\alpha$ to investigate the sensitivity of generated fields to this parameter. 

For each of the regions, three different values of $N$ are used for $\mathscr{L}_N$ to assess the effect of various amounts of spatial smoothing on the FS results. Additionally, a non-frequency separation GAN (non-FS GAN) and pure CNN are also trained for each region. The non-FS GAN provides all frequencies to both the content and adversarial loss term in Equation \ref{eq:generator_sr}. The pure CNN is configured identically to the non-FS GAN but with the adversarial loss excluded from its objective function. Consideration of the pure CNN tests the role of the adversarial loss on the generated wind fields and can be viewed as a limit case of FS whereby all frequencies are delegated to the content loss, and no frequencies are delegated to the adversarial loss.

\begin{table*}
\centering
    \caption{Summary of hyperparameters. 
    Following \cite{gulrajani_improved_2017}, we define one GAN epoch as updating the Critic once every mini-batch (while the Generator is updated every fifth mini-batch), but define one pure CNN epoch as updating the Generator once every mini-batch.}
    \label{table:table1}
\resizebox{\textwidth}{!}{
    \begin{tabular}{@{\qquad}ccccccc@{\qquad}ccc@{\qquad}ccc}
      \toprule
      \multicolumn{7}{c}{\textbf{Training Hyperparameters}} & \multicolumn{3}{c}{\textbf{Adam Optimizer}} & \multicolumn{3}{c}{\textbf{Frequency Separation Avg2DPool}} \\
      & $\lambda$ & $\alpha$ & & Epochs & Batch Size & Critic iterations & Learning rate & $\beta_1$ & $\beta_2$ & Filter size ($N$) & Stride & Reflection Padding \\
      \midrule
      FS GANs & 10.0 & 500 & & 1000 & 64 & 5 & $1\times10^{-5}$ & 0.9 & 0.99 & 5, 9, 13 & 1 & 2, 4, 6 \\
      \midrule
      partial FS GANs & 10.0 & 50, 500 & & 1000 & 64 & 5 & $1\times10^{-5}$ & 0.9 & 0.99 & 5, 9, 13 & 1 & \textbf{--} \\
      \midrule
      non-FS GANs & 10.0 & 500 & & 1000 & 64 & 5 & $1\times10^{-5}$ & 0.9 & 0.99 & \textbf{--} & \textbf{--} & \textbf{--} \\
      \midrule
      Pure CNNs & 10.0 & 1 & & 1000 & 64 & 5 & $1\times10^{-5}$ & 0.9 & 0.99 & \textbf{--} & \textbf{--} & \textbf{--} \\

      \bottomrule
    \end{tabular}
    }
\end{table*}

\section{Results}
\label{section:results}

\subsection{Experiment 1: Frequency Separation}

\subsubsection{Visual Quality of Generated Fields}
\begin{figure*}
    \noindent\includegraphics[width=\textwidth]{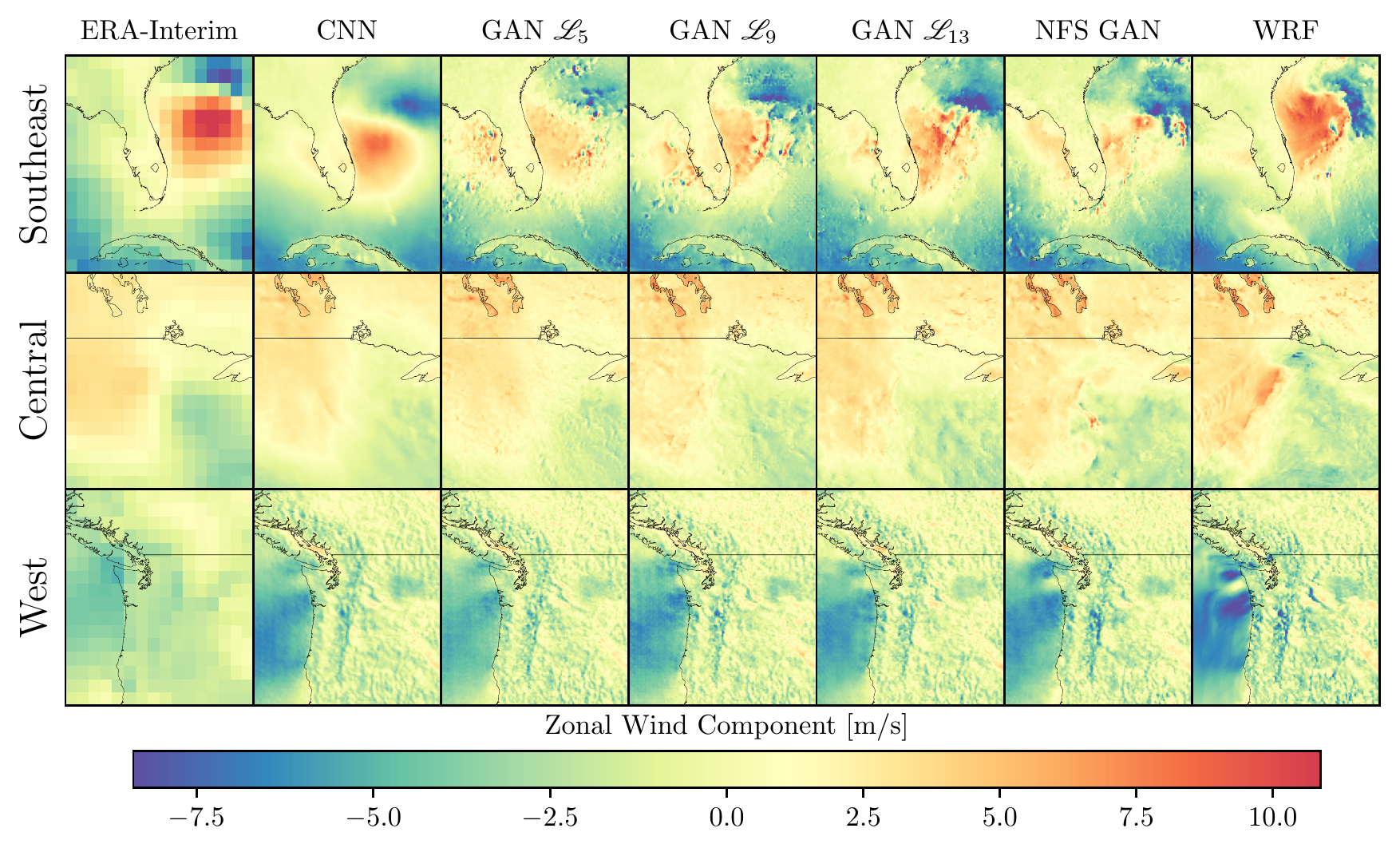}
    \caption{Example fields of testing data and generated $u10$. Each field corresponds to the same time step, on October 5th, 2000 at 12:00:00 UTC. The top row shows the Southeast region, the middle row is the Central region, and the bottom row is the West region. From left to right: the $u10$ field from LR ERA-Interim, the pure CNN, FS GAN with a $\mathscr{L}_N$ kernel, non-FS GAN, followed by WRF HRCONUS. For this particular time-step, Tropical Storm Leslie (a subtropical cyclone forming over eastern Florida) is prominent in the Southeast region; inland features (such as lakes) are seen to influence the wind patterns in the Central region; and strong land-sea contrast is shown in the West region with clear evidence of the influence of topography on the flow. Importantly, these realizations demonstrate that despite each GAN being conditioned on the same coarse ERA-Interim fields, they produce fine-scale structures at different locations than the WRF HRCONUS, and are placing fine-scale weather where it could be, but not necessarily exactly where it is in the WRF HRCONUS fields.}
    \label{fig:fig4}
\end{figure*}

\begin{figure*}
    \noindent\includegraphics[width=\textwidth]{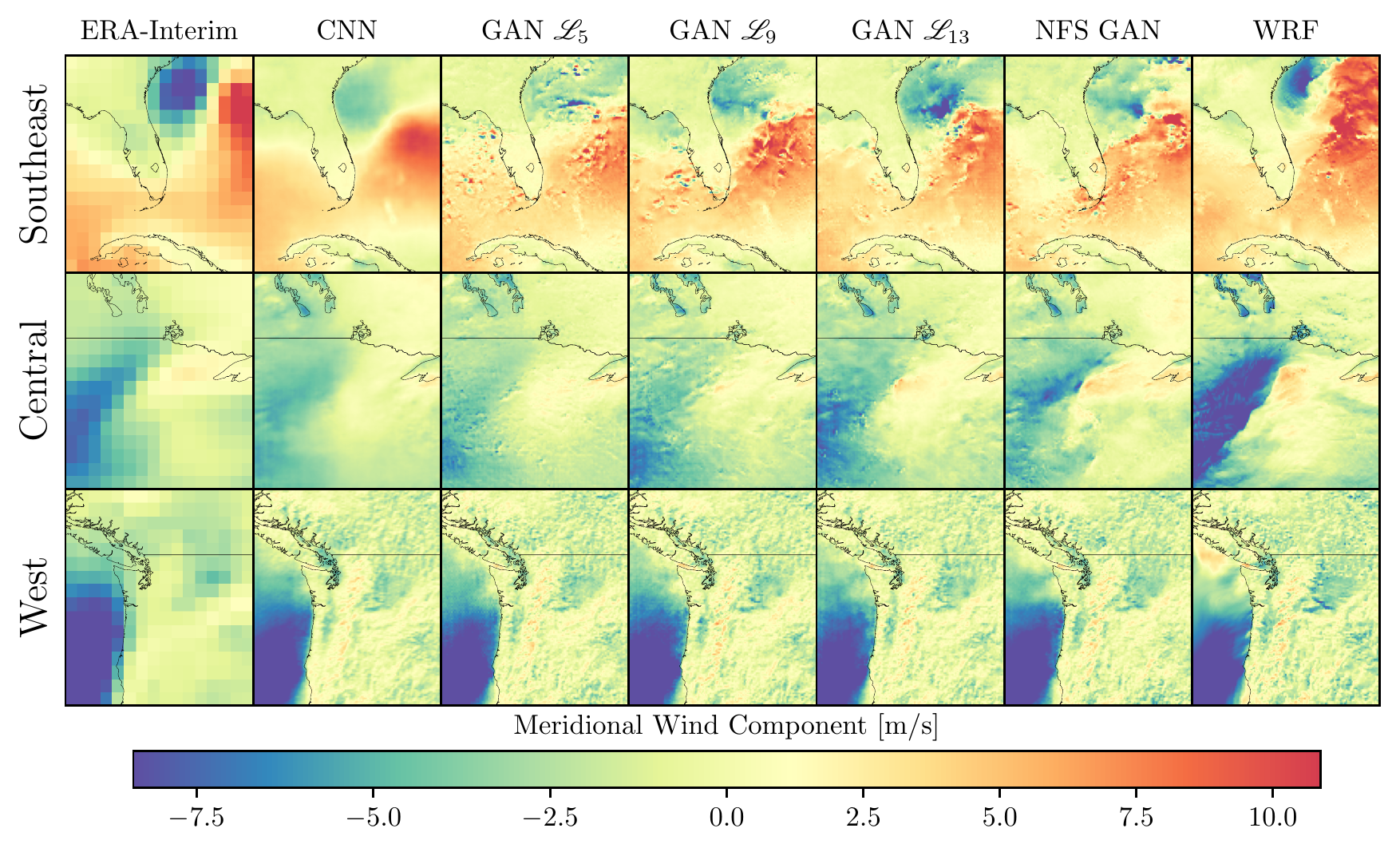}
    \caption{As in Figure \ref{fig:fig4} but for $v10$.}
    \label{fig:fig5}
\end{figure*}

A representative set of the $u10$ and $v10$ wind field maps produced using the SR models on October 5th, 2000 at 12:00 UTC is shown in Figure \ref{fig:fig4} and \ref{fig:fig5} respectively. While there are broad consistencies at large scales between WRF HRCONUS and ERA-Interim, differences in the locations of certain structures are present in the fields, illustrating the non-idealized nature of the pairing and the internal variability of WRF. For example, the negative $u10$ wind feature in the northeast part of the Southeast region is oriented slightly differently in WRF HRCONUS and ERA-Interim. 

Fields produced by the CNN do not contain the fine-scale variability seen in the WRF HRCONUS field for the Southeast and Central region. This fact is most obvious in the Southeast region where fine-scale convective features are not produced by the pure CNN and the fields are too smooth. When comparing individual realizations, we cannot conclude that the lack of details is worse (given that ``smooth'' realizations may be physically realizable). However, this ``smoothness'' is observed systematically over several realizations (Figure S1 - S6) and over each region demonstrating that the CNN is limited in the spatial structures it can produce. The objective function based on content loss alone does not allow the CNN to ``hallucinate'' fine-scale features because it constrains realization pairs to be similar, rather than sampling from distributions as when the adversarial loss is included. This effect can also be observed in the Central region, although to a lesser extent. The West region shows generally good agreement between the CNN and WRF HRCONUS, in particular with the inland topographical features. However, the CNN for the West region is lacking some of the sharp and well-defined topographical details found in WRF HRCONUS.

GANs with and without FS show little perceptual difference for the West and Central regions; both show an improvement in fine spatial structure over the pure CNN. In the GAN with $\mathscr{L}_5$ FS for the Southeast region, there is a noticeable reduction in power in the medium and low spatial frequencies surrounding organized spatial structures in the northeast part of the domain (e.g., $u10$ in Figure \ref{fig:fig4}, $v10$ in Figure \ref{fig:fig5}). This effect can be seen as isolated fine-scale features in the $\mathscr{L}_5$ FS GANs when compared to WRF HRCONUS. GANs with $\mathscr{L}_9$, $\mathscr{L}_{13}$, and non-FS show little perceptual difference in quality in the realizations. The increase in the perceptual quality of the GAN-generated fields can be attributed to the Critic's ability to sample from the distribution of fine-scale features. A larger set of example patterns exhibiting similar features to those discussed above are presented in Figures S1 - S6.

\subsubsection{Evolution of Performance Metrics While Training}
Several metrics were recorded during the training of the SR models. Among them are the MAE, MSE, Multi-scale Structural Similarity Index (MS-SSIM), and the Wasserstein distance. MS-SSIM is a metric comparing images across multiple spatial scales taken from the computer vision field,  designed to correlate well with perceived image quality in image reconstruction tasks.

\begin{figure*}
    \noindent\includegraphics[width=\textwidth]{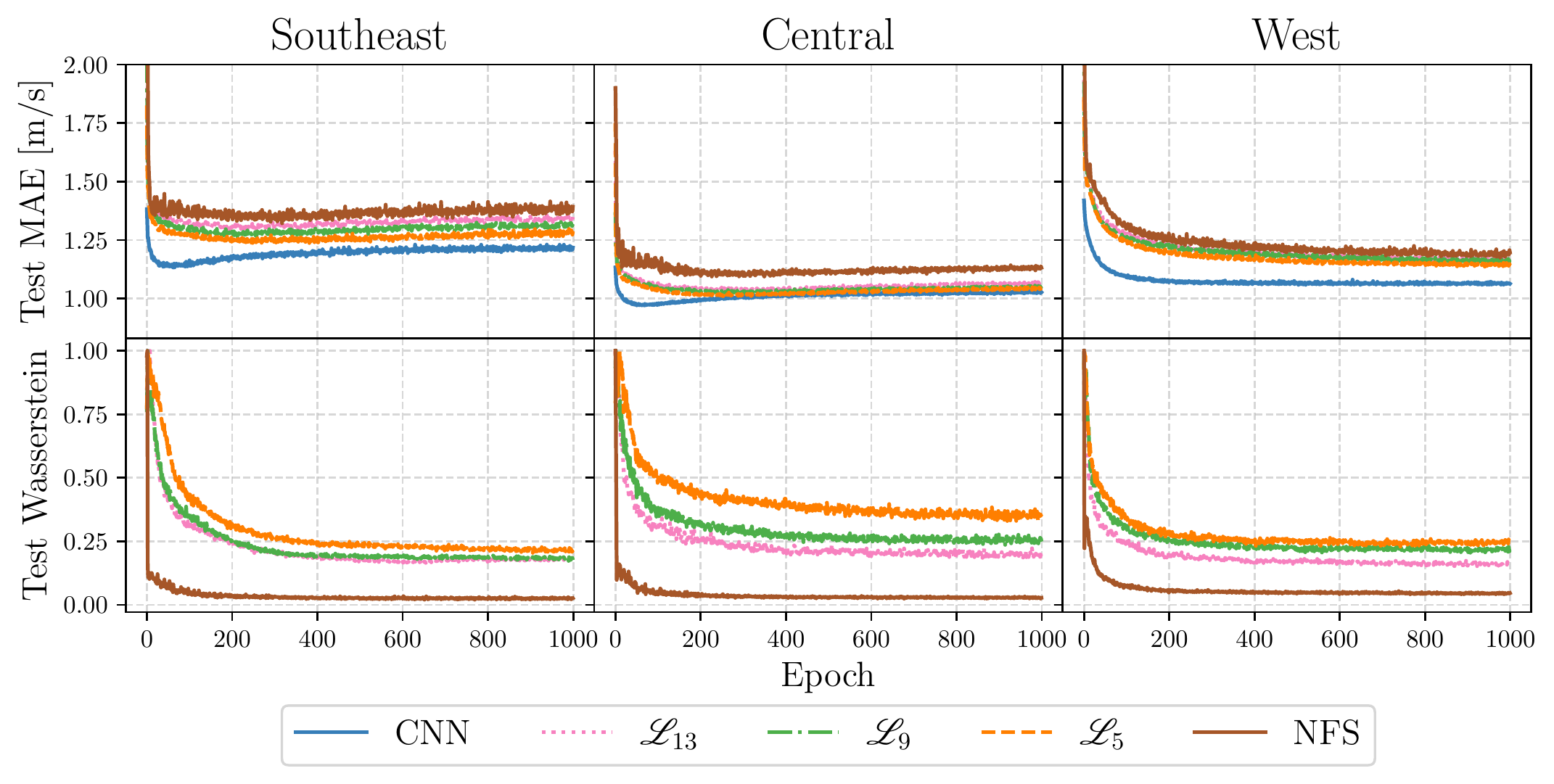}
    \caption{Evolution of the MAE (top row) and the Wasserstein distance (bottom row) for 1000 epochs on the test data. Each SR model is indicated in the legend. The MAE is calculated pixel-wise for all pixels combined from the $u10$ and $v10$ HR fields. Individual curves in the Wasserstein distance evolution cannot be compared, since applying FS changes the distributions to which the Wasserstein distance estimation is applied. Note that no Wasserstein distance is calculated for the pure CNN since the adversarial loss is not used in this model. The Wasserstein distance estimate is scaled by the initial distance as its shape deserves attention, not the actual values.}
    \label{fig:fig6}
\end{figure*}

Figure \ref{fig:fig6} shows the evolution of the MAE and the Wasserstein distance on the test sets during the training process while Figure S7 shows the training evolution on both the test and train sets for MAE, MSE, MS-SSIM, and the Wasserstein distance. While training, the MAE, MSE, and MS-SSIM are computed by comparing pairs of realizations in mini-batches, while the Wasserstein distance is approximated between the entire set of realizations in the mini-batches.

The MAE over the test data reaches a minimum value after $\sim$200 epochs for the Southeast and Central regions. For the pure CNNs, this minimum is more pronounced and occurs earlier than the GANs because the Generator is updated more frequently (see Table \ref{table:table1}). The presence of a local minimum in MAE is indicative of overfitting of the Generator in these two regions since no minimum is found in the evolution of MAE on the training set (Figure S7). At late epochs, the test set MAE does not grow substantially, so overfitting as measured by the MAE is minimal. Interestingly, evidence of overfitting is only present in the MAE/MSE, not the Wasserstein distance. No evidence of overfitting is found in the West region.  The slightly larger MAE at later epochs for the Southeast and Central regions may be indicative of the Generator learning large-scale differences between ERA-Interim and WRF HRCONUS in the training set only, while not generalizing to the test set. The topic of large-scale differences will be discussed in more detail later in this section. 

For the evolution of the MAE, MSE, and to some extent the MS-SSIM (Figure S7), there is a robust ordering of the performance of the different SR models across the regions. The best-performing model in the MAE sense is the pure CNN, followed by the $\mathscr{L}_5$, $\mathscr{L}_9$, and $\mathscr{L}_{13}$ FS GANs, and finally, the non-FS GAN shows the largest MAE (Figure \ref{fig:fig6}). This ordering can be explained by examining the role of the components of the Generator's objective function in Equation \ref{eq:generator_sr}. 

By construction, pure CNNs minimize the MAE and as such are expected to perform the best among all models with regard to this metric. The smoothness of the conditional median reduces the impact of the double-penalty problem on the generated fields.  The CNN performs similarly well in terms of MSE and MS-SSIM, both of which are also performance measures of realization convergence \citep{sampat_complex_2009}. For the FS GANs, when there is less smoothing, the optimization problem tends to be more similar to the pure CNNs since a large range of frequencies are delegated to the content loss. This explains why the $\mathscr{L}_5$ kernel has the lowest MAE, and why including any FS leads to a lower MAE than the non-FS GAN. When there is more smoothing, a larger range of frequencies are delegated to the adversarial loss, which allows for the generation of fine-scale weather across a larger range of frequencies which ends up increasing the effects of the double-penalty problem.  For the non-FS GAN, the adversarial loss makes use of the full range of frequency scales of the fields; it is not limited to evaluating the high-frequency distribution only. Non-FS GAN has more freedom to conditionally generate variability across frequency scales which increases the MAE because of the double-penalty problem.

\subsubsection{Radially Averaged Power Spectra} 
\begin{figure*}[t]
    \noindent
    \includegraphics[width=\textwidth]{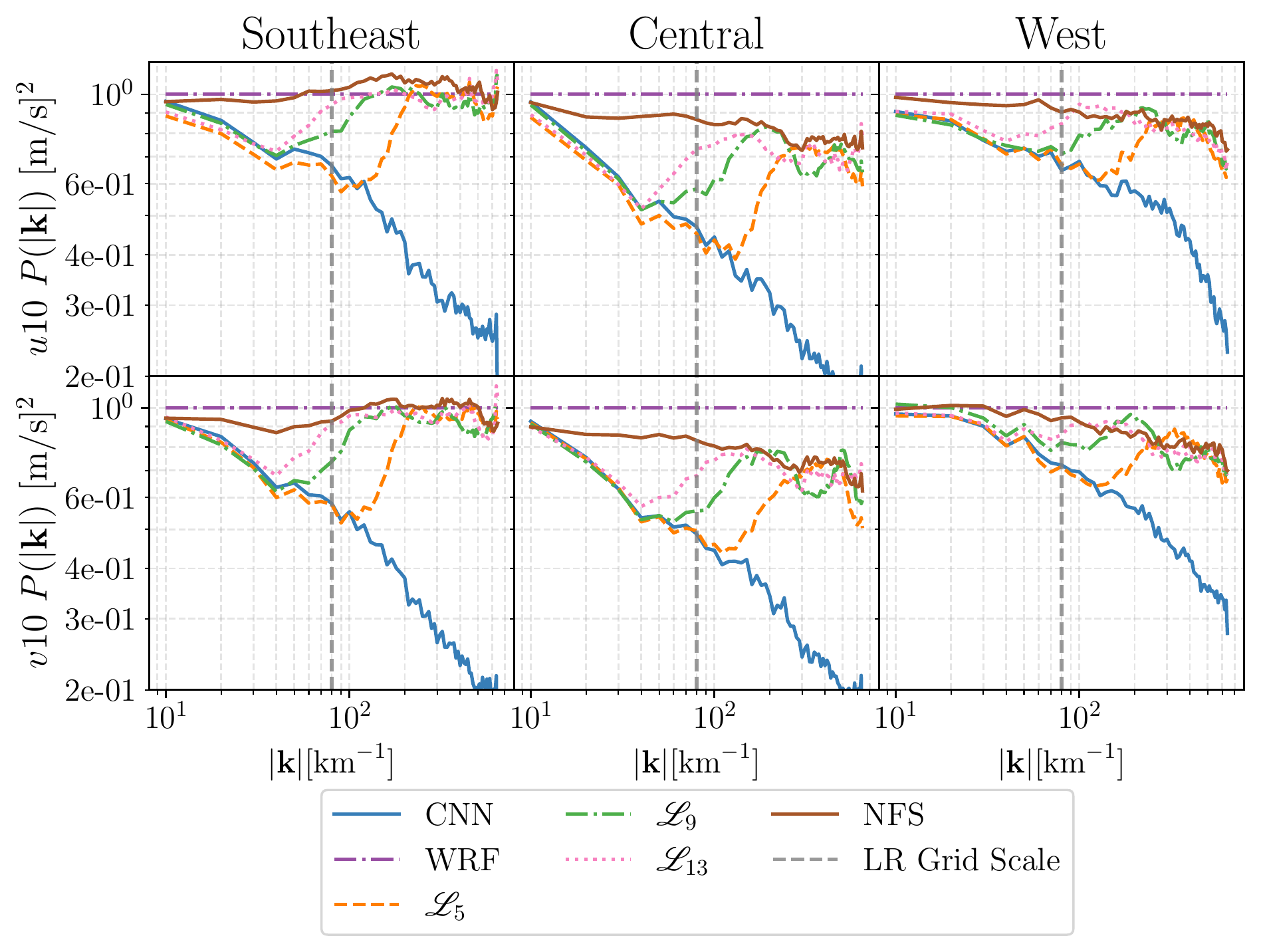}
    \caption{RAPSD ratio of the SR models relative to WRF HRCONUS represented as lines. The top row shows the $u10$ component, and the bottom row shows the $v10$ component. Each column is a separate region.}
    \label{fig:fig7}
\end{figure*}

Quantifying the perceptual quality of generated realizations using metrics that align with ``true'' realizations is challenging due to the double-penalty problem. Instead, we shift our focus to evaluating the statistical characteristics of the generated fields. Specifically, we analyze the spatial correlation structures of HR wind fields using power spectra, as suggested by previous studies \citep{singh_downscaling_2019, stengel_adversarial_2020, kashinath_physics-informed_2021}. Each wind component is assessed separately, and we calculate the radially-averaged power spectral density (RAPSD) following the naming convention in \cite{harris_generative_2022} after their application of RAPSD to SR precipitation fields. The SR to WRF HRCONUS RAPSD ratios are depicted in Figure \ref{fig:fig7}.

The RAPSD of the FS GANs shows how the results of the optimization problem change when the spatial frequencies are separated. The CNN shows strong low variance biases at fine scales, consistent with the over-smooth quality of the generated fields.  Each FS GAN shows similar power at large scales to the pure CNN (since the range of wavenumbers for both is optimized using the MAE) until the frequencies are separated, and the spectra break from the pure CNN, and join the non-FS GAN spectra at higher wavenumbers (since these wavenumbers were optimized using the adversarial component). The non-FS GAN spectra show that the SR models more accurately match WRF HRCONUS when the adversarial loss is provided with the full range of frequencies, despite the increase in pixel-wise errors when doing so.

\subsection{Experiment 2: Partial Frequency Separation}

\begin{figure*}[t]
    \noindent
    \includegraphics[width=\textwidth]{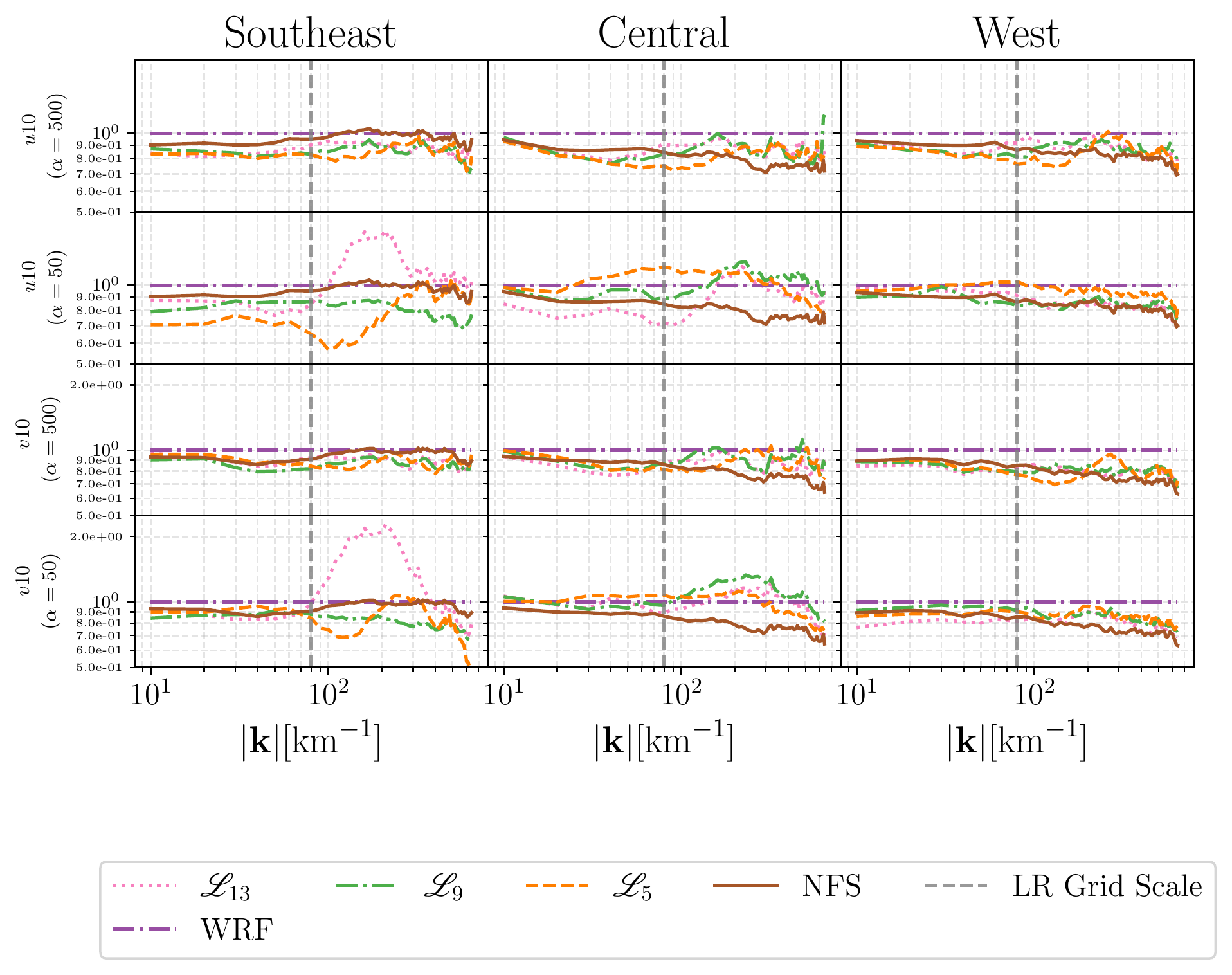}
    \caption{RAPSD ratio of the partial FS GAN SR models relative to WRF HRCONUS represented as lines. The top two rows show the $u10$ component, and the bottom two rows show the $v10$ component. Each column is a separate region.}
    \label{fig:fig8}
\end{figure*}

We build on the FS GAN result that the variability of the fields generated by SR models improves when the adversarial loss receives all frequencies, and consider  partial FS GAN. We also adjust the hyperparameter $\alpha$ -- the relative weight of the content and adversarial loss terms -- to examine its role in optimizing the variability of the generated fields. The resulting partial FS GAN RAPSD ratios, with $\alpha = 50$, and $\alpha = 500$, are presented in Figure \ref{fig:fig8}.

The partial FS GAN spectra are largely similar to the non-FS GAN spectra, however, there are more fluctuations in the RAPSD ratio for $\alpha = 50$. For example, $\mathscr{L}_{13}$ for the Southeast region shows an isolated high-power bias at intermediate to large wavenumbers (a similar isolated low-power bias is observed for $\mathscr{L}_5$). We hypothesize that the fluctuations in the RAPSD ratio for $\alpha = 50$ may come from artifacts in the generated fields due to the reduced importance of the content loss in the optimization. 

Increasing the value of $\alpha$ for the Central and West region reduces the variability in the RAPSD, and enhances the small-scale power. The results of the partial FS GANs, with different values of $\alpha$ and $N$, show that the RAPSD is strongly influenced by the role of the content loss in the optimization depending on the region of focus. 

\subsection{Comparing and Summarizing Model Performance}
We summarize model performance using a range of different metrics (Table \ref{table:table2}). The MAE and MS-SSIM are included, as well as biases in the mean, standard deviation, and 90th percentile of the wind speed for each region. The spatial maps of the biases in these statistics are reported in Figures S8 - S16 for the pure CNN, FS GANs, non-FS GAN, and partial FS GANs (for both values of $\alpha$). Wind speed bias is selected as a stringent test since low-variance biases in the wind components result in biases in the mean of the wind speed. Systematic low-variance biases are represented as negative spatial averages for these metrics -- consistent with the general low-power bias in the RAPSD.

Table \ref{table:table2} summarizes the result that the adversarial loss introduces fine-scale variability that contributes to the double-penalty problem.  Similar to \cite{harris_generative_2022}, we find the MS-SSIM not very useful for evaluating the generated fields. We hypothesize that the MS-SSIM may be more sensitive to noise and artifacts (common to images), rather than the potentially non-congruent fine-scale convective features of the wind patterns like those that contribute to the double-penalty problem. The spatial means of the wind speed biases show that GANs generally outperform the pure CNN, especially for the standard deviation and 90th percentile. This is not a surprising result; the GANs are introducing variability consistent with WRF into the generated fields and are better able to represent these climatological statistics. 

To compare the RAPSD of each SR model, we use the Median Symmetric Accuracy (MSA) metric \citep{morley_measures_2018}:

\begin{equation}
    \xi = 100 \left( \textnormal{exp} \left[M_{|\textbf{k}|} \left( \left| \log{ \left(\frac{P(|\textbf{k}|)}{P(|\textbf{k}|)_{WRF}}\right)}\right|\right)\right]-1\right)
\end{equation}

\noindent where $M_{|\textbf{k}|}$ represents the median of the test-set-averaged RAPSD computed over each wavenumber. Values of $\xi$ are reported in Table \ref{table:table2} for each wind component. The partial FS GANs reduce values of $\xi$ for the West and Central regions, with the exception of the Southeast region where values of $\xi$ are larger.

The MAE of persistence (MAEP), which summarizes the difference between realizations (either from WRF or the SR models) at each time step and those 6 hours prior, is also provided in Table \ref{table:table2}. Values of MAE are lower than MAEP$_{WRF}$ and persistence of the models on the 6-hourly timescale is consistent between the SR models (MAEP$_{SR}$) and WRF  (MAEP$_{WRF}$). This provides additional evidence that the models are producing realistic results. 

\begin{table*}
    \centering
\caption{Summary of metrics computed for each SR model over test data. The MAE and MS-SSIM are computed for all pixels in the $u10$ and $v10$ fields. Spatial averages of the differences (i.e. the bias) between the generated and ``true'' climatological mean, standard deviation, and 90th percentile over the test set of the wind speed are reported as $\mu$, $\sigma$, and $Q^{90}$ respectively. The MSA, $\xi$, is also included as well as the MAE of persistence (MAEP) between 6-hourly WRF timesteps. Bold text indicates the optimally performing model for the given metric. \label{table:table2}}

\begin{adjustbox}{width=\textwidth,center=\textwidth}

    \begin{tabular}{ccccccccccc} \hline
(a) Southeast & MAE [m/s] & MS-SSIM & $\mu$ [m/s] & $\sigma$ [m/s] & $Q^{90}$ [m/s] & $\xi_{u10}$ [\%] & $\xi_{v10}$ [\%] & MAEP$_{WRF}$ [m/s] & MAEP$_{SR}$ [m/s] &\\ \hline
CNN &\textbf{1.230} & \textbf{0.868} & -0.217 & -0.127 & -0.376 & 220.642 & 281.405 & 1.728 & {1.396} & \\
NFS GAN &1.433 & 0.848 & -0.043 & \textbf{-0.007} & \textbf{-0.072} & 3.383 & \textbf{2.426} & 1.728 & \textbf{1.738} & \\
FS $\mathscr{{L}}_5$ &1.303 & \textbf{0.868} & -0.228 & -0.081 & -0.339 & 6.931 & 6.558 & 1.728 & 1.484 & \\
FS $\mathscr{{L}}_9$ &1.325 & 0.851 & -0.196 & -0.054 & -0.297 & 7.666 & 10.096 & 1.728 & 1.526 & \\
FS $\mathscr{{L}}_{13}$ &1.360 & 0.843 & -0.158 & -0.029 & -0.202 & \textbf{2.916} & 5.260 & 1.728 & 1.573 & \\
PFS ($\alpha = 500$) $\mathscr{{L}}_{13}$ &1.413 & 0.849 & -0.107 & -0.071 & -0.230 & 14.706 & 12.564 & 1.728 & 1.675 & \\
PFS ($\alpha = 500$) $\mathscr{{L}}_9$ &1.414 & 0.847 & -0.199 & -0.111 & -0.372 & 16.498 & 16.384 & 1.728 & 1.656 & \\
PFS ($\alpha = 500$) $\mathscr{{L}}_5$ &1.382 & 0.856 & -0.120 & -0.094 & -0.264 & 18.095 & 14.716 & 1.728 & 1.622 & \\
PFS ($\alpha = 50$) $\mathscr{{L}}_{13}$ &1.458 & 0.765 & \textbf{-0.012} & 0.018 & -0.081 & 11.456 & 27.159 & 1.728 & 1.610 & \\
PFS ($\alpha = 50$) $\mathscr{{L}}_9$ &1.390 & 0.845 & -0.378 & -0.215 & -0.718 & 25.833 & 26.554 & 1.728 & 1.622 & \\
PFS ($\alpha = 50$) $\mathscr{{L}}_5$ &1.385 & 0.836 & -0.382 & -0.179 & -0.693 & 11.425 & 21.727 & 1.728 & 1.459 & \\
\hline
(b) Central & MAE [m/s] & MS-SSIM & $\mu$ [m/s] & $\sigma$ [m/s] & $Q^{90}$ [m/s] & $\xi_{u10}$ [\%] & $\xi_{v10}$ [\%] & MAEP$_{WRF}$ [m/s] & MAEP$_{SR}$ [m/s] &\\ \hline
CNN &\textbf{1.027} & 0.841 & -0.265 & -0.170 & -0.551 & 339.855 & 305.648 & 1.558 & {1.258} & \\
NFS GAN &1.123 & 0.840 & -0.112 & -0.090 & -0.290 & 31.851 & 31.849 & 1.558 & 1.440 & \\
FS $\mathscr{{L}}_5$ &1.041 & \textbf{0.854} & -0.243 & -0.128 & -0.481 & 48.166 & 56.369 & 1.558 & 1.259 & \\
FS $\mathscr{{L}}_9$ &1.051 & 0.842 & -0.198 & -0.117 & -0.412 & 44.538 & 51.693 & 1.558 & 1.298 & \\
FS $\mathscr{{L}}_{13}$ &1.060 & 0.847 & -0.244 & -0.127 & -0.465 & 42.171 & 47.377 & 1.558 & 1.312 & \\
PFS ($\alpha = 500$) $\mathscr{{L}}_{13}$ &1.158 & 0.830 & -0.150 & -0.060 & -0.278 & 14.729 & 18.195 & 1.558 & 1.457 & \\
PFS ($\alpha = 500$) $\mathscr{{L}}_9$ &1.159 & 0.815 & -0.086 & -0.045 & -0.183 & 18.693 & 10.644 & 1.558 & 1.473 & \\
PFS ($\alpha = 500$) $\mathscr{{L}}_5$ &1.126 & 0.838 & -0.078 & -0.060 & -0.196 & 20.098 & 14.098 & 1.558 & 1.438 & \\
PFS ($\alpha = 50$) $\mathscr{{L}}_{13}$ &1.178 & 0.821 & -0.085 & \textbf{-0.006} & -0.082 & 8.910 & \textbf{7.488} & 1.558 & 1.362 & \\
PFS ($\alpha = 50$) $\mathscr{{L}}_9$ &1.183 & 0.836 & 0.037 & -0.034 & \textbf{-0.022} & \textbf{7.105} & 11.796 & 1.558 & 1.406 & \\
PFS ($\alpha = 50$) $\mathscr{{L}}_5$ &1.191 & 0.827 & \textbf{-0.024} & -0.025 & -0.082 & 11.396 & 10.463 & 1.558 & \textbf{1.484} & \\
\hline
(c) West & MAE [m/s] & MS-SSIM & $\mu$ [m/s] & $\sigma$ [m/s] & $Q^{90}$ [m/s] & $\xi_{u10}$ [\%] & $\xi_{v10}$ [\%] & MAEP$_{WRF}$ [m/s] & MAEP$_{SR}$ [m/s] &\\ \hline
CNN &\textbf{1.070} & \textbf{0.896} & -0.405 & -0.204 & -0.669 & 102.759 & 120.488 & 1.677 & {1.263} & \\
NFS GAN &1.235 & 0.880 & -0.286 & -0.119 & -0.434 & 21.322 & 36.113 & 1.677 & 1.530 & \\
FS $\mathscr{{L}}_5$ &1.149 & 0.884 & -0.288 & -0.115 & -0.431 & 30.373 & 28.782 & 1.677 & 1.375 & \\
FS $\mathscr{{L}}_9$ &1.169 & 0.883 & -0.201 & -0.067 & -0.280 & 23.916 & 28.806 & 1.677 & 1.423 & \\
FS $\mathscr{{L}}_{13}$ &1.173 & 0.881 & -0.218 & -0.064 & -0.296 & 21.871 & 26.170 & 1.677 & 1.454 & \\
PFS ($\alpha = 500$) $\mathscr{{L}}_{13}$ &1.373 & 0.860 & -0.322 & -0.097 & -0.440 & 15.269 & 24.663 & 1.677 & 1.612 & \\
PFS ($\alpha = 500$) $\mathscr{{L}}_9$ &1.361 & 0.865 & -0.289 & -0.092 & -0.389 & 15.585 & 23.799 & 1.677 & 1.594 & \\
PFS ($\alpha = 500$) $\mathscr{{L}}_5$ &1.321 & 0.873 & -0.293 & -0.092 & -0.400 & 19.212 & 26.610 & 1.677 & 1.558 & \\
PFS ($\alpha = 50$) $\mathscr{{L}}_{13}$ &1.421 & 0.851 & -0.358 & -0.111 & -0.498 & 20.040 & 26.676 & 1.677 & \textbf{1.674} & \\
PFS ($\alpha = 50$) $\mathscr{{L}}_9$ &1.412 & 0.852 & \textbf{-0.182} & \textbf{-0.033} & \textbf{-0.222} & 18.205 & 17.478 & 1.677 & 1.724 & \\
PFS ($\alpha = 50$) $\mathscr{{L}}_5$ &1.421 & 0.855 & -0.217 & -0.041 & -0.256 & \textbf{7.415} & \textbf{16.493} & 1.677 & 1.692 & \\
\hline
    \end{tabular}
\end{adjustbox}
\end{table*}

\subsection{Experiment 3: Low-resolution Covariates} 
\label{section:exp3}
\subsubsection{Idealized Covariates}
\begin{figure*}[t]
    \noindent
    \includegraphics[width=\textwidth]{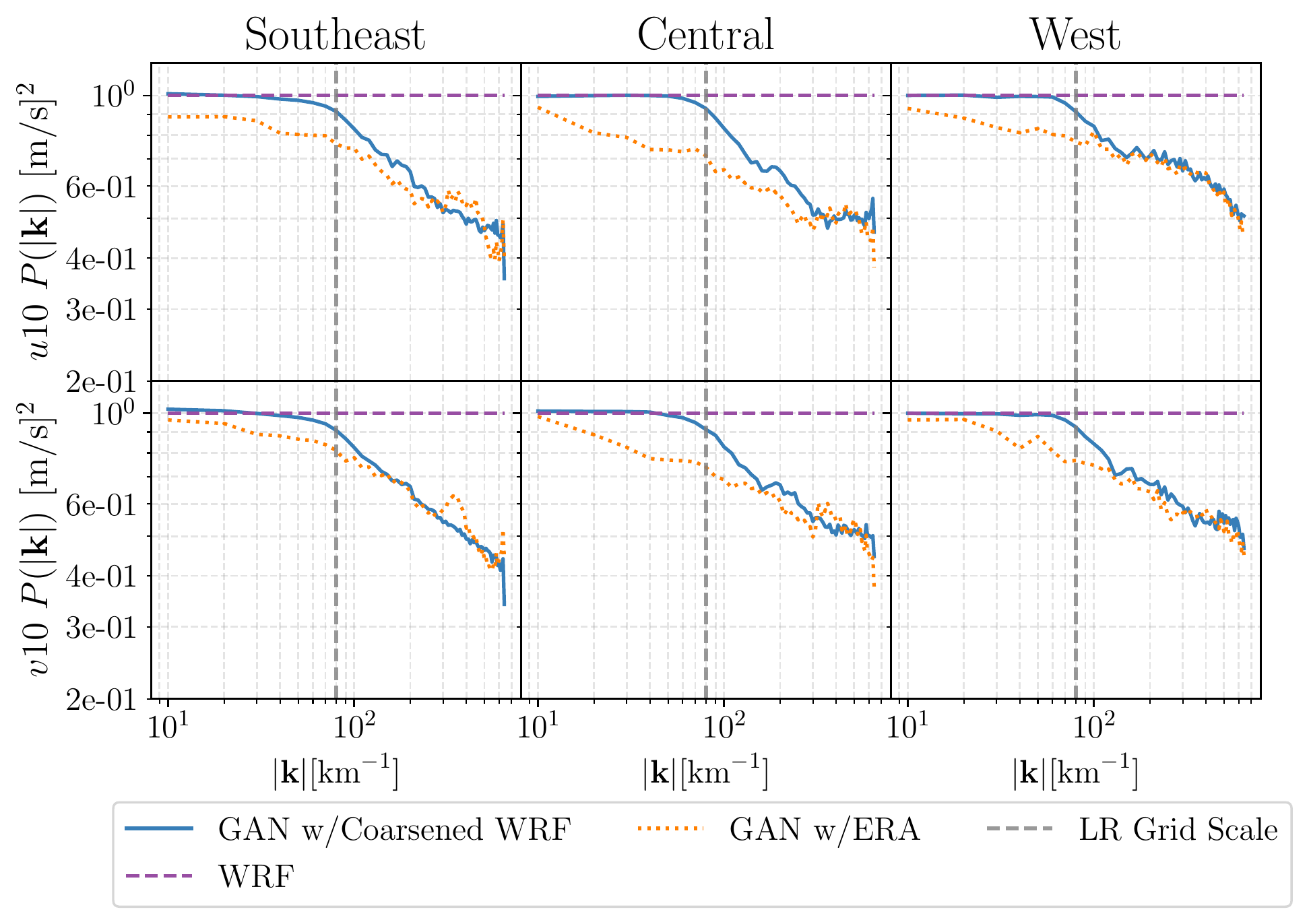}
    \caption{Ratio of RAPSD of the SR models with idealized (coarsened WRF) and non-idealized (ERA) covariates (exclusively $u10$ and $v10$) represented as different lines. The top row shows the $u10$ component, and the bottom row shows the $v10$ component. Each column is a separate region.}
    \label{fig:fig9}
\end{figure*}

Using idealized coarse covariates resulted in a training evolution of the MAE and MSE without signs of overfitting (not shown). The MAE and MSE  plateau at late epochs, supporting the hypothesis that the SR models are overfitting large-scale differences in the location of spatial features between ERA-Interim and WRF HRCONUS in the training set. Figure \ref{fig:fig9} shows the RAPSD ratio (relative to WRF HRCONUS) of $u10$ and $v10$ wind fields with the idealized GAN (GAN with Coarsened WRF) and the GAN trained with  $u10$ and $v10$ ERA-Interim covariates (GAN with ERA).

Due to large-scale differences between WRF HRCONUS and ERA-Interim, a low-power bias at small wavenumbers is found in Figure \ref{fig:fig9} for GANs using ERA-Interim covariates. This bias almost entirely vanishes using the idealized covariates. At high frequencies, there is less of a difference seen between the coarsened WRF GAN and ERA GAN with just $u10$ and $v10$.

\subsubsection{Additional Covariates}
\begin{figure*}[t]
    \noindent
    \includegraphics[width=\textwidth]{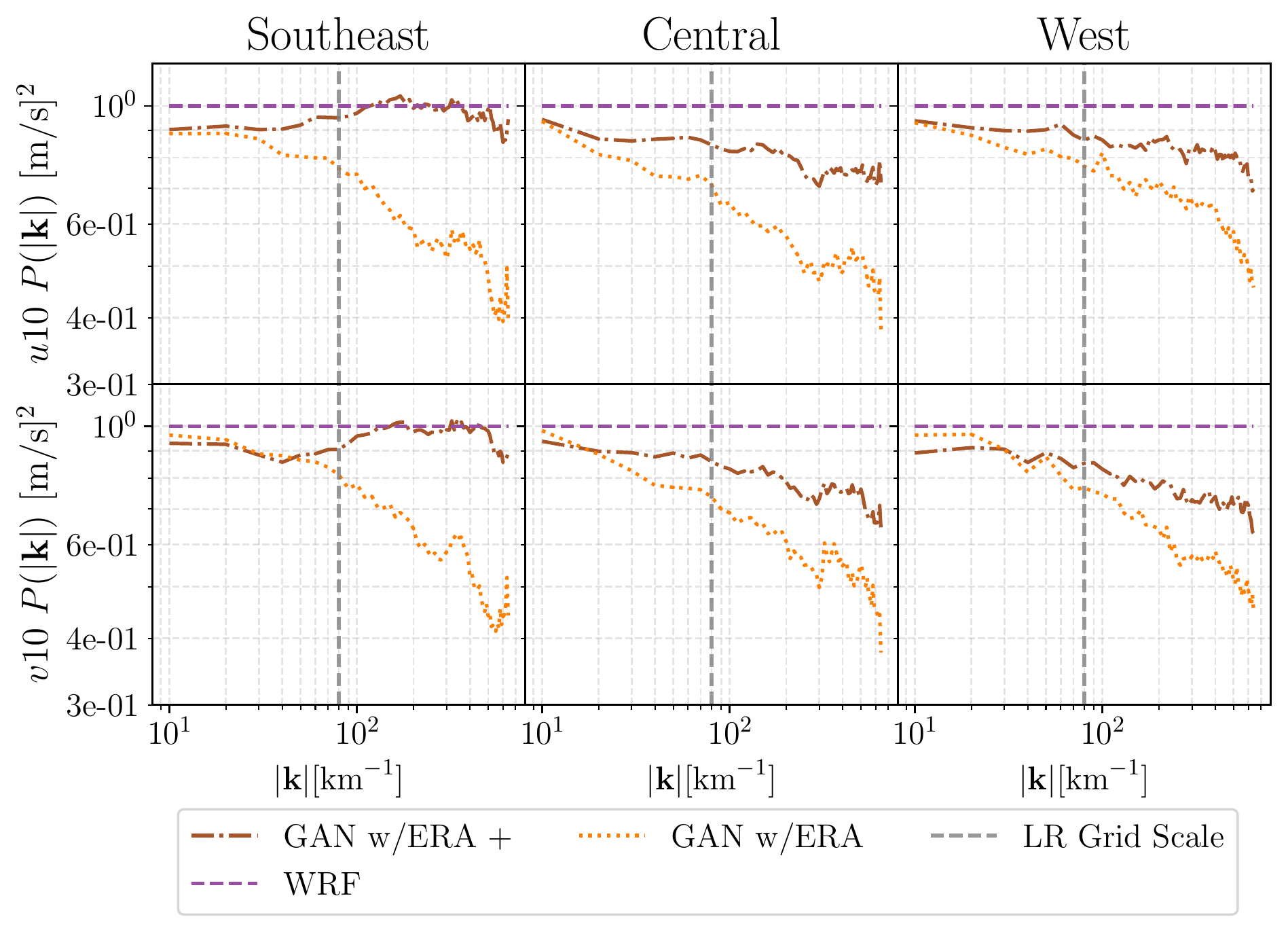}
    \caption{Ratio of RAPSD of the GANs trained with exclusively ERA $u10$ and $v10$ compared to the GAN with all of the additional ERA covariates (GAN w/ERA +, same as non-FS GAN) represented as different lines. The top row shows the $u10$ component, and the bottom row shows the $v10$ component. Each column is a separate region.}
    \label{fig:fig10}
\end{figure*}

Figure \ref{fig:fig10} shows the power spectra of the non-FS GAN from earlier with all seven covariate fields (GAN with ERA +, where ``+'' is shorthand to indicate that these GANs were trained with the additional covariates discussed in Section \ref{section:training}) and compares it to the spectra of the GAN with only ERA $u10$ and $v10$. GAN with ERA + shows a minor improvement in this bias at large scales and a significant improvement in the small scales. This demonstrates that the additional covariates are robustly improving the Generator's ability to produce high spatial frequency information consistent with WRF for both the $u10$ and $v10$ fields for each region. 

\begin{figure*}[t]
    \noindent
    \includegraphics[width=\textwidth]{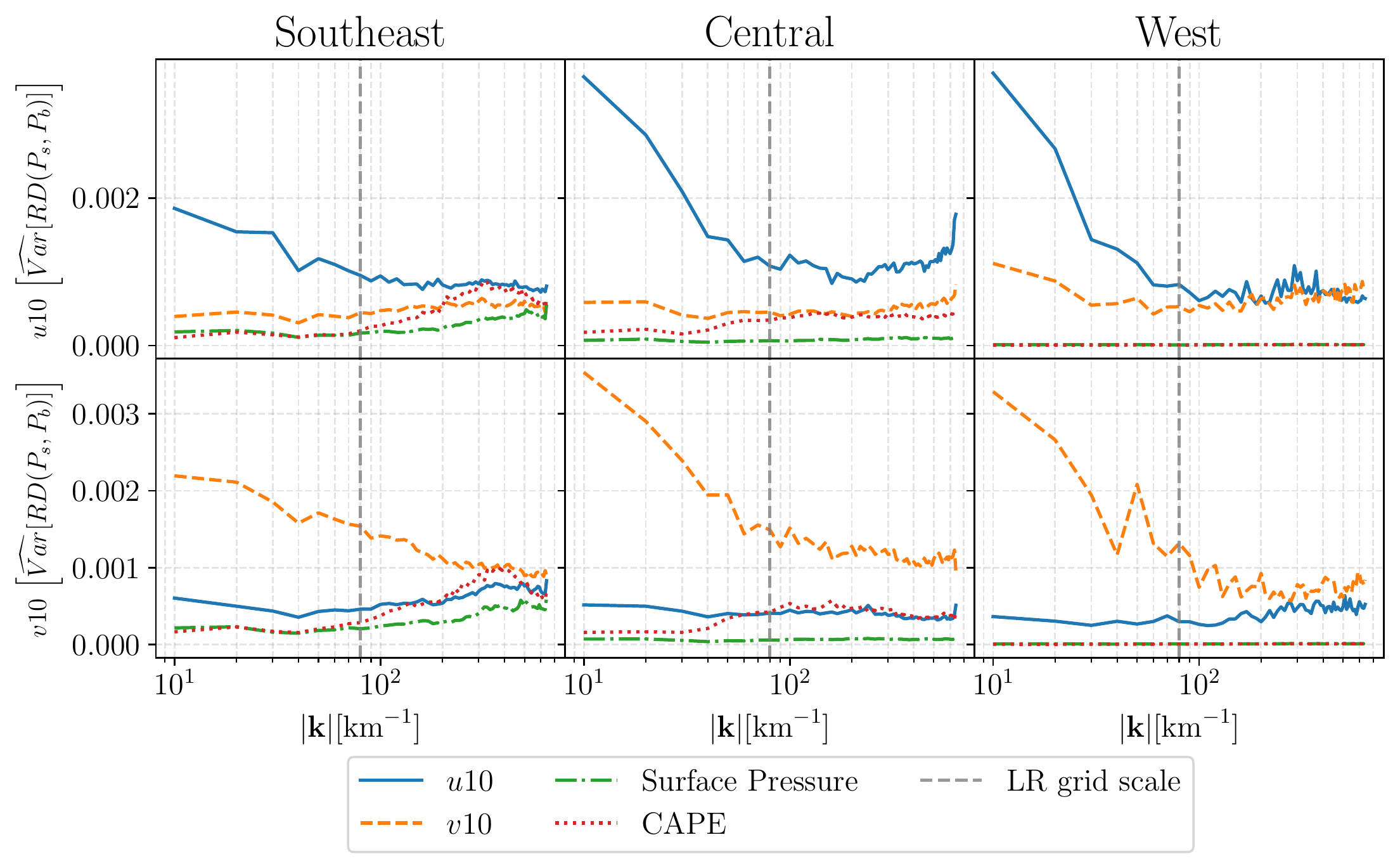}
    \caption{Sensitivity of the HR field power spectrum in response to modified LR conditioning fields. Each line represents a different covariate, with the variance reported at different wavenumbers. The top row is for $u10$ wind, and the bottom is for the $v10$ wind. Each column is a different region.}
    \label{fig:fig11}
\end{figure*}

The results of the single-pass permutation importance experiment are summarized in Figure \ref{fig:fig11}. We find that  $u10$ and $v10$  are most sensitive to changes to the respective coarse $u10$ and $v10$ covariates, especially at large scales. This is not surprising given that the large scales in WRF HRCONUS are synchronous with ERA-Interim, and so the networks are making direct use of this large-scale information. This same result can be seen for each region and each wind component.

Interestingly, the degree to which each wind component is sensitive to the other (i.e. the sensitivity of HR $v10$ fields to LR $u10$ fields, and vice versa) is less than that seen for some of the other covariates, such as CAPE in the Southeast region.  CAPE is highly correlated with convective processes that dominate the high spatial frequencies of the generated wind components \citep{houze_jr_mesoscale_2004} in the Southeast region where fine-scale convective features are common. Surface pressure also plays a moderate role in the Southeast region, possibly correlating with weather systems accompanied by small-scale variability caused by squall lines or multi-cell storms. 

The Central region shows sensitivity to $u10$, $v10$, and CAPE. Like the Southeast region, convective features are also common in the Central region, which explains the observed sensitivity to CAPE. The frequency of convective systems in WRF HRCONUS in the Central region is not expected to be quite as high as in the Southeast region, resulting in a slightly lower relative sensitivity to CAPE. 

For the West region, generated fields are most sensitive to $u10$ and $v10$ but are not sensitive to CAPE and surface pressure. This result can also be understood in the context of the West region's climatology, where convective storms are rare. The strong influences of land-sea boundaries and complex topography are better predicted by the coarse $u10$ and $v10$ fields themselves. 

\section{Discussion}
\label{section:discussion}
\subsection{Overview}
GANs for SR show impressive capabilities in generating fine-scale variability that is similar in distribution to the ``true'' variability simulated by the convection-permitting model. Extensive dynamical downscaling by convection-permitting models is operationally infeasible due to computational costs, which makes statistical downscaling using GANs a very attractive and practical alternative and work to date bodes well for their operational feasibility.

We present three experiments aimed at understanding our research questions aimed at the SR objective function and LR covariates in the SR configurations. While we do not propose a ``best'' performing model, we design experiments that provide potential avenues for fine-tuning future models. A discussion of these three experiments and future avenues is included below.

\subsection{Experiments 1: Frequency Separation and Experiment 2: Partial Frequency Separation}

When using FS, results vary for metrics that evaluate the convergence of realizations depending on the value of $N$. Non-FS GAN captures the variability of WRF HRCONUS well because the adversarial loss considers variability across all scales, unlike FS GANs, which only capture certain scales. Although FS GANs demonstrate lower MAE, they sacrifice perceptual realism and variability. Spatial correlation measures like RAPSD better reflect the perceptual accuracy of generated fields.

Partial FS can substantially influence the generated spectra for each region (Table \ref{table:table2}) by mimicking the use of the conditional mean/median in the content loss in stochastic approaches. As such, compared to stochastic GANs (that require an ensemble) partial FS can significantly decrease the computational requirements. Notably, partial FS GANs can generate more large-scale variability when the content loss is applied to low frequencies only. Potential improvements provided by partial FS GANs compared to non-FS GANs depend on the relative weighting of content and adversarial losses, which requires further study to determine an optimal approach.

The results of both FS experiments helped to address the research question (i): \textit{How do the generated outputs change when we manipulate the objective functions taken directly from the computer vision literature?} Namely, the experiments demonstrated a tension between the convergence of realizations and convergence of distributions that is at the core of the Generator's objective function in SR, and also offer useful directions for future tuning to ease this tension.

\subsection{Experiment 3: Low-resolution Covariates}

\subsubsection{Low-power Biases}
There exists a stubborn low-power bias between WRF HRCONUS and the non-FS GAN (GAN with ERA +) (Figure \ref{fig:fig10}) at small wavenumbers even with additional covariates and an adversarial loss computed over all frequencies. We provide evidence that this low-power bias originates from differences in the placement of large-scale spatial features between ERA-Interim and WRF HRCONUS and show that idealized coarsening of the HR fields to produce the LR fields for training dramatically reduces it. This result suggests that since the HR-generated fields inherit large-scale information from the LR fields, differences between ERA-Interim and WRF HRCONUS can manifest in the generated fields as low-power biases at large scales.

In existing studies, stochastic methods have demonstrated differences in the performance of the GANs when trained with idealized \citep{leinonen_stochastic_2020}, and non-idealized \citep{price_increasing_2022, harris_generative_2022} LR/HR pairs. Similar to the present study, \cite{harris_generative_2022} performed an idealized experiment using covariates derived from the HR fields, which revealed that idealized covariates improve the calibration and continuous ranked probability score (CRPS). Moreover, \cite{harris_generative_2022} attributed large-scale differences between the LR/HR pairs as the main limiting factor in the GAN performance, and saw a large improvement in CRPS and calibration when the GANs ingested coarsened HR fields from the same target HR dataset. For deterministic approaches, non-idealized pairs limit the large-scale variability in the generated fields as represented by a low-power bias at large scales in the RAPSD. We hypothesize that this difference in deterministic settings comes from differences in internal variability between ERA-Interim and WRF causing the misalignment of features on shared scales. Conversely, the absence of significant biases in power at large scales for the idealized pairs can be attributed to the lack of internal variability between the LR and HR fields on shared scales. To the extent to which systematic differences exist between the non-idealized HR and LR pairs,  a bias-correction methodology, similar to \cite{price_increasing_2022}, or including HR topographical information during training, could improve skill.

The results of performing the non-idealized vs. idealized experiment, particularly when examining the power spectra help address research question (ii): \textit{What capacity do the networks have to deal with non-idealized LR/HR pairs?} 

\subsubsection{Additional Covariates}
To our knowledge, this study represents the first application of SR to climate fields that demonstrates the importance of including additional physically-relevant LR variables, beyond LR versions of the HR target variables, as covariates. Including these additional covariates reduces low-power bias across all frequencies, particularly at high frequencies (Figure \ref{fig:fig10}). This finding is important for designing SR models for climate and weather fields, as the GANs mirror the physical relationships between covariates and target variables.

The methodology we use to assess the importance of specific covariates finds results consistent with important physical processes in the regions considered.  This methodology assumes independence of covariates. One caveat of the above method is that if the covariates are not independent, certain combinations of covariates may be more important than individual covariates. It should also be noted that the method does not apply to invariant covariates like surface roughness length, topography, or land-sea mask, which may play a crucial role in representing high-frequency variability. Future work could explore the elimination of these covariates in analyzing the RAPSD of new GANs to determine their importance, and also perform multi-pass permutation importance to measure combined effects \citep{MakingtheBlackBoxMoreTransparentUnderstandingthePhysicalImplicationsofMachineLearning}. Furthermore, HR versions of invariant covariates could also significantly improve model skills. GAN SR with wind variables could leverage HR topography, surface roughness, and a land-sea mask (although a land-sea mask might provide redundant information to topographical information). A similar approach has been used previously in \cite{harris_generative_2022}, and quite successfully in \cite{sha_deep-learning-based_2020}. In the generator network, HR invariant information could be embedded in a set of additional LR covariates \citep{harris_generative_2022}, or it could also be concatenated after the upsampling blocks in the generator network for dimensional consistency. One interesting follow-up study to ours would be to measure the importance of invariant HR covariates for GAN SR.

The results of including additional covariates, as well as the power spectra breakdown of their importance, addresses research question (iii):\textit{ What role do select LR covariates play in super-resolved near-surface wind fields?}

\subsection{Future GANs}
Our deterministic approach examines how the low-power bias is represented in the spectra of the generated fields, and, importantly, how appropriately chosen covariates and hyperparameters can reduce this bias. We hypothesize that one additional way to reduce RAPSD power biases may be to introduce a loss/regularization term that directly evaluates the RAPSD of batches of generated and ``true'' fields. Such an approach was introduced in \cite{kashinath_physics-informed_2021} by replacing the adversarial loss entirely with this spectral loss. This approach performs well, but its implementation in \cite{kashinath_physics-informed_2021} comes with caveats -- i.e. it was tested with idealized covariates and a scale gap of 4$\times$. Moreover, leaving out the adversarial loss may inhibit the ability of the model to sample diverse realizations from the conditional distribution. Rather than replace components of the Generator's loss function entirely, a spectral loss could serve an auxiliary role that supports realistically generated variability across spatial scales in addition to the content and adversarial components.

Given that the generated fields suffer from these low-power biases, future avenues might also explore (1) the extent to which SR model behaviour might change if trained using LR fields from different climate models, or trained with one LR model and evaluated with a different LR model; and (2) how well our SR models will extrapolate when provided with LR fields from future climate model projections.

\section{Conclusions}
\label{section:conclusions}

The SRGANs used with the WGAN-GP framework show promising potential and feasibility for the downscaling of multivariate wind patterns, indicating their potential usefulness in a variety of practical applications. While we do not propose ``best'' performing models, our results shed light on the SR task when applied to climate fields.

Using SR for statistical downscaling means generating fine-scale features that could exist, rather than those which may have actually existed, resulting in challenges in selecting appropriate error metrics. Using RAPSD, we demonstrated scale-dependent biases in the generated variability allowing us to more easily compare the SR models. We emphasize that selecting appropriate metrics is vital for the comparability of future SR approaches.

The role of covariates was closely examined in the RAPSD of the generated fields. Specifically, we showed that internal variability in HR fields can result in a low-power bias at large scales in generated fields. We also showed that carefully chosen covariates help reduce low-power biases at all spatial scales, but especially in the fine-scale features. To further investigate the role of our chosen covariates, a sensitivity experiment was conducted to demonstrate the value added to spatial structures in the generated fields by each of the covariates. The importance of the covariates differed between regions, consistent with the relative importance of CAPE in producing small-scale wind variability.

Additionally, we adopted frequency separation (FS) from the computer vision field. While FS did not  result in a more skillful GAN (in terms of the power spectra of the generated fields),  it did reveal how modifying the Generator’s objective function changed the RAPSD of the generated fields. Specifically, we demonstrate the important role the adversarial loss has in generating variability across spatial scales. For deterministic SR, we discuss how the content loss and adversarial loss are implicitly oppositional in their objective to express variability. While stochastic SR can mitigate this problem using ensembles, we introduce partial FS as a simpler and more computationally efficient option. We also show evidence of the sensitivity of generated spatial structures to the hyperparameters $\alpha$ and $N$ in partial FS.  A central result of this analysis is the importance to the generated fields of the two kinds of loss terms used in the objective function - adversarial and content - and the scales to which these are applied.

\acknowledgments
AHM acknowledges the support of the Natural Sciences and Engineering Research Council of Canada (NSERC) (funding reference RGPIN-2019-204986).  We acknowledge helpful discussions with David John Gagne II and Merc\`e Casas-Prat. We would also like to express our gratitude to the three reviewers for their constructive feedback, which greatly enhanced our manuscript.

\datastatement
We have organized our code using two avenues to reproduce our results. (1) The underlying code used for training the Wasserstein GAN models is archived at doi:10.5281/zenodo.7604242; and (2), due to the challenge of using complicated software with dynamic dependencies, we made efforts to isolate our software environment and data so that our analysis is reproducible. As such, we developed a Docker image (nannau/annau-2023) hosted on Docker Hub and include the corresponding documentation with the source code at doi:10.5281/zenodo.7604267 \citep{merkel_docker_2014}.

\bibliographystyle{ametsocV6}
\bibliography{annau2023}

\begin{thebibliography}{60}
\providecommand{\natexlab}[1]{#1}
\providecommand{\url}[1]{\texttt{#1}}
\renewcommand{\UrlFont}{\rmfamily}
\providecommand{\urlprefix}{URL }
\expandafter\ifx\csname urlstyle\endcsname\relax
  \providecommand{\doi}[1]{https://doi.org/\discretionary{}{}{}#1}\else
  \providecommand{\doi}{https://doi.org/\discretionary{}{}{}\begingroup \urlstyle{rm}\Url}\fi
\providecommand{\eprint}[2][]{\url{#2}}

\bibitem[{Adewoyin et~al.(2021)Adewoyin, Dueben, Watson, He,, and Dutta}]{adewoyin_tru-net_2021}
Adewoyin, R.~A., P.~Dueben, P.~Watson, Y.~He, and R.~Dutta, 2021: {TRU}-{NET}: a deep learning approach to high resolution prediction of rainfall. \textit{Machine Learning}, \textbf{110~(8)}, 2035--2062, \doi{10.1007/s10994-021-06022-6}, \urlprefix\url{https://doi.org/10.1007/s10994-021-06022-6}.

\bibitem[{Arjovsky et~al.(2017)Arjovsky, Chintala,, and Bottou}]{arjovsky_wasserstein_2017}
Arjovsky, M., S.~Chintala, and L.~Bottou, 2017: Wasserstein {GAN}. \textit{arXiv:1701.07875 [cs, stat]}, \urlprefix\url{http://arxiv.org/abs/1701.07875}, arXiv: 1701.07875.

\bibitem[{Ban et~al.(2015)Ban, Schmidli,, and Schär}]{ban_heavy_2015}
Ban, N., J.~Schmidli, and C.~Schär, 2015: Heavy precipitation in a changing climate: {Does} short‐term summer precipitation increase faster? \textit{Geophysical Research Letters}, \textbf{42~(4)}, 1165--1172, iSBN: 0094-8276 Publisher: Wiley Online Library.

\bibitem[{Bessac et~al.(2019)Bessac, Monahan, Christensen,, and Weitzel}]{bessac_stochastic_2019}
Bessac, J., A.~H. Monahan, H.~M. Christensen, and N.~Weitzel, 2019: Stochastic {Parameterization} of {Subgrid}-{Scale} {Velocity} {Enhancement} of {Sea} {Surface} {Fluxes}. \textit{Monthly Weather Review}, \textbf{147~(5)}, 1447 -- 1469, \doi{10.1175/MWR-D-18-0384.1}, \urlprefix\url{https://journals.ametsoc.org/view/journals/mwre/147/5/mwr-d- 18-0384.1.xml}, place: Boston MA, USA Publisher: American Meteorological Society.

\bibitem[{Cannon(2008)}]{cannon_probabilistic_2008}
Cannon, A.~J., 2008: Probabilistic {Multisite} {Precipitation} {Downscaling} by an {Expanded} {Bernoulli}–{Gamma} {Density} {Network}. \textit{Journal of Hydrometeorology}, \textbf{9~(6)}, 1284--1300, \doi{10.1175/2008JHM960.1}, \urlprefix\url{http://journals.ametsoc.org/view/journals/hydr/9/6/ 2008jhm960_1.xml}, publisher: American Meteorological Society Section: Journal of Hydrometeorology.

\bibitem[{Cheng et~al.(2020)Cheng, Liu, Xu, Shen,, and Kuang}]{cheng_generating_2020}
Cheng, J., J.~Liu, Z.~Xu, C.~Shen, and Q.~Kuang, 2020: Generating {High}-{Resolution} {Climate} {Prediction} through {Generative} {Adversarial} {Network}. \textit{Procedia Computer Science}, \textbf{174}, 123--127, \doi{10.1016/j.procs.2020.06.067}, \urlprefix\url{https://linkinghub.elsevier.com/retrieve/pii/ S1877050920315817}.

\bibitem[{Dee et~al.(2011)}]{dee_era-interim_2011}
Dee, D.~P., and Coauthors, 2011: The {ERA}-{Interim} reanalysis: configuration and performance of the data assimilation system. \textit{Quarterly Journal of the Royal Meteorological Society}, \textbf{137~(656)}, 553--597, \doi{10.1002/qj.828}, \urlprefix\url{https://onlinelibrary.wiley.com/doi/abs/10.1002/qj.828}, eprint: https://onlinelibrary.wiley.com/doi/pdf/10.1002/qj.828.

\bibitem[{Dong et~al.(2014)Dong, Loy, He,, and Tang}]{dong_learning_2014}
Dong, C., C.~C. Loy, K.~He, and X.~Tang, 2014: Learning a {Deep} {Convolutional} {Network} for {Image} {Super}-{Resolution}. \textit{Computer {Vision} – {ECCV} 2014}, D.~Fleet, T.~Pajdla, B.~Schiele, and T.~Tuytelaars, Eds., Springer International Publishing, Cham, 184--199, Lecture {Notes} in {Computer} {Science}, \doi{10.1007/978-3-319-10593-2_13}.

\bibitem[{Dong et~al.(2015)Dong, Loy, He,, and Tang}]{dong_image_2015}
Dong, C., C.~C. Loy, K.~He, and X.~Tang, 2015: Image {Super}-{Resolution} {Using} {Deep} {Convolutional} {Networks}. \textit{arXiv:1501.00092 [cs]}, \urlprefix\url{http://arxiv.org/abs/1501.00092}, arXiv: 1501.00092.

\bibitem[{Frei et~al.(2003)Frei, Christensen, Déqué, Jacob, Jones,, and Vidale}]{frei_daily_2003}
Frei, C., J.~H. Christensen, M.~Déqué, D.~Jacob, R.~G. Jones, and P.~L. Vidale, 2003: Daily precipitation statistics in regional climate models: {Evaluation} and intercomparison for the {European} {Alps}. \textit{Journal of Geophysical Research: Atmospheres}, \textbf{108~(D3)}, iSBN: 0148-0227 Publisher: Wiley Online Library.

\bibitem[{Fritsche et~al.(2019)Fritsche, Gu,, and Timofte}]{fritsche_frequency_2019}
Fritsche, M., S.~Gu, and R.~Timofte, 2019: Frequency {Separation} for {Real}-{World} {Super}-{Resolution}. \textit{arXiv:1911.07850 [cs, eess]}, \urlprefix\url{http://arxiv.org/abs/1911.07850}, arXiv: 1911.07850.

\bibitem[{Gardner and Dorling(1998)Gardner, and Dorling}]{gardner_artificial_1998}
Gardner, M.~W., and S.~R. Dorling, 1998: Artificial neural networks (the multilayer perceptron)—a review of applications in the atmospheric sciences. \textit{Atmospheric Environment}, \textbf{32~(14)}, 2627--2636, \doi{10.1016/S1352-2310(97)00447-0}, \urlprefix\url{https://www.sciencedirect.com/science/article/pii/ S1352231097004470}.

\bibitem[{Goodfellow et~al.(2014)Goodfellow, Pouget-Abadie, Mirza, Xu, Warde-Farley, Ozair, Courville,, and Bengio}]{goodfellow_generative_2014}
Goodfellow, I.~J., J.~Pouget-Abadie, M.~Mirza, B.~Xu, D.~Warde-Farley, S.~Ozair, A.~Courville, and Y.~Bengio, 2014: Generative {Adversarial} {Networks}. \textit{arXiv:1406.2661 [cs, stat]}, \urlprefix\url{http://arxiv.org/abs/1406.2661}, arXiv: 1406.2661.

\bibitem[{Gulrajani et~al.(2017)Gulrajani, Ahmed, Arjovsky, Dumoulin,, and Courville}]{gulrajani_improved_2017}
Gulrajani, I., F.~Ahmed, M.~Arjovsky, V.~Dumoulin, and A.~Courville, 2017: Improved {Training} of {Wasserstein} {GANs}. \textit{arXiv:1704.00028 [cs, stat]}, \urlprefix\url{http://arxiv.org/abs/1704.00028}, arXiv: 1704.00028.

\bibitem[{Harris et~al.(2020)}]{harris_array_2020}
Harris, C.~R., and Coauthors, 2020: Array programming with {NumPy}. \textit{Nature}, \textbf{585~(7825)}, 357--362, \doi{10.1038/s41586-020-2649-2}, \urlprefix\url{https://doi.org/10.1038/s41586-020-2649-2}, publisher: Springer Science and Business Media LLC.

\bibitem[{Harris et~al.(2022)Harris, McRae, Chantry, Dueben,, and Palmer}]{harris_generative_2022}
Harris, L., A.~T.~T. McRae, M.~Chantry, P.~D. Dueben, and T.~N. Palmer, 2022: A generative deep learning approach to stochastic downscaling of precipitation forecasts. \textit{Journal of Advances in Modeling Earth Systems}, \textbf{14~(10)}, e2022MS003\,120, \doi{https://doi.org/10.1029/2022MS003120}, \urlprefix\url{https://agupubs.onlinelibrary.wiley.com/doi/abs/10.1029/2022MS003120}, e2022MS003120 2022MS003120, \eprint{https://agupubs.onlinelibrary.wiley.com/doi/pdf/10.1029/2022MS003120}.

\bibitem[{Hersbach et~al.(2020)}]{hersbach_era5_2020}
Hersbach, H., and Coauthors, 2020: The {ERA5} global reanalysis. \textit{Quarterly Journal of the Royal Meteorological Society}, \textbf{146~(730)}, 1999--2049, \doi{10.1002/qj.3803}, \urlprefix\url{https://doi.org/10.1002/qj.3803}, publisher: John Wiley \& Sons, Ltd.

\bibitem[{Houze~Jr(2004)}]{houze_jr_mesoscale_2004}
Houze~Jr, R.~A., 2004: Mesoscale convective systems. \textit{Reviews of geophysics (1985)}, \textbf{42~(4)}, RG4003--n/a, \doi{10.1029/2004RG000150}, edition: Houze, R. A.Jr. (2004), Mesoscale convective systems, Rev. Geophys., 42, RG4003, doi:10.1029/2004RG000150. Publisher: American Geophysical Union.

\bibitem[{Innocenti et~al.(2019)Innocenti, Mailhot, Frigon, Cannon,, and Leduc}]{innocenti_observed_2019}
Innocenti, S., A.~Mailhot, A.~Frigon, A.~J. Cannon, and M.~Leduc, 2019: Observed and {Simulated} {Precipitation} over {Northeastern} {North} {America}: {How} {Do} {Daily} and {Subdaily} {Extremes} {Scale} in {Space} and {Time}? \textit{Journal of Climate}, \textbf{32~(24)}, 8563--8582, \doi{10.1175/JCLI-D-19-0021.1}, \urlprefix\url{https://journals.ametsoc.org/view/journals/clim/32/24/jcli- d-19-0021.1.xml}, publisher: American Meteorological Society Section: Journal of Climate.

\bibitem[{Judt(2018)}]{judt_insights_2018}
Judt, F., 2018: Insights into {Atmospheric} {Predictability} through {Global} {Convection}-{Permitting} {Model} {Simulations}. \textit{Journal of the Atmospheric Sciences}, \textbf{75~(5)}, 1477 -- 1497, \doi{10.1175/JAS-D-17-0343.1}, \urlprefix\url{https://journals.ametsoc.org/view/journals/atsc/75/5/jas-d- 17-0343.1.xml}, place: Boston MA, USA Publisher: American Meteorological Society.

\bibitem[{Karpathy(2022)}]{karpathy_cs231n_2022}
Karpathy, A., 2022: {CS231n} {Convolutional} {Neural} {Networks} for {Visual} {Recognition}. \urlprefix\url{https://cs231n.github.io/convolutional-networks/#add}.

\bibitem[{Kashinath et~al.(2021)}]{kashinath_physics-informed_2021}
Kashinath, K., and Coauthors, 2021: Physics-informed machine learning: case studies for weather and climate modelling. \textit{Philosophical Transactions of the Royal Society A: Mathematical, Physical and Engineering Sciences}, \textbf{379~(2194)}, 20200\,093, \doi{10.1098/rsta.2020.0093}, \urlprefix\url{https://royalsocietypublishing.org/doi/10.1098/ rsta.2020.0093}.

\bibitem[{Kharin et~al.(2007)Kharin, Zwiers, Zhang,, and Hegerl}]{kharin_changes_2007}
Kharin, V.~V., F.~W. Zwiers, X.~Zhang, and G.~C. Hegerl, 2007: Changes in {Temperature} and {Precipitation} {Extremes} in the {IPCC} {Ensemble} of {Global} {Coupled} {Model} {Simulations}. \textit{Journal of Climate}, \textbf{20~(8)}, 1419--1444, \doi{10.1175/JCLI4066.1}, \urlprefix\url{https://journals.ametsoc.org/view/journals/clim/20/8/ jcli4066.1.xml}, publisher: American Meteorological Society Section: Journal of Climate.

\bibitem[{Kingma and Ba(2017)Kingma, and Ba}]{kingma_adam_2017}
Kingma, D.~P., and J.~Ba, 2017: Adam: {A} {Method} for {Stochastic} {Optimization}. \textit{arXiv:1412.6980 [cs]}, \urlprefix\url{http://arxiv.org/abs/1412.6980}, arXiv: 1412.6980.

\bibitem[{Kopparla et~al.(2013)Kopparla, Fischer, Hannay,, and Knutti}]{kopparla_improved_2013}
Kopparla, P., E.~M. Fischer, C.~Hannay, and R.~Knutti, 2013: Improved simulation of extreme precipitation in a high-resolution atmosphere model. \textit{Geophysical Research Letters}, \textbf{40~(21)}, 5803--5808, \doi{10.1002/2013GL057866}, \urlprefix\url{https://onlinelibrary.wiley.com/doi/abs/10.1002/ 2013GL057866}, \_eprint: https://onlinelibrary.wiley.com/doi/pdf/10.1002/2013GL057866.

\bibitem[{Krizhevsky et~al.(2012)Krizhevsky, Sutskever,, and Hinton}]{krizhevsky_imagenet_2012}
Krizhevsky, A., I.~Sutskever, and G.~E. Hinton, 2012: {ImageNet} {Classification} with {Deep} {Convolutional} {Neural} {Networks}. \textit{Advances in {Neural} {Information} {Processing} {Systems}}, Curran Associates, Inc., Vol.~25, \urlprefix\url{https://proceedings.neurips.cc/paper/2012/hash/ c399862d3b9d6b76c8436e924a68c45b-Abstract.html}.

\bibitem[{Kumar et~al.(2021)Kumar, Chattopadhyay, Singh, Chaudhari, Kodari,, and Barve}]{kumar_deep_2021}
Kumar, B., R.~Chattopadhyay, M.~Singh, N.~Chaudhari, K.~Kodari, and A.~Barve, 2021: Deep learning–based downscaling of summer monsoon rainfall data over {Indian} region. \textit{Theoretical and Applied Climatology}, \textbf{143~(3)}, 1145--1156, \doi{10.1007/s00704-020-03489-6}, \urlprefix\url{https://doi.org/10.1007/s00704-020-03489-6}.

\bibitem[{Ledig et~al.(2017)}]{ledig_photo-realistic_2017}
Ledig, C., and Coauthors, 2017: Photo-{Realistic} {Single} {Image} {Super}-{Resolution} {Using} a {Generative} {Adversarial} {Network}. \textit{arXiv:1609.04802 [cs, stat]}, \urlprefix\url{http://arxiv.org/abs/1609.04802}, arXiv: 1609.04802.

\bibitem[{Leinonen et~al.(2020)Leinonen, Nerini,, and Berne}]{leinonen_stochastic_2020}
Leinonen, J., D.~Nerini, and A.~Berne, 2020: Stochastic {Super}-{Resolution} for {Downscaling} {Time}-{Evolving} {Atmospheric} {Fields} {With} a {Generative} {Adversarial} {Network}. \textit{IEEE Transactions on Geoscience and Remote Sensing}, 1--13, \doi{10.1109/TGRS.2020.3032790}, \urlprefix\url{https://ieeexplore.ieee.org/document/9246532/}.

\bibitem[{Li et~al.(2018)Li, Zhang, Cannon, Murdock, Sobie, Zwiers, Anderson,, and Qian}]{li_indices_2018}
Li, G., X.~Zhang, A.~J. Cannon, T.~Murdock, S.~Sobie, F.~Zwiers, K.~Anderson, and B.~Qian, 2018: Indices of {Canada}’s future climate for general and agricultural adaptation applications. \textit{Climatic Change}, \textbf{148~(1)}, 249--263, \doi{10.1007/s10584-018-2199-x}, \urlprefix\url{https://doi.org/10.1007/s10584-018-2199-x}.

\bibitem[{Liu et~al.(2017)}]{liu_continental-scale_2017}
Liu, C., and Coauthors, 2017: Continental-scale convection-permitting modeling of the current and future climate of {North} {America}. \textit{Climate Dynamics}, \textbf{49~(1)}, 71--95, \doi{10.1007/s00382-016-3327-9}, \urlprefix\url{https://doi.org/10.1007/s00382-016-3327-9}.

\bibitem[{Lucas-Picher et~al.(2008)Lucas-Picher, Caya, de~Elía,, and Laprise}]{lucas-picher_investigation_2008}
Lucas-Picher, P., D.~Caya, R.~de~Elía, and R.~Laprise, 2008: Investigation of regional climate models' internal variability with a ten-member ensemble of 10-year simulations over a large domain. \textit{Climate dynamics}, \textbf{31~(7-8)}, 927--940, place: Berlin/Heidelberg Publisher: Berlin/Heidelberg : Springer-Verlag.

\bibitem[{Maraun et~al.(2010)}]{maraun_precipitation_2010}
Maraun, D., and Coauthors, 2010: Precipitation downscaling under climate change: {Recent} developments to bridge the gap between dynamical models and the end user. \textit{Reviews of Geophysics}, \textbf{48~(3)}, \doi{https://doi.org/10.1029/2009RG000314}, \urlprefix\url{https://agupubs.onlinelibrary.wiley.com/doi/abs/10.1029/ 2009RG000314}, \_eprint: https://agupubs.onlinelibrary.wiley.com/doi/pdf/10.1029/2009RG000314.

\bibitem[{McGovern et~al.(2019)McGovern, Lagerquist, Gagne, Jergensen, Elmore, Homeyer,, and Smith}]{MakingtheBlackBoxMoreTransparentUnderstandingthePhysicalImplicationsofMachineLearning}
McGovern, A., R.~Lagerquist, D.~J. Gagne, G.~E. Jergensen, K.~L. Elmore, C.~R. Homeyer, and T.~Smith, 2019: Making the black box more transparent: Understanding the physical implications of machine learning. \textit{Bulletin of the American Meteorological Society}, \textbf{100~(11)}, 2175 -- 2199, \doi{https://doi.org/10.1175/BAMS-D-18-0195.1}, \urlprefix\url{https://journals.ametsoc.org/view/journals/bams/100/11/bams-d-18-0195.1.xml}.

\bibitem[{Merkel(2014)}]{merkel_docker_2014}
Merkel, D., 2014: Docker: lightweight linux containers for consistent development and deployment. \textit{Linux journal}, \textbf{2014~(239)}, 2.

\bibitem[{Michaelides(2008)}]{michaelides_precipitation_2008}
Michaelides, S.~C., 2008: \textit{Precipitation: {Advances} in {Measurement}, {Estimation} and {Prediction}}. 1st ed., Springer Berlin Heidelberg, Berlin, Heidelberg, \doi{10.1007/978-3-540-77655-0}.

\bibitem[{Mirza and Osindero(2014)Mirza, and Osindero}]{mirza_conditional_2014}
Mirza, M., and S.~Osindero, 2014: Conditional {Generative} {Adversarial} {Nets}. \textit{arXiv:1411.1784 [cs, stat]}, \urlprefix\url{http://arxiv.org/abs/1411.1784}, arXiv: 1411.1784.

\bibitem[{Morley et~al.(2018)Morley, Brito,, and Welling}]{morley_measures_2018}
Morley, S.~K., T.~V. Brito, and D.~T. Welling, 2018: Measures of {Model} {Performance} {Based} {On} the {Log} {Accuracy} {Ratio}. \textit{Space Weather}, \textbf{16~(1)}, 69--88, \doi{https://doi.org/10.1002/2017SW001669}, \urlprefix\url{https://agupubs.onlinelibrary.wiley.com/doi/abs/10.1002/ 2017SW001669}, \_eprint: https://agupubs.onlinelibrary.wiley.com/doi/pdf/10.1002/2017SW001669.

\bibitem[{Prein et~al.(2016)}]{prein_precipitation_2016}
Prein, A.~F., and Coauthors, 2016: Precipitation in the {EURO}-{CORDEX} 0.11 deg and 0.44 deg simulations: high resolution, high benefits? \textit{Climate Dynamics}, \textbf{46~(1)}, 383--412, \doi{10.1007/s00382-015-2589-y}, \urlprefix\url{https://doi.org/10.1007/s00382-015-2589-y}.

\bibitem[{Price and Rasp(2022)Price, and Rasp}]{price_increasing_2022}
Price, I., and S.~Rasp, 2022: Increasing the accuracy and resolution of precipitation forecasts using deep generative models. \textit{International {Conference} on {Artificial} {Intelligence} and {Statistics}}, PMLR, 10\,555--10\,571.

\bibitem[{Rasmussen and Liu(2017)Rasmussen, and Liu}]{rasmussen_high_2017}
Rasmussen, R., and C.~Liu, 2017: High {Resolution} {WRF} {Simulations} of the {Current} and {Future} {Climate} of {North} {America}. \doi{10.5065/D6V40SXP}, \urlprefix\url{https://rda.ucar.edu/datasets/ds612.0/}, publisher: UCAR/NCAR - Research Data Archive Type: dataset.

\bibitem[{Rossa et~al.(2008)Rossa, Nurmi,, and Ebert}]{rossa_overview_2008}
Rossa, A., P.~Nurmi, and E.~Ebert, 2008: Overview of methods for the verification of quantitative precipitation forecasts. \textit{Precipitation: {Advances} in {Measurement}, {Estimation} and {Prediction}}, Springer Berlin Heidelberg, Berlin, Heidelberg, 419--452.

\bibitem[{Sampat et~al.(2009)Sampat, Wang, Gupta, Bovik,, and Markey}]{sampat_complex_2009}
Sampat, M.~P., Z.~Wang, S.~Gupta, A.~C. Bovik, and M.~K. Markey, 2009: Complex {Wavelet} {Structural} {Similarity}: {A} {New} {Image} {Similarity} {Index}. \textit{IEEE Transactions on Image Processing}, \textbf{18~(11)}, 2385--2401, \doi{10.1109/TIP.2009.2025923}.

\bibitem[{Schlager et~al.(2019)Schlager, Kirchengast, Fuchsberger, Kann,, and Truhetz}]{schlager_spatial_2019}
Schlager, C., G.~Kirchengast, J.~Fuchsberger, A.~Kann, and H.~Truhetz, 2019: A spatial evaluation of high-resolution wind fields from empirical and dynamical modeling in hilly and mountainous terrain. \textit{Geoscientific Model Development}, \textbf{12~(7)}, 2855--2873, \doi{10.5194/gmd-12-2855-2019}, \urlprefix\url{https://gmd.copernicus.org/articles/12/2855/2019/}.

\bibitem[{Sha et~al.(2020)Sha, II, West,, and Stull}]{sha_deep-learning-based_2020}
Sha, Y., D.~J.~G. II, G.~West, and R.~Stull, 2020: Deep-{Learning}-{Based} {Gridded} {Downscaling} of {Surface} {Meteorological} {Variables} in {Complex} {Terrain}. {Part} {I}: {Daily} {Maximum} and {Minimum} 2-m {Temperature}. \textit{Journal of Applied Meteorology and Climatology}, \textbf{59~(12)}, 2057 -- 2073, \doi{10.1175/JAMC-D-20-0057.1}, \urlprefix\url{https://journals.ametsoc.org/view/journals/apme/59/12/jamc- d-20-0057.1.xml}, place: Boston MA, USA Publisher: American Meteorological Society.

\bibitem[{Shocher et~al.(2018)Shocher, Cohen,, and Irani}]{shocher_zero-shot_2018}
Shocher, A., N.~Cohen, and M.~Irani, 2018: “zero-shot” super-resolution using deep internal learning. \textit{Proceedings of the {IEEE} conference on computer vision and pattern recognition}, 3118--3126.

\bibitem[{Sillmann et~al.(2013)Sillmann, Kharin, Zhang, Zwiers,, and Bronaugh}]{sillmann_climate_2013}
Sillmann, J., V.~V. Kharin, X.~Zhang, F.~W. Zwiers, and D.~Bronaugh, 2013: Climate extremes indices in the {CMIP5} multimodel ensemble: {Part} 1. {Model} evaluation in the present climate. \textit{Journal of Geophysical Research: Atmospheres}, \textbf{118~(4)}, 1716--1733, \doi{10.1002/jgrd.50203}, \urlprefix\url{https://onlinelibrary.wiley.com/doi/abs/10.1002/jgrd.50203}, \_eprint: https://onlinelibrary.wiley.com/doi/pdf/10.1002/jgrd.50203.

\bibitem[{Singh et~al.(2019)Singh, Albert,, and White}]{singh_downscaling_2019}
Singh, A., A.~Albert, and B.~White, 2019: Downscaling {Numerical} {Weather} {Models} with {GANs}. \textit{9th International Workshop on Climate Informatics}, École Normale Supérieure, Paris, France, 4.

\bibitem[{Sobie and Murdock(2017)Sobie, and Murdock}]{sobie_high-resolution_2017}
Sobie, S.~R., and T.~Q. Murdock, 2017: High-{Resolution} {Statistical} {Downscaling} in {Southwestern} {British} {Columbia}. \textit{Journal of Applied Meteorology and Climatology}, \textbf{56~(6)}, 1625--1641, \doi{10.1175/JAMC-D-16-0287.1}, \urlprefix\url{https://journals.ametsoc.org/view/journals/apme/56/6/jamc-d- 16-0287.1.xml}, publisher: American Meteorological Society Section: Journal of Applied Meteorology and Climatology.

\bibitem[{Song et~al.(2020)Song, Her, Shin, Cho, Paudel, Khare, Obeysekera,, and Martinez}]{song_evaluating_2020}
Song, J.-H., Y.~Her, S.~Shin, J.~Cho, R.~Paudel, Y.~P. Khare, J.~Obeysekera, and C.~J. Martinez, 2020: Evaluating the performance of climate models in reproducing the hydrological characteristics of rainfall events. \textit{Hydrological sciences journal}, \textbf{65~(9)}, 1490--1511, \doi{10.1080/02626667.2020.1750616}, publisher: Taylor \& Francis.

\bibitem[{Stengel et~al.(2020)Stengel, Glaws, Hettinger,, and King}]{stengel_adversarial_2020}
Stengel, K., A.~Glaws, D.~Hettinger, and R.~N. King, 2020: Adversarial super-resolution of climatological wind and solar data. \textit{Proceedings of the National Academy of Sciences}, \textbf{117~(29)}, 16\,805--16\,815, \doi{10.1073/pnas.1918964117}, \urlprefix\url{https://www.pnas.org/content/117/29/16805}, publisher: National Academy of Sciences Section: Physical Sciences.

\bibitem[{Stephens et~al.(2010)}]{stephens_dreary_2010}
Stephens, G.~L., and Coauthors, 2010: Dreary state of precipitation in global models. \textit{Journal of Geophysical Research: Atmospheres}, \textbf{115~(D24)}, \doi{10.1029/2010JD014532}, \urlprefix\url{https://onlinelibrary.wiley.com/doi/abs/10.1029/ 2010JD014532}, \_eprint: https://onlinelibrary.wiley.com/doi/pdf/10.1029/2010JD014532.

\bibitem[{Torma et~al.(2015)Torma, Giorgi,, and Coppola}]{torma_added_2015}
Torma, C., F.~Giorgi, and E.~Coppola, 2015: Added value of regional climate modeling over areas characterized by complex terrain-{Precipitation} over the {Alps}. \textit{Journal of geophysical research. Atmospheres}, \textbf{120~(9)}, 3957--3972, \doi{10.1002/2014JD022781}, edition: Torma, Cs., F. Giorgi, and E. Coppola (2015), Added value of regional climate modeling over areas characterized by complex terrain-Precipitation over the Alps, J. Geophys. Res. Atmos., 120, 3957-3972. doi: 10.1002/2014JD022781. Publisher: Blackwell Publishing Ltd.

\bibitem[{Wang et~al.(2021)Wang, Tian, Lowe, Kalin,, and Lehrter}]{wang_deep_2021}
Wang, F., D.~Tian, L.~Lowe, L.~Kalin, and J.~Lehrter, 2021: Deep {Learning} for {Daily} {Precipitation} and {Temperature} {Downscaling}. \textit{Water Resources Research}, \textbf{57~(4)}, e2020WR029\,308, \doi{10.1029/2020WR029308}, \urlprefix\url{https://doi.org/10.1029/2020WR029308}, publisher: John Wiley \& Sons, Ltd.

\bibitem[{Wang et~al.(2018)}]{wang_esrgan_2018}
Wang, X., and Coauthors, 2018: {ESRGAN}: {Enhanced} {Super}-{Resolution} {Generative} {Adversarial} {Networks}. \textit{arXiv:1809.00219 [cs]}, \urlprefix\url{http://arxiv.org/abs/1809.00219}, arXiv: 1809.00219.

\bibitem[{Whiteman(2000)}]{whiteman_mountain_2000}
Whiteman, C.~D., 2000: Mountain {Climates} of {North} {America}. \textit{Mountain {Climates} of {North} {America}}, Oxford University Press, \doi{10.1093/oso/9780195132717.003.0008}, \urlprefix\url{https://oxford.universitypressscholarship.com/10.1093/oso/ 9780195132717.001.0001/isbn-9780195132717-book-part-8}.

\bibitem[{Wilby and Wigley(1997)Wilby, and Wigley}]{wilby_downscaling_1997}
Wilby, R.~L., and T.~M.~L. Wigley, 1997: Downscaling general circulation model output: a review of methods and limitations. \textit{Progress in Physical Geography}, \textbf{21}, 530 -- 548.

\bibitem[{Zhang et~al.(2018)Zhang, Tian, Kong, Zhong,, and Fu}]{zhang_residual_2018}
Zhang, Y., Y.~Tian, Y.~Kong, B.~Zhong, and Y.~Fu, 2018: Residual {Dense} {Network} for {Image} {Super}-{Resolution}. \textit{arXiv:1802.08797 [cs]}, \urlprefix\url{http://arxiv.org/abs/1802.08797}, arXiv: 1802.08797.

\bibitem[{Zhang et~al.(2020)Zhang, Zhang, DiVerdi, Wang, Echevarria,, and Fu}]{zhang_texture_2020}
Zhang, Y., Z.~Zhang, S.~DiVerdi, Z.~Wang, J.~Echevarria, and Y.~Fu, 2020: Texture hallucination for large-factor painting super-resolution. \textit{European {Conference} on {Computer} {Vision}}, Springer, 209--225.

\bibitem[{Zhu et~al.(2020)Zhu, Zhang, Zhang, Liu, Shen,, and Zhao}]{zhu_gan-based_2020}
Zhu, X., L.~Zhang, L.~Zhang, X.~Liu, Y.~Shen, and S.~Zhao, 2020: {GAN}-{Based} {Image} {Super}-{Resolution} with a {Novel} {Quality} {Loss}. \textit{Mathematical Problems in Engineering}, \textbf{2020}, e5217\,429, \doi{10.1155/2020/5217429}, \urlprefix\url{https://www.hindawi.com/journals/mpe/2020/5217429/}, publisher: Hindawi.

\end{thebibliography}


\begin{thebibliography}{17}
\providecommand{\natexlab}[1]{#1}
\providecommand{\url}[1]{\texttt{#1}}
\renewcommand{\UrlFont}{\rmfamily}
\providecommand{\urlprefix}{URL }
\expandafter\ifx\csname urlstyle\endcsname\relax
  \providecommand{\doi}[1]{https://doi.org/\discretionary{}{}{}#1}\else
  \providecommand{\doi}{https://doi.org/\discretionary{}{}{}\begingroup \urlstyle{rm}\Url}\fi
\providecommand{\eprint}[2][]{\url{#2}}

\bibitem[{Arjovsky et~al.(2017)Arjovsky, Chintala,, and Bottou}]{arjovsky_wasserstein_2017}
Arjovsky, M., S.~Chintala, and L.~Bottou, 2017: Wasserstein {GAN}. \textit{arXiv:1701.07875 [cs, stat]}, \urlprefix\url{http://arxiv.org/abs/1701.07875}, arXiv: 1701.07875.

\bibitem[{Bessac et~al.(2019)Bessac, Monahan, Christensen,, and Weitzel}]{bessac_stochastic_2019}
Bessac, J., A.~H. Monahan, H.~M. Christensen, and N.~Weitzel, 2019: Stochastic {Parameterization} of {Subgrid}-{Scale} {Velocity} {Enhancement} of {Sea} {Surface} {Fluxes}. \textit{Monthly Weather Review}, \textbf{147~(5)}, 1447 -- 1469, \doi{10.1175/MWR-D-18-0384.1}, \urlprefix\url{https://journals.ametsoc.org/view/journals/mwre/147/5/mwr-d- 18-0384.1.xml}, place: Boston MA, USA Publisher: American Meteorological Society.

\bibitem[{Cheng et~al.(2020)Cheng, Liu, Xu, Shen,, and Kuang}]{cheng_generating_2020}
Cheng, J., J.~Liu, Z.~Xu, C.~Shen, and Q.~Kuang, 2020: Generating {High}-{Resolution} {Climate} {Prediction} through {Generative} {Adversarial} {Network}. \textit{Procedia Computer Science}, \textbf{174}, 123--127, \doi{10.1016/j.procs.2020.06.067}, \urlprefix\url{https://linkinghub.elsevier.com/retrieve/pii/ S1877050920315817}.

\bibitem[{Dong et~al.(2015)Dong, Loy, He,, and Tang}]{dong_image_2015}
Dong, C., C.~C. Loy, K.~He, and X.~Tang, 2015: Image {Super}-{Resolution} {Using} {Deep} {Convolutional} {Networks}. \textit{arXiv:1501.00092 [cs]}, \urlprefix\url{http://arxiv.org/abs/1501.00092}, arXiv: 1501.00092.

\bibitem[{Fritsche et~al.(2019)Fritsche, Gu,, and Timofte}]{fritsche_frequency_2019}
Fritsche, M., S.~Gu, and R.~Timofte, 2019: Frequency {Separation} for {Real}-{World} {Super}-{Resolution}. \textit{arXiv:1911.07850 [cs, eess]}, \urlprefix\url{http://arxiv.org/abs/1911.07850}, arXiv: 1911.07850.

\bibitem[{Gulrajani et~al.(2017)Gulrajani, Ahmed, Arjovsky, Dumoulin,, and Courville}]{gulrajani_improved_2017}
Gulrajani, I., F.~Ahmed, M.~Arjovsky, V.~Dumoulin, and A.~Courville, 2017: Improved {Training} of {Wasserstein} {GANs}. \textit{arXiv:1704.00028 [cs, stat]}, \urlprefix\url{http://arxiv.org/abs/1704.00028}, arXiv: 1704.00028.

\bibitem[{Kumar et~al.(2021)Kumar, Chattopadhyay, Singh, Chaudhari, Kodari,, and Barve}]{kumar_deep_2021}
Kumar, B., R.~Chattopadhyay, M.~Singh, N.~Chaudhari, K.~Kodari, and A.~Barve, 2021: Deep learning–based downscaling of summer monsoon rainfall data over {Indian} region. \textit{Theoretical and Applied Climatology}, \textbf{143~(3)}, 1145--1156, \doi{10.1007/s00704-020-03489-6}, \urlprefix\url{https://doi.org/10.1007/s00704-020-03489-6}.

\bibitem[{Ledig et~al.(2017)}]{ledig_photo-realistic_2017}
Ledig, C., and Coauthors, 2017: Photo-{Realistic} {Single} {Image} {Super}-{Resolution} {Using} a {Generative} {Adversarial} {Network}. \textit{arXiv:1609.04802 [cs, stat]}, \urlprefix\url{http://arxiv.org/abs/1609.04802}, arXiv: 1609.04802.

\bibitem[{Lucas-Picher et~al.(2008)Lucas-Picher, Caya, de~Elía,, and Laprise}]{lucas-picher_investigation_2008}
Lucas-Picher, P., D.~Caya, R.~de~Elía, and R.~Laprise, 2008: Investigation of regional climate models' internal variability with a ten-member ensemble of 10-year simulations over a large domain. \textit{Climate dynamics}, \textbf{31~(7-8)}, 927--940, place: Berlin/Heidelberg Publisher: Berlin/Heidelberg : Springer-Verlag.

\bibitem[{McGovern et~al.(2019)McGovern, Lagerquist, Gagne, Jergensen, Elmore, Homeyer,, and Smith}]{MakingtheBlackBoxMoreTransparentUnderstandingthePhysicalImplicationsofMachineLearning}
McGovern, A., R.~Lagerquist, D.~J. Gagne, G.~E. Jergensen, K.~L. Elmore, C.~R. Homeyer, and T.~Smith, 2019: Making the black box more transparent: Understanding the physical implications of machine learning. \textit{Bulletin of the American Meteorological Society}, \textbf{100~(11)}, 2175 -- 2199, \doi{https://doi.org/10.1175/BAMS-D-18-0195.1}, \urlprefix\url{https://journals.ametsoc.org/view/journals/bams/100/11/bams-d-18-0195.1.xml}.

\bibitem[{Mirza and Osindero(2014)Mirza, and Osindero}]{mirza_conditional_2014}
Mirza, M., and S.~Osindero, 2014: Conditional {Generative} {Adversarial} {Nets}. \textit{arXiv:1411.1784 [cs, stat]}, \urlprefix\url{http://arxiv.org/abs/1411.1784}, arXiv: 1411.1784.

\bibitem[{Sha et~al.(2020)Sha, II, West,, and Stull}]{sha_deep-learning-based_2020}
Sha, Y., D.~J.~G. II, G.~West, and R.~Stull, 2020: Deep-{Learning}-{Based} {Gridded} {Downscaling} of {Surface} {Meteorological} {Variables} in {Complex} {Terrain}. {Part} {I}: {Daily} {Maximum} and {Minimum} 2-m {Temperature}. \textit{Journal of Applied Meteorology and Climatology}, \textbf{59~(12)}, 2057 -- 2073, \doi{10.1175/JAMC-D-20-0057.1}, \urlprefix\url{https://journals.ametsoc.org/view/journals/apme/59/12/jamc- d-20-0057.1.xml}, place: Boston MA, USA Publisher: American Meteorological Society.

\bibitem[{Singh et~al.(2019)Singh, Albert,, and White}]{singh_downscaling_2019}
Singh, A., A.~Albert, and B.~White, 2019: Downscaling {Numerical} {Weather} {Models} with {GANs}. \textit{9th International Workshop on Climate Informatics}, École Normale Supérieure, Paris, France, 4.

\bibitem[{Stengel et~al.(2020)Stengel, Glaws, Hettinger,, and King}]{stengel_adversarial_2020}
Stengel, K., A.~Glaws, D.~Hettinger, and R.~N. King, 2020: Adversarial super-resolution of climatological wind and solar data. \textit{Proceedings of the National Academy of Sciences}, \textbf{117~(29)}, 16\,805--16\,815, \doi{10.1073/pnas.1918964117}, \urlprefix\url{https://www.pnas.org/content/117/29/16805}, publisher: National Academy of Sciences Section: Physical Sciences.

\bibitem[{Wang et~al.(2021)Wang, Tian, Lowe, Kalin,, and Lehrter}]{wang_deep_2021}
Wang, F., D.~Tian, L.~Lowe, L.~Kalin, and J.~Lehrter, 2021: Deep {Learning} for {Daily} {Precipitation} and {Temperature} {Downscaling}. \textit{Water Resources Research}, \textbf{57~(4)}, e2020WR029\,308, \doi{10.1029/2020WR029308}, \urlprefix\url{https://doi.org/10.1029/2020WR029308}, publisher: John Wiley \& Sons, Ltd.

\bibitem[{Wang et~al.(2018)}]{wang_esrgan_2018}
Wang, X., and Coauthors, 2018: {ESRGAN}: {Enhanced} {Super}-{Resolution} {Generative} {Adversarial} {Networks}. \textit{arXiv:1809.00219 [cs]}, \urlprefix\url{http://arxiv.org/abs/1809.00219}, arXiv: 1809.00219.

\bibitem[{Zhang et~al.(2020)Zhang, Zhang, DiVerdi, Wang, Echevarria,, and Fu}]{zhang_texture_2020}
Zhang, Y., Z.~Zhang, S.~DiVerdi, Z.~Wang, J.~Echevarria, and Y.~Fu, 2020: Texture hallucination for large-factor painting super-resolution. \textit{European {Conference} on {Computer} {Vision}}, Springer, 209--225.

\end{thebibliography}

\end{document}


\appendixtitle{Supplemental Material}
\section{Acroynms}
Below is a list of acronyms and their definition we use:
\begin{itemize}
    \item SR - super-resolution
    \item CNN - convolutional neural networks
    \item GAN - generative adversarial network
    \item LR/HR - low/high resolution
    \item FS - frequency separation
    \item WGAN-GP - wasserstein GAN with gradient penalty
    \item SRGAN - super-resolution GAN
    \item MAE - mean absolute error
    \item MSE - mean-squared error
    \item WRF - weather research and forecasting
    \item HRCONUS - high-resolution contiguous united states
    \item CAPE - convective available potential energy
    \item ERA - ECMWF reanalysis
    \item ERA5 - ECMWF reanalysis v5
    \item GPU - graphical processing unit
    \item MS-SSIM - multi-scale structural similarity index
    \item PNSR - peak signal-to-noise ratio
    \item RAPSD - radially averaged power spectral density
    \item MSA - median symmetric accuracy
    \item RD - relative differences
\end{itemize}

\section{Super-Resolution and Statistical Downscaling}
\label{section:challenge_of_sr}
In the computer vision field, SR is concerned with enhancing the resolution of images in a deterministic manner. That is, the SR method generates unique arrangements of HR patterns provided unique arrangements of LR patterns. Similarly, statistical downscaling is often performed deterministically. However, since both SR and statistical downscaling are concerned with enhancing spatial resolution, they are both inherently ill-posed in the presence of a scale gap; the set of possible arrangements of HR patterns associated with fixed arrangements of LR patterns is infinite.

For HR images, this indeterminacy of fine-scale features can be considered an undesirable limitation of SR because only one true reference image exists. However, in the context of statistical downscaling, a range of HR realizations are generally physically possible for a given LR state.  For example, in convection-permitting models driven with LR models as boundary conditions, different realizations of fine-scale features are produced if provided different initial conditions \citep{lucas-picher_investigation_2008}. Since initial conditions are not perfectly known, an ensemble of convection-permitting models produces a distribution of fine-scale features that are physically consistent with the LR state. It follows that generating a distribution of fine-scale features (e.g. weather in climate fields) through statistical downscaling that could exist in climate fields is a desired outcome of SR.

Generating these fine-scale features using SR has been likened to \textit{hallucinating} details that could exist, but do not necessarily exactly match ``true'' HR features in the training data \citep{zhang_texture_2020}. Although deterministic SR does not sample from the full set of HR arrangements associated with given LR fields, it does generate one possible arrangement. 



Another consideration of SR and statistical downscaling lies in the relationship between the LR and HR patterns in the training and test sets. Implementations of SR in  computer vision and statistical downscaling studies frequently coarsen HR images to produce ``ideal'' LR inputs for training \citep{dong_image_2015, ledig_photo-realistic_2017, wang_esrgan_2018, singh_downscaling_2019, sha_deep-learning-based_2020, cheng_generating_2020, stengel_adversarial_2020, wang_deep_2021, kumar_deep_2021}. Idealized coarsening is generally not appropriate for climate downscaling since LR and HR patterns may be produced separately in different models and have differences in structures in overlapping scales -- even if the high-resolution model is being driven by the low-resolution one. These differences can result from internal variability of the high-resolution models or from differences in model biases.  

Finally, we note that when the scale separation is large, there is a larger range of HR fields that LR patterns can produce \citep{bessac_stochastic_2019}. We consider a linear scale factor of eight, resulting in the LR scales having weaker control on the HR scales than the widely used scale factor of four in the computer vision field \citep{ledig_photo-realistic_2017, wang_esrgan_2018, fritsche_frequency_2019}.

\section{Wasserstein Distance}
\label{section:GANs_for_SR}

The success of GANs for SR can be attributed to minimizing a distributional distance between ``true'' and generated HR samples; rather than, for example, only the MAE or MSE. \cite{arjovsky_wasserstein_2017}, emphasizing how the choice of distance metric can improve GAN behaviour, proposed Wasserstein GANs (WGANs). In WGANs, the Wasserstein distance between the ``true'' and generated distributions ($\PX_r$ and $\PX_g$ respectively) is estimated using the \textit{Critic} (formerly the Discriminator) and minimized through training the Generator. The Wasserstein distance has several advantages as a distributional distance; using it in GANs leads to stability and improves their ability to learn target distributions \citep{arjovsky_wasserstein_2017, gulrajani_improved_2017}. WGAN from \cite{arjovsky_wasserstein_2017} was superseded by WGAN with gradient penalty (WGAN-GP) \citep{gulrajani_improved_2017}, which uses a softer and more practical approach to constrain gradients of the Critic (a technical requirement of WGANs).

A conditional GAN approach \citep{mirza_conditional_2014} was adopted by \cite{ledig_photo-realistic_2017} in super-resolution GAN (SRGAN), where LR samples condition the Generator to produce HR samples. The Critic and Generator each have objective functions that, when minimized, optimize their respective weights. The Critic's objective function has a Wasserstein distance component and a gradient penalty term

\begin{equation}\label{eq:critic_sr}
    \mathcal{L}_C =  \underbrace{\EX_{G(\textbf{x}) \sim \PX_g} [C(G(\textbf{x}))] - \EX_{\textbf{y} \sim \PX_r} [C(\textbf{y})]}_{\textnormal{Wasserstein component}} +\underbrace{\lambda\EX_{\hat{\textbf{x}} \sim \PX_{\hat{\textbf{x}}}}[(||\nabla_{\hat{\textbf{x}}} C(\hat{\textbf{x}})||_2 -1 )^2)]}_{\textnormal{Gradient Penalty}}
\end{equation}

\noindent where \textbf{x} are LR covariates, \textbf{y} are ``true'' HR fields, $\lambda$ is a tuneable hyperparameter, and $\hat{\textbf{x}}$ is a randomly synthesized field interpolated on a straight line between ``true'' and generated samples as

\begin{equation}\label{eq:grad_pen}
    \hat{\textbf{x}} = \epsilon \textbf{y} + (1-\epsilon) G(\textbf{x})
\end{equation}

\noindent where $\epsilon$ is a random number drawn from uniform noise, $\epsilon \sim U[0, 1]$, and $G(\textbf{x})$ is the generated field with LR covariates synchronous with $\textbf{y}$. Importantly, Equation \ref{eq:critic_sr} optimizes the Critic to estimate the Wasserstein distance. A further discussion of the gradient penalty term can be found in \cite{gulrajani_improved_2017}. For the Generator, the objective function is:

\begin{equation}\label{eq:generator_sr}
\mathcal{L}_G =  \underbrace{ - \EX_{G(\textbf{x}) \sim \PX_g} [C(G(\textbf{x}))]}_{\textnormal{Adversarial component/loss}} + \underbrace{\alpha\EX_{\textbf{y} \sim \PX_r} l_c(\textbf{y}, G(\textbf{x}))}_{\textnormal{Content loss}}
\end{equation}

\noindent where $\alpha$ is a hyperparameter that weights the relative importance of the content loss, $l_c$, and the adversarial loss. The content loss is meant to guide the generated fields towards $\textbf{y}$ (it is desirable that generated and true realizations agree on larger common scales) and is necessary for stability while training the SR models. This present work uses the mean absolute error (MAE) for the content loss as follows:

\begin{equation}\label{eq:content_loss}
    l_c(\textbf{y}, G(\textbf{x})) = \frac{1}{WH}\sum_{i=1}^H\sum_{j=1}^W |\textbf{y}_{i, j} - G(\textbf{x})_{i, j}|
\end{equation}

\noindent where $W$ and $H$ are the width and height of the field in pixels.

The adversarial component of Equation \ref{eq:generator_sr} does not compare grid points between ``true'' and generated fields but is directly related to the distributional distance between collections (i.e. batches) of fields measured by $C$. Because $\mathcal{L}_G$ contains both a distributional and MAE component, even if $G(\textbf{x})$ is produced without stochastic conditioning information, the adversarial component in Equation \ref{eq:generator_sr} is responsible for generating similar variability to $\textbf{y}$, putting features where they \textit{could} be, but not necessarily where they are in the ``true'' HR fields.

\subsection{Frequency Separation Implementation}

If $\mathscr{L}$ represents a low pass filter, the low-frequency and high-frequency fields are determined with the following:

\begin{equation}\label{eq:freq_sep_high}
\begin{split}
    \textbf{y}_h & = \textbf{y} - \mathscr{L}(\textbf{y}) \\
    G(\textbf{x})_h & = G(\textbf{x}) - \mathscr{L}(G(\textbf{x}))
\end{split}
\end{equation}

\noindent where the $h$ subscript indicates the high-frequency field after applying $\mathscr{L}$ to the HR field.
The objective function for the Critic is then modified as:

\begin{equation}\label{eq:critic_wass_loss_fs}
    \mathcal{L}_C = \mathbb{E}_{G(\textbf{x})_h \sim \PX_{g, h}}[C(G(\textbf{x})_h)] - \mathbb{E}_{\textbf{y}_h \sim \PX_{r, h}}[C(\textbf{y}_h)] + \lambda \mathbb{E}_{\hat{\textbf{x}}_h \sim \mathbb{P}_{\hat{\textbf{x}}_h}}[(||\nabla_{\hat{\textbf{x}}_h} C(\hat{\textbf{x}}_h)||_2 - 1)^2]
\end{equation}

\noindent where $\PX_{r,h}$ and $\PX_{g,h}$ are the distributions of the high-frequency ``true'' and generated fields respectively, and $\hat{\textbf{x}}_h$ represents random samples from the high-frequency fields only, i.e. $\hat{\textbf{x}}_h = \epsilon \textbf{y}_h + (1-\epsilon) G(\textbf{x})_h$. Similarly, the Generator's objective is modified as:

\begin{equation}\label{eq:generator_wass_loss_fs}
    \mathcal{L}_G =  \underbrace{ - \EX_{G(\textbf{x})_h \sim \PX_{g,h}} [C(G(\textbf{x})_h)]}_{\textnormal{Adversarial component}} + \underbrace{\alpha \EX_{\textbf{y} \sim \PX_r} l_c(\mathscr{L}(\textbf{y}), \mathscr{L}(G(\textbf{x})))}_{\textnormal{Content loss}}
\end{equation}

For fields with FS, a reflection pad corresponding to the half-width of the averaging kernel is applied to preserve the same size between the low-frequency and original fields. Reflection padding extends the size of the image by mirroring pixels close to its edges and offers more continuous patterns for smoothing than, for example, zero padding. 

\section{Covariate Sensitivity}

To further explore the role covariates play in the SR models, an experiment is devised to perturb covariate fields of the already-trained non-FS GAN, and measure across wavenumbers the resulting changes in the spectra. The experiment proceeds as follows:

\begin{enumerate}
    \item Identify a covariate to modify and replace each time step with a randomly sampled one from somewhere else in the test set. This modifies the selected covariate to have six correct covariate fields, and one randomly sampled one from another time-step in the set (preserving the distribution but not the statistical dependence with other fields).
    \item Generate HR fields with the modified covariate input and compute the power spectrum for each time step. Compute an unmodified power spectrum of the field as a baseline.
    \item Repeat Steps 1 and 2 25 times to generate many HR fields with randomly sampled covariates.
    \item Calculate the relative differences, $RD$, between the RAPSD of the modified, $P_s(\mathbf{k})$, and unmodified baseline, $P_b(\mathbf{k})$ as 

    \begin{equation}\label{eq:sensitivity}
        RD(P_s, P_b) = \frac{\log{P_{s}(\mathbf{k})} - \log{P_{b}(\mathbf{k})}}{\log{P_b(\mathbf{k})}}
    \end{equation}

    \item Compute the variance in $RD$ as $\Var{[RD(P_s, P_b)]}$ at each wavenumber as a measure of model sensitivity.
    \item Perform Steps 1 - 5 for $u10$, $v10$, surface pressure, and CAPE.
\end{enumerate}

\noindent The above experiment is known as singe-pass \textit{permutation importance} and is a common feature importance experiment \citep{MakingtheBlackBoxMoreTransparentUnderstandingthePhysicalImplicationsofMachineLearning}.
\clearpage
\newpage

\section{Additional Figures}

\begin{figure}[b]
    \centering
    \noindent
    \includegraphics[width=\textwidth]{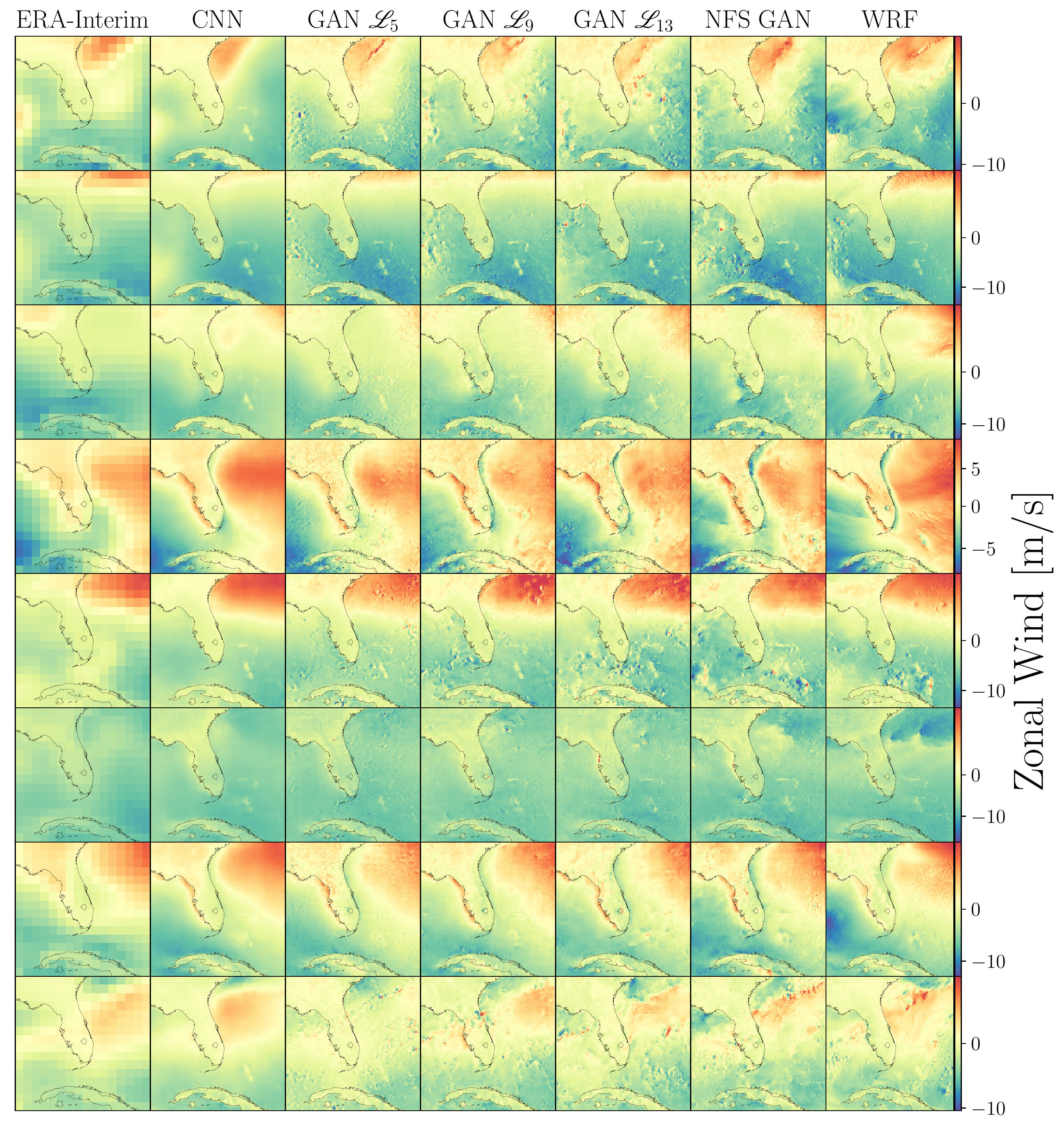}
    \caption{Example realizations of $u10$ wind for the Southeast region. Each row represents a randomized time step from the test set, while each column is a different model considered in this work.}
    \label{fig:S1}
\end{figure}

\begin{figure}[t]
    \centering
    \noindent
    \includegraphics[width=\textwidth]{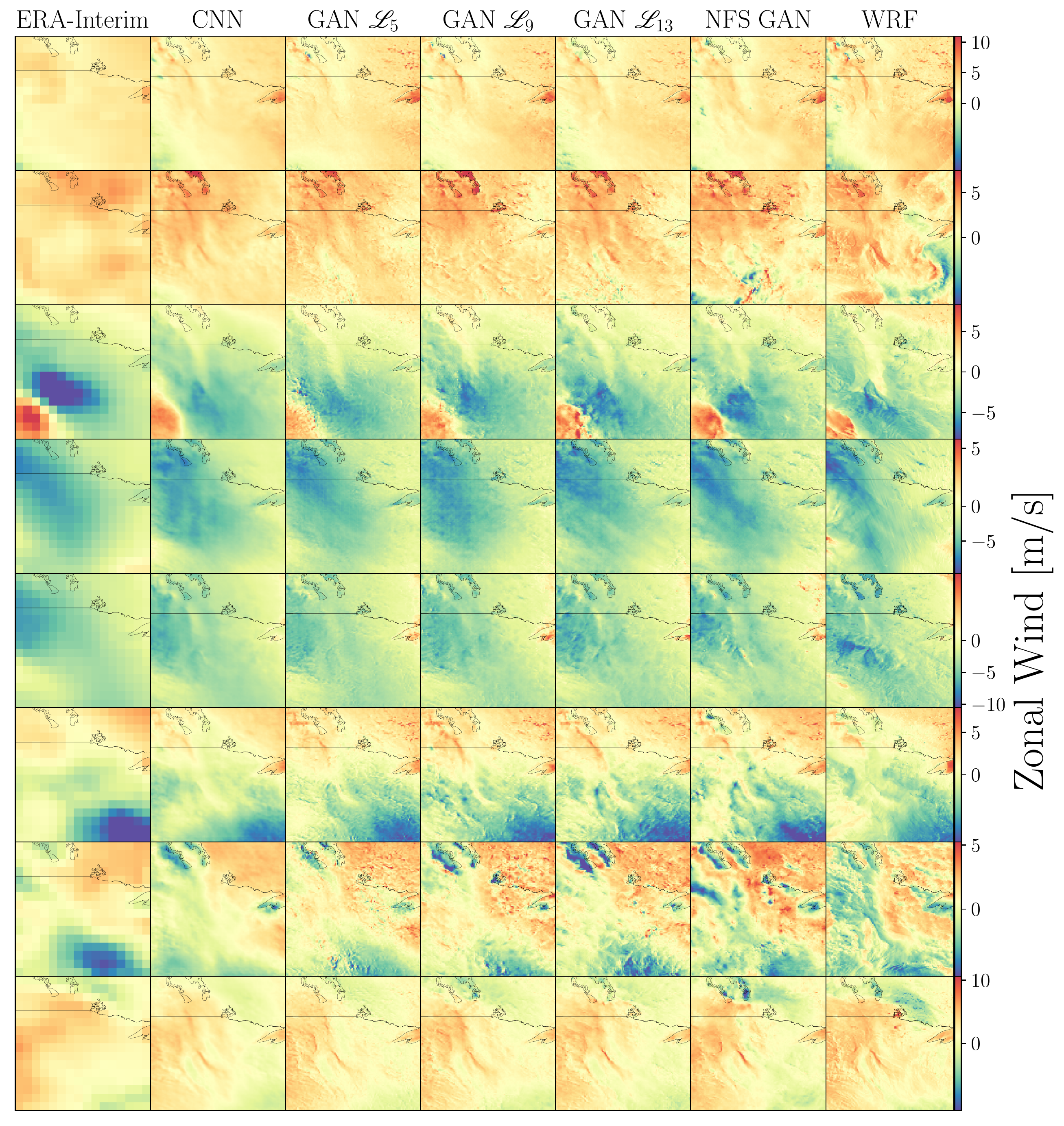}
    \caption{Example realizations of $u10$ wind for the Central region. Each row represents a randomized time step from the test set, while each column is a different model considered in this work.}
    \label{fig:S2}
\end{figure}

\begin{figure}[t]
    \centering
    \noindent
    \includegraphics[width=\textwidth]{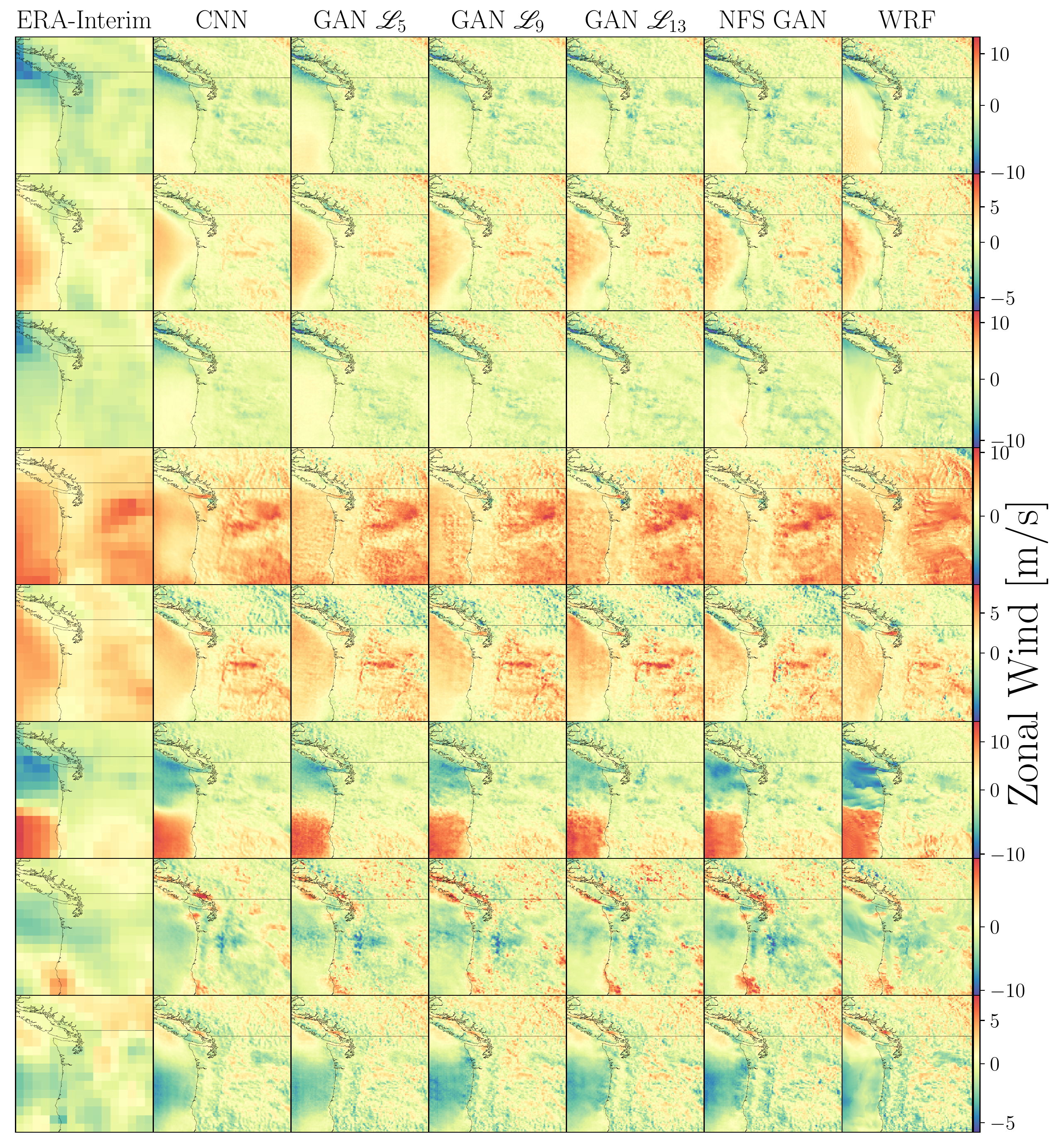}
    \caption{Example realizations of $u10$ wind for the West region. Each row represents a randomized time step from the test set, while each column is a different model considered in this work.}
    \label{fig:S3}
\end{figure}

\begin{figure}[t]
    \centering
    \noindent
    \includegraphics[width=\textwidth]{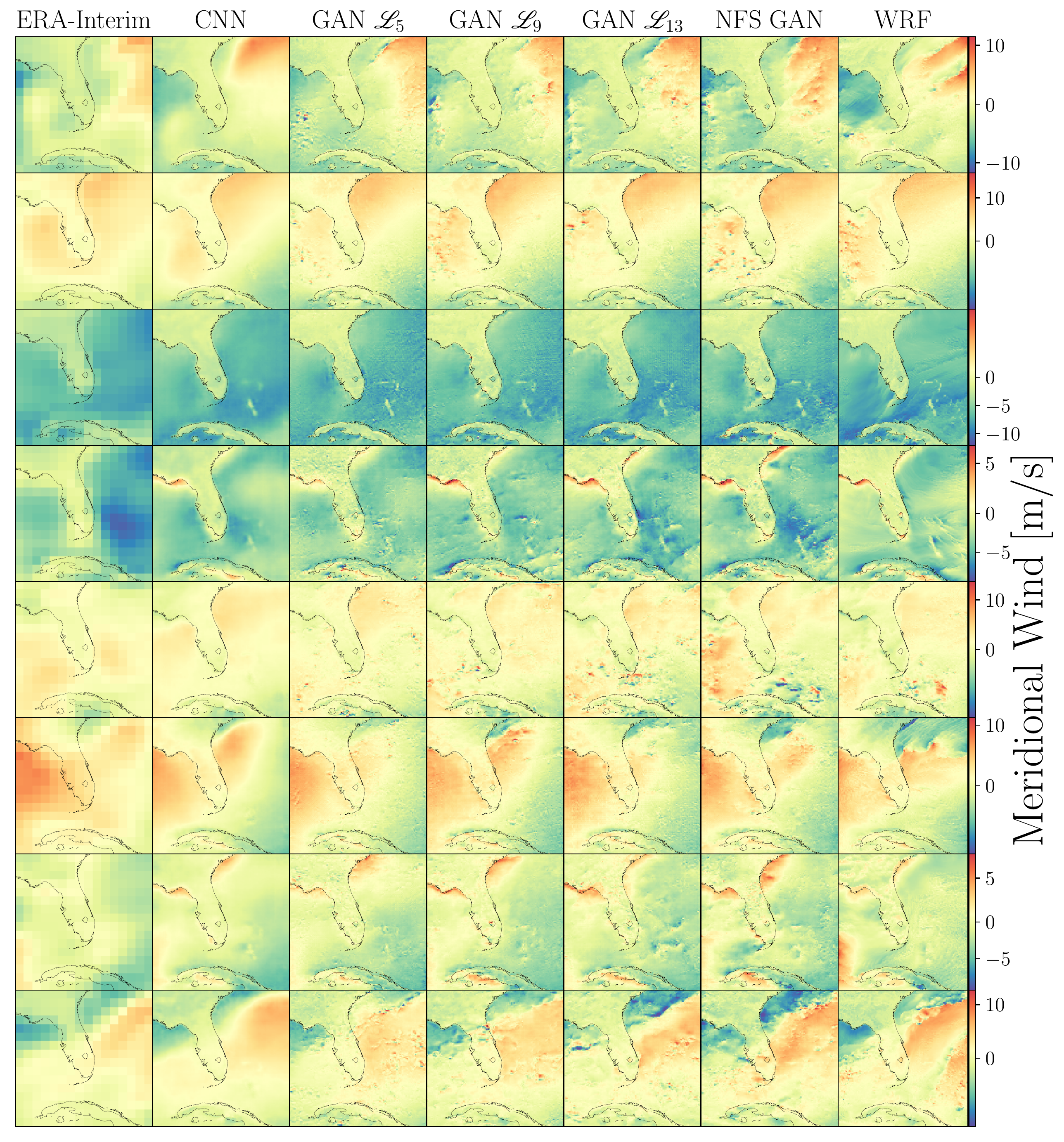}
    \caption{Example realizations of $v10$ wind for the Southeast region. Each row represents a randomized time step from the test set, while each column is a different model considered in this work.}
    \label{fig:S4}
\end{figure}

\begin{figure}[t]
    \centering
    \noindent
    \includegraphics[width=\textwidth]{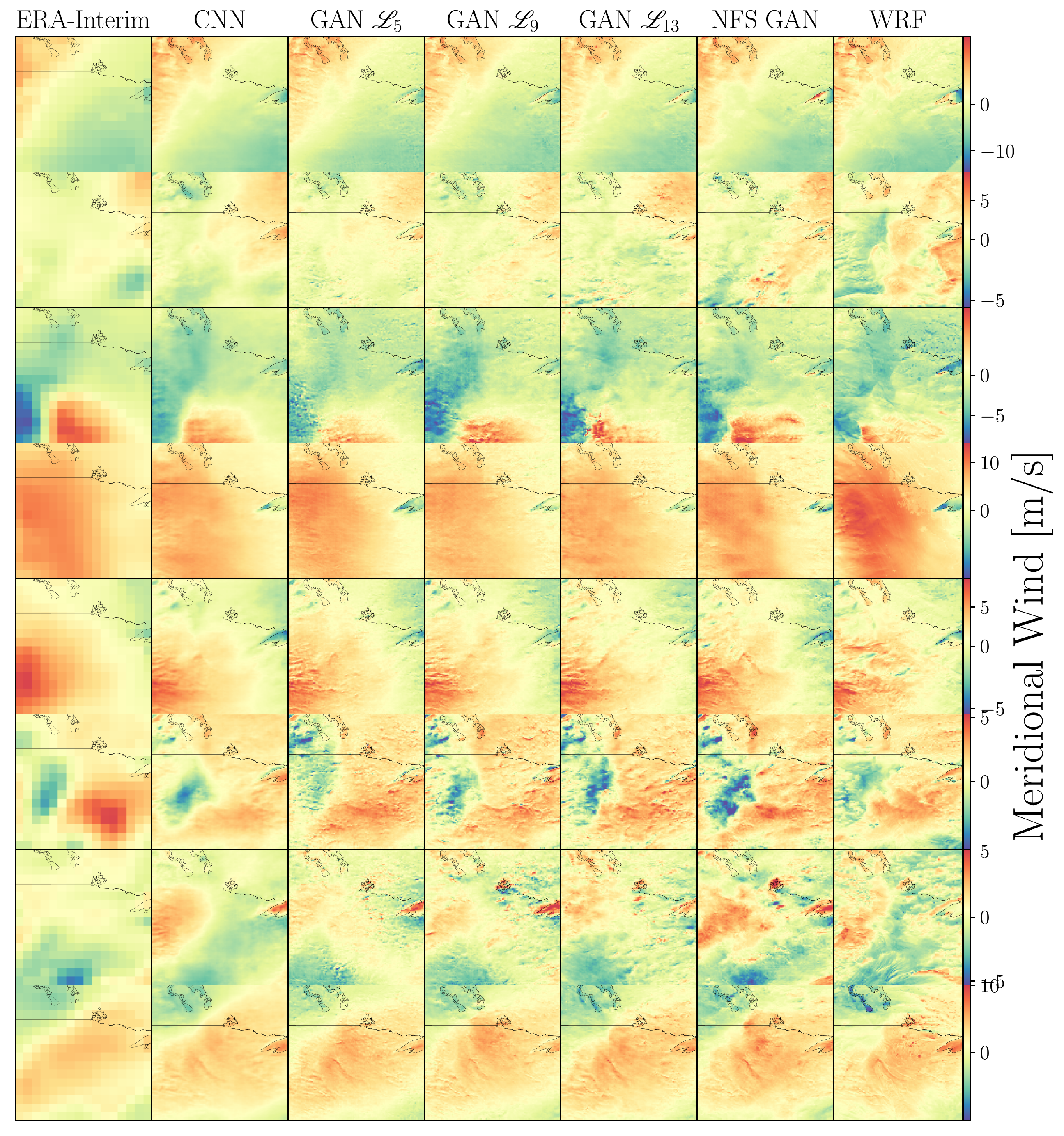}
    \caption{Example realizations of $v10$ wind for the Central region. Each row represents a randomized time step from the test set, while each column is a different model considered in this work.}
    \label{fig:S5}
\end{figure}

\begin{figure}[t]
    \centering
    \noindent
    \includegraphics[width=\textwidth]{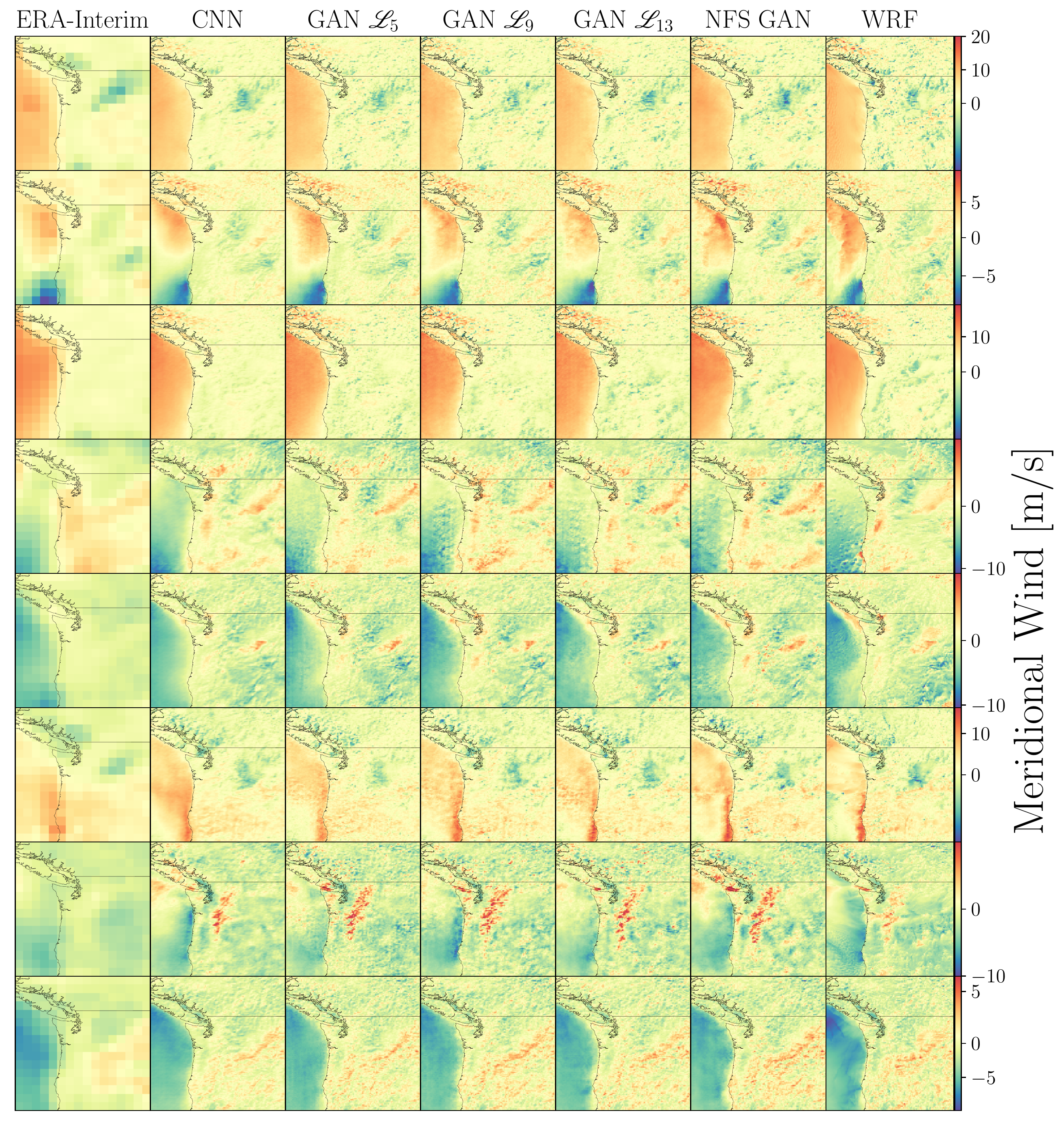}
    \caption{Example realizations of $v10$ wind for the West region. Each row represents a randomized time step from the test set, while each column is a different model considered in this work.}
    \label{fig:S6}
\end{figure}

\begin{figure}[t]
    \centering
    \noindent
    \includegraphics[width=0.7\textwidth]{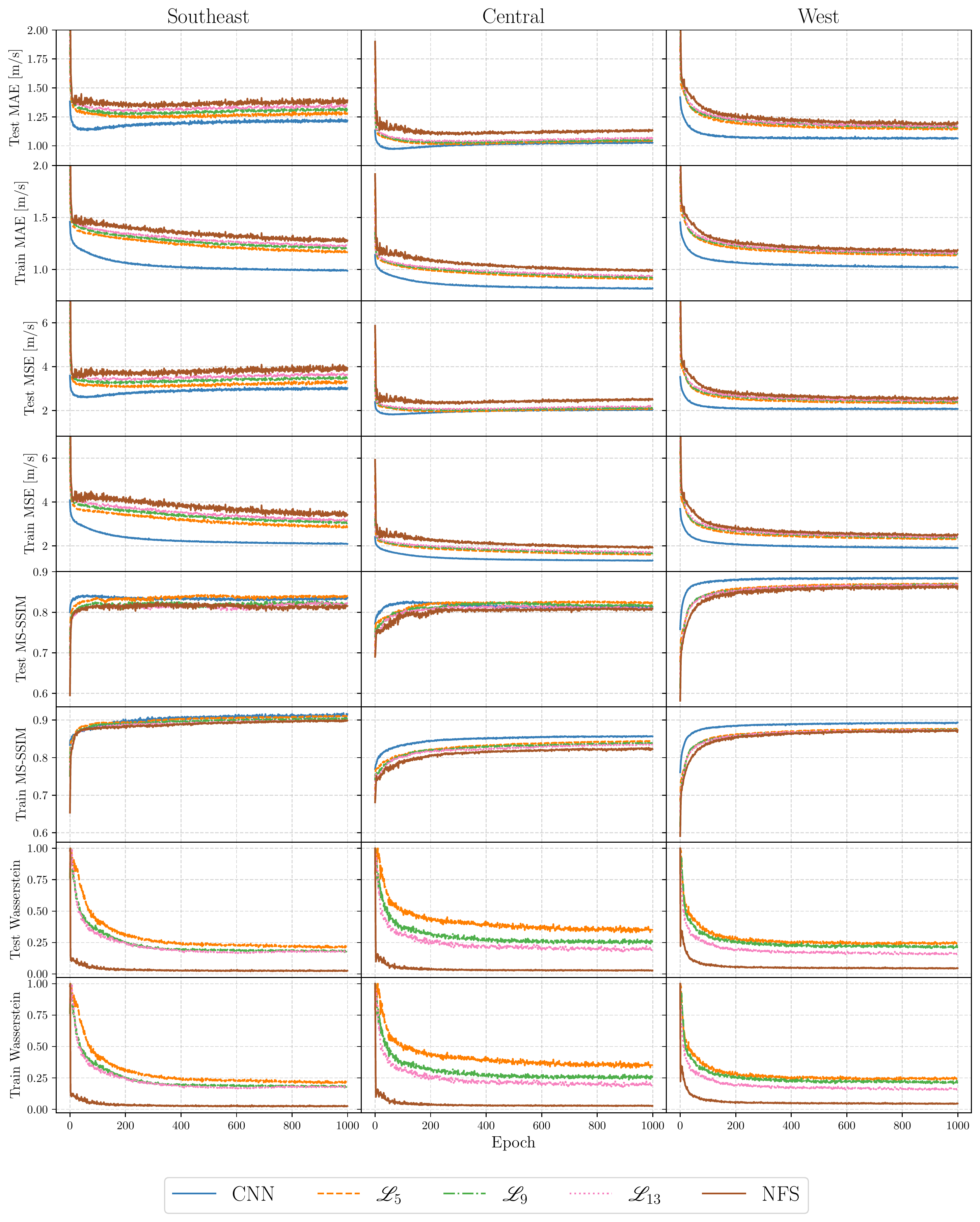}
    \caption{Training evolution of the MAE, MSE, MS-SSIM, and Wasserstein distance for 1000 epochs on both the test and training data. Each SR model is indicated in the legend. The MAE, MSE, and MS-SSIM are calculated grid-wise with combined $u10$ and $v10$ HR fields. The Wasserstein distance estimate is scaled by the initial distance. Note that no Wasserstein distance is calculated for the pure CNN, since no Critic network was trained.}
    \label{fig:S7}
\end{figure}

\begin{figure}[t]
    \centering
    \includegraphics[width=\textwidth]{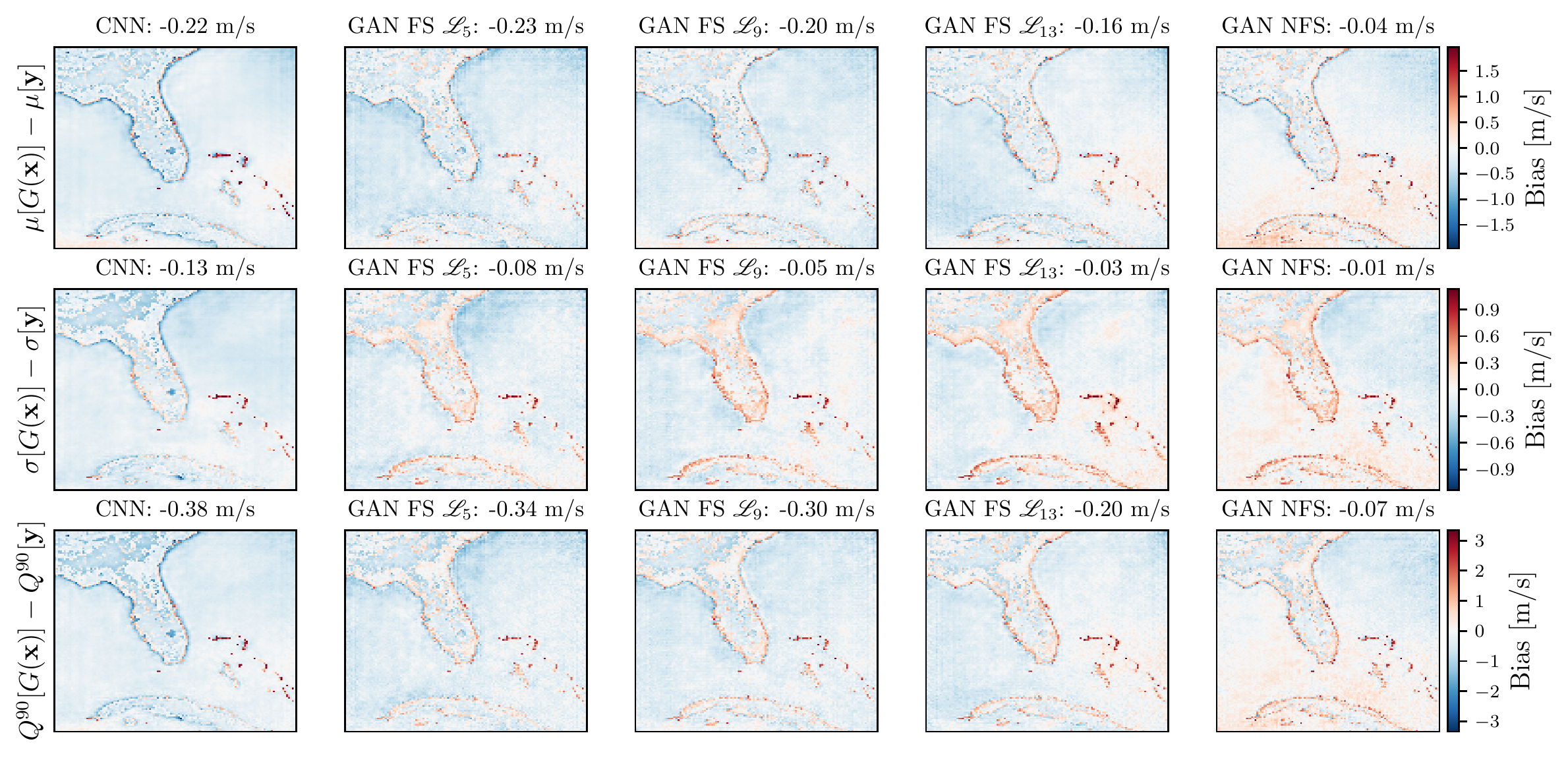}
    \caption{Map of the Southeast region SR wind speed bias in the mean (top row), standard deviation (middle row), and 90th percentile (bottom row) for each SR model. Statistics are calculated along the temporal dimension of the test set. The spatial mean of the bias is reported in the title of each panel.}
    \label{fig:S8}
\end{figure}

\begin{figure}[t]
    \centering
    \includegraphics[width=\textwidth]{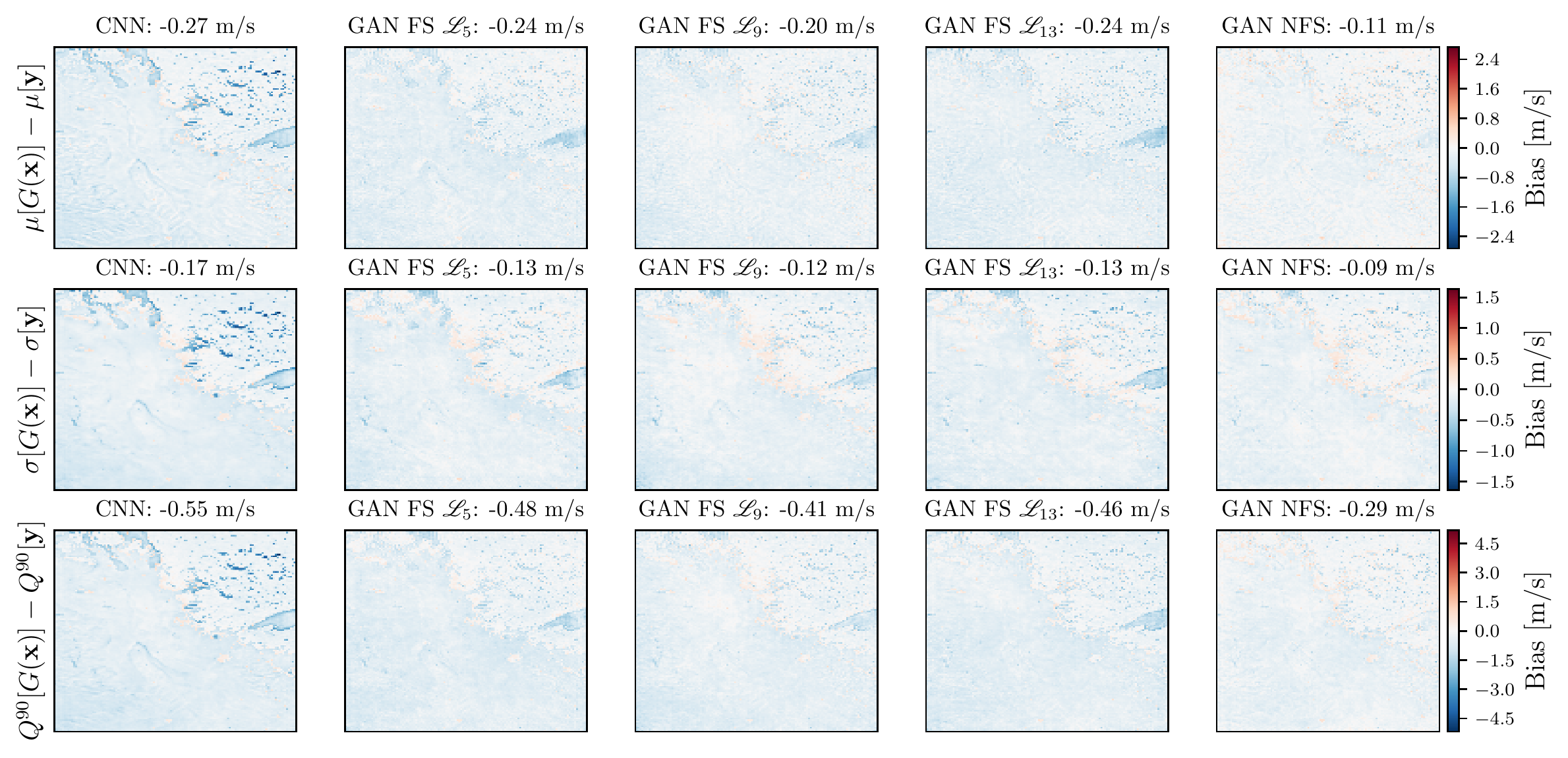}
    \caption{Map of the Central region SR wind speed bias in the mean (top row), standard deviation (middle row), and 90th percentile (bottom row) for each SR model. Statistics are calculated along the temporal dimension of the test set. The spatial mean of the bias is reported in the title of each panel.}
    \label{fig:S9}
\end{figure}

\begin{figure}[t]
    \centering
    \includegraphics[width=\textwidth]{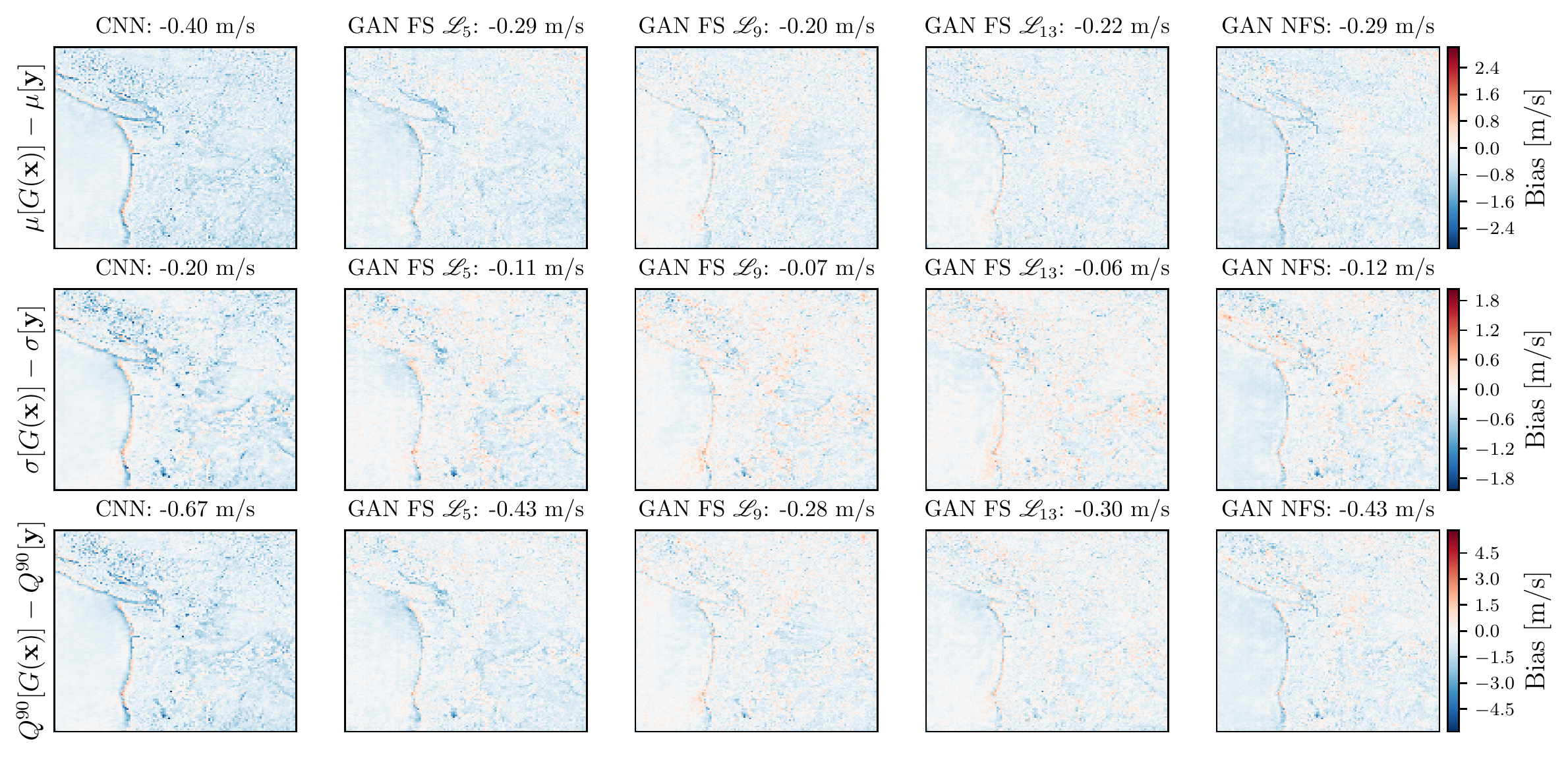}
    \caption{Map of the West region SR wind speed bias in the mean (top row), standard deviation (middle row), and 90th percentile (bottom row) for each SR model. Statistics are calculated along the temporal dimension of the test set. The spatial mean of the bias is reported in the title of each panel.}
    \label{fig:S10}
\end{figure}

\begin{figure}[t]
    \centering
    \includegraphics[width=\textwidth]{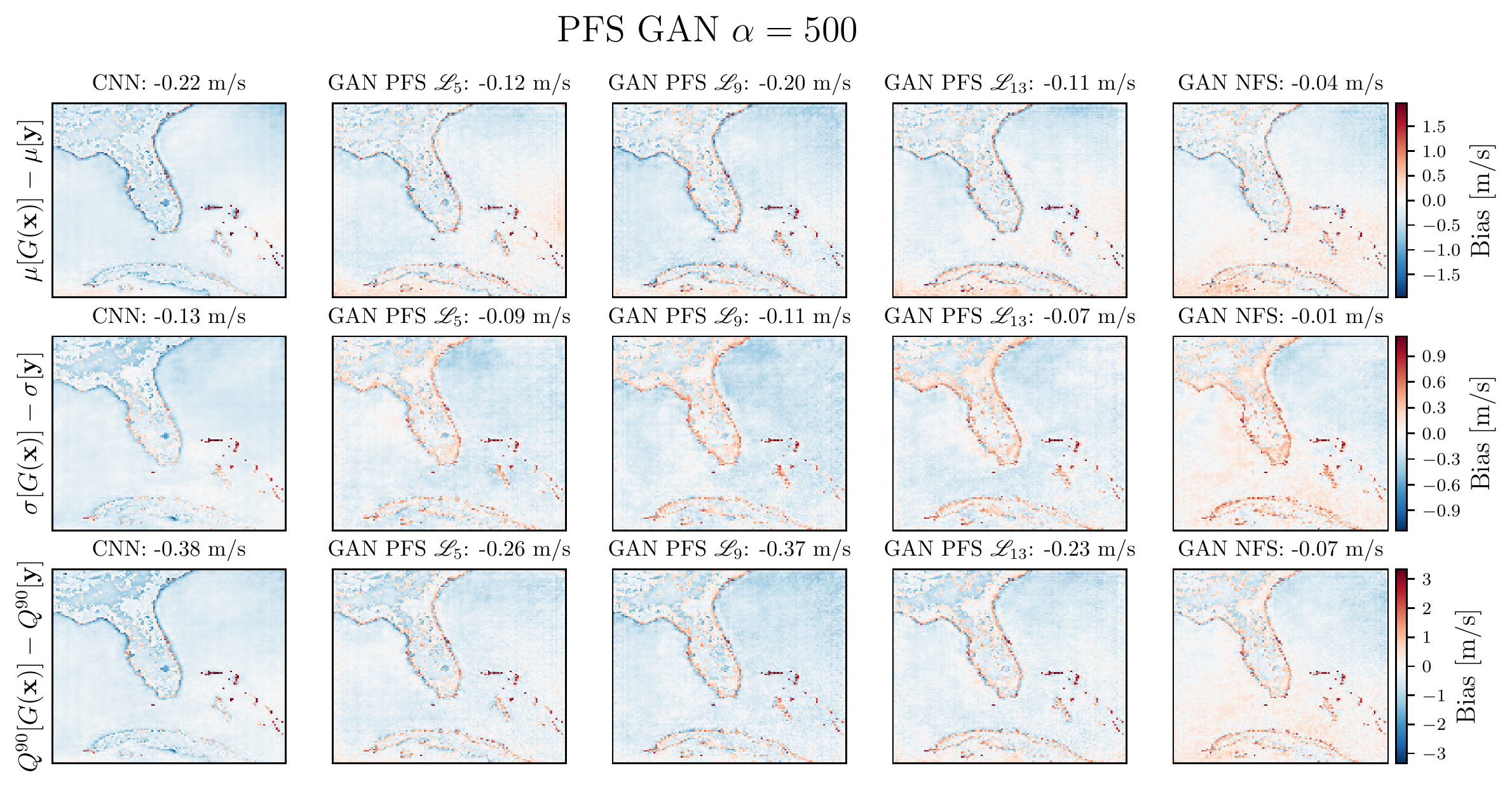}
    \caption{Map of the Southeast region partial FS GAN ($\alpha = 500$) wind speed bias in the mean (top row), standard deviation (middle row), and 90th percentile (bottom row) for each SR model. Statistics are calculated along the temporal dimension of the test set. The spatial mean of the bias is reported in the title of each panel.}
    \label{fig:S11}
\end{figure}

\begin{figure}[t]
    \centering
    \includegraphics[width=\textwidth]{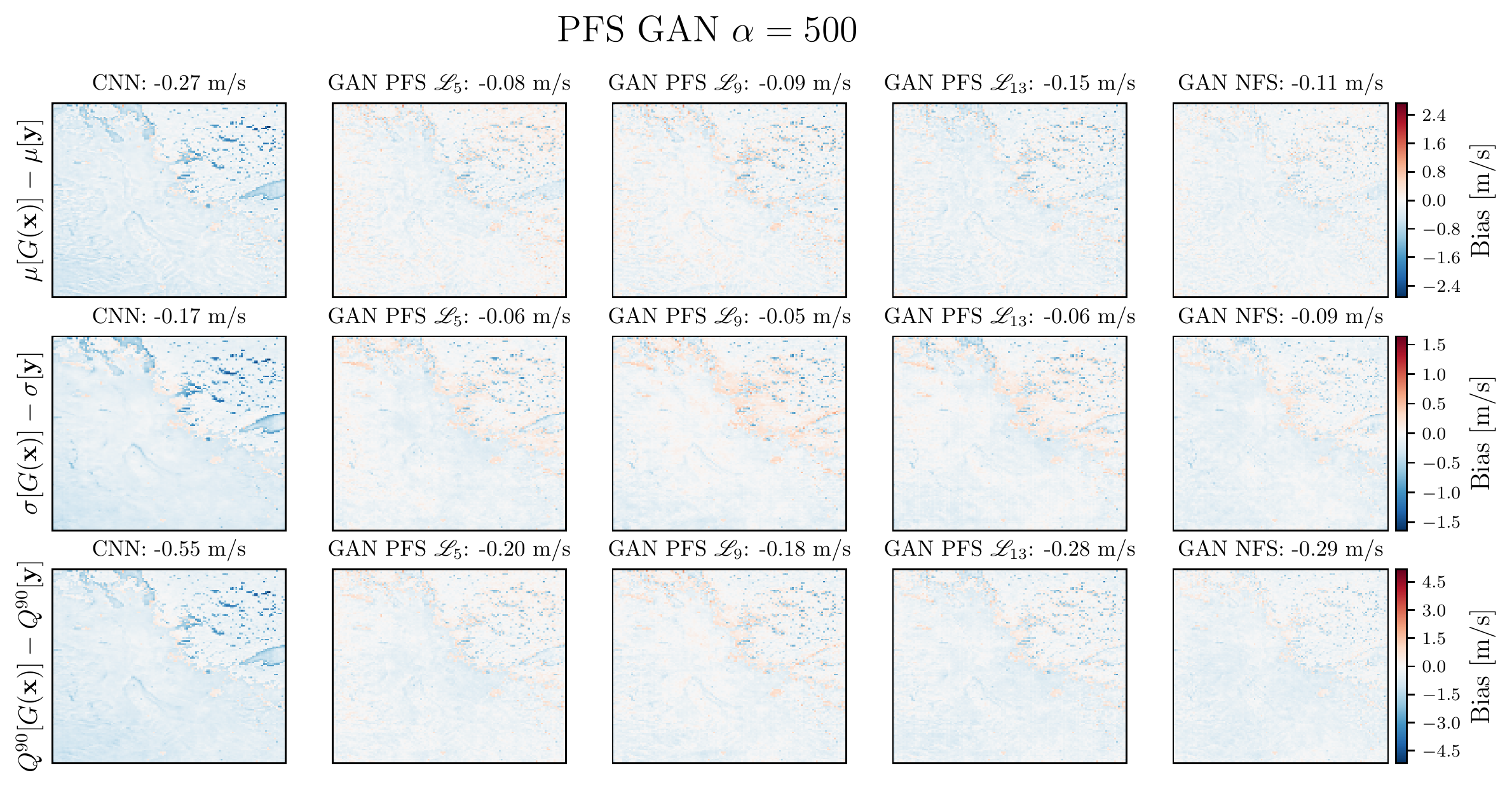}
    \caption{Map of the Central region partial FS GAN ($\alpha = 500$) and other SR wind speed bias in the mean (top row), standard deviation (middle row), and 90th percentile (bottom row) for each SR model. Statistics are calculated along the temporal dimension of the test set. The spatial mean of the bias is reported in the title of each panel.}
    \label{fig:S12}
\end{figure}

\begin{figure}[t]
    \centering
    \includegraphics[width=\textwidth]{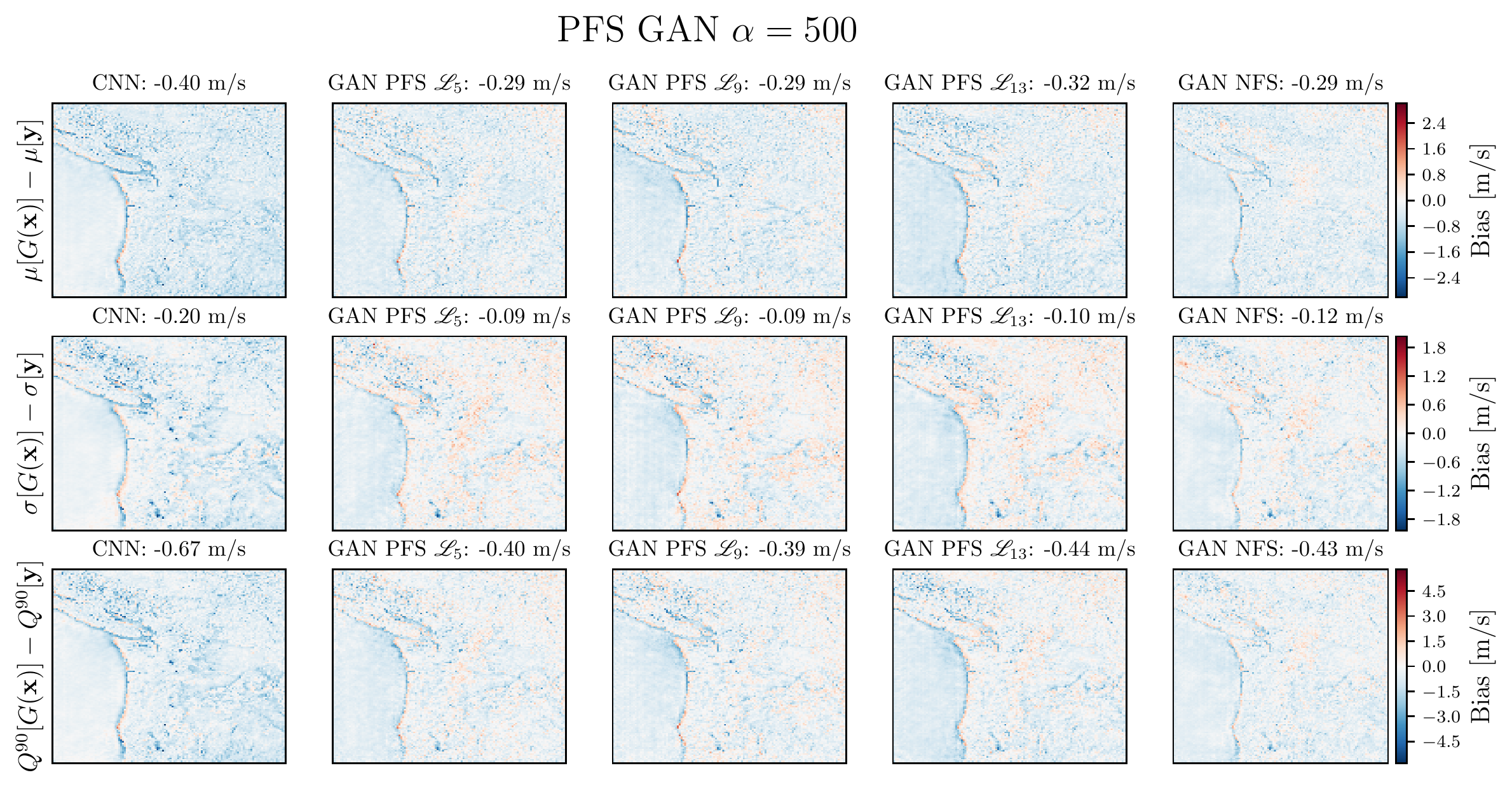}
    \caption{Map of the West region partial FS GAN ($\alpha = 500$) and other SR  wind speed bias in the mean (top row), standard deviation (middle row), and 90th percentile (bottom row) for each SR model. Statistics are calculated along the temporal dimension of the test set. The spatial mean of the bias is reported in the title of each panel.}
    \label{fig:S13}
\end{figure}

\begin{figure}[t]
    \centering
    \includegraphics[width=\textwidth]{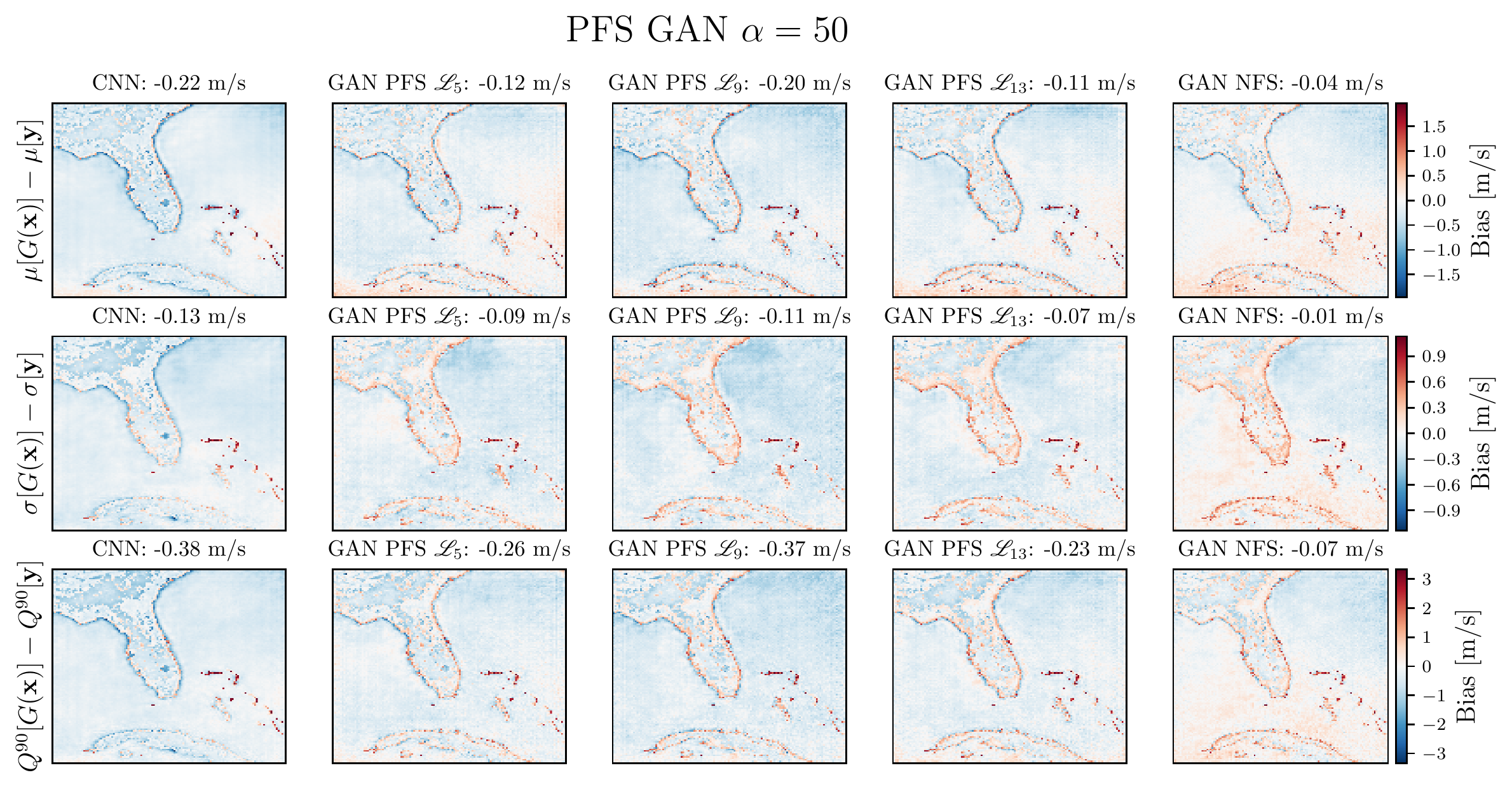}
    \caption{Map of the Southeast region partial FS GAN ($\alpha = 50$) and other SR  wind speed bias in the mean (top row), standard deviation (middle row), and 90th percentile (bottom row) for each SR model. Statistics are calculated along the temporal dimension of the test set. The spatial mean of the bias is reported in the title of each panel.}
    \label{fig:S14}
\end{figure}

\begin{figure}[t]
    \centering
    \includegraphics[width=\textwidth]{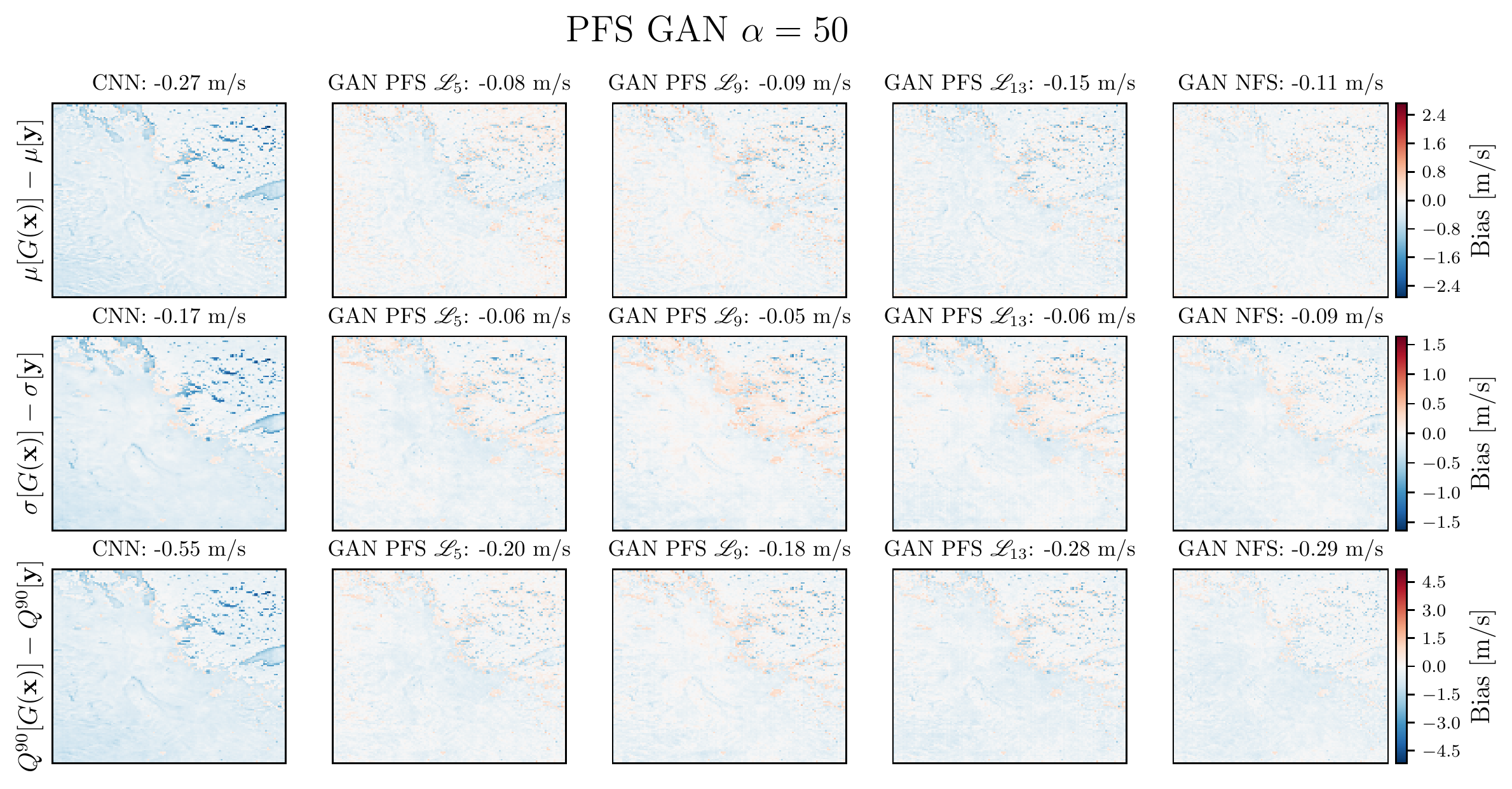}
    \caption{Map of the Central region partial FS GAN ($\alpha = 50$) and other SR  wind speed bias in the mean (top row), standard deviation (middle row), and 90th percentile (bottom row) for each SR model. Statistics are calculated along the temporal dimension of the test set. The spatial mean of the bias is reported in the title of each panel.}
    \label{fig:S15}
\end{figure}

\begin{figure}[t]
    \centering
    \includegraphics[width=\textwidth]{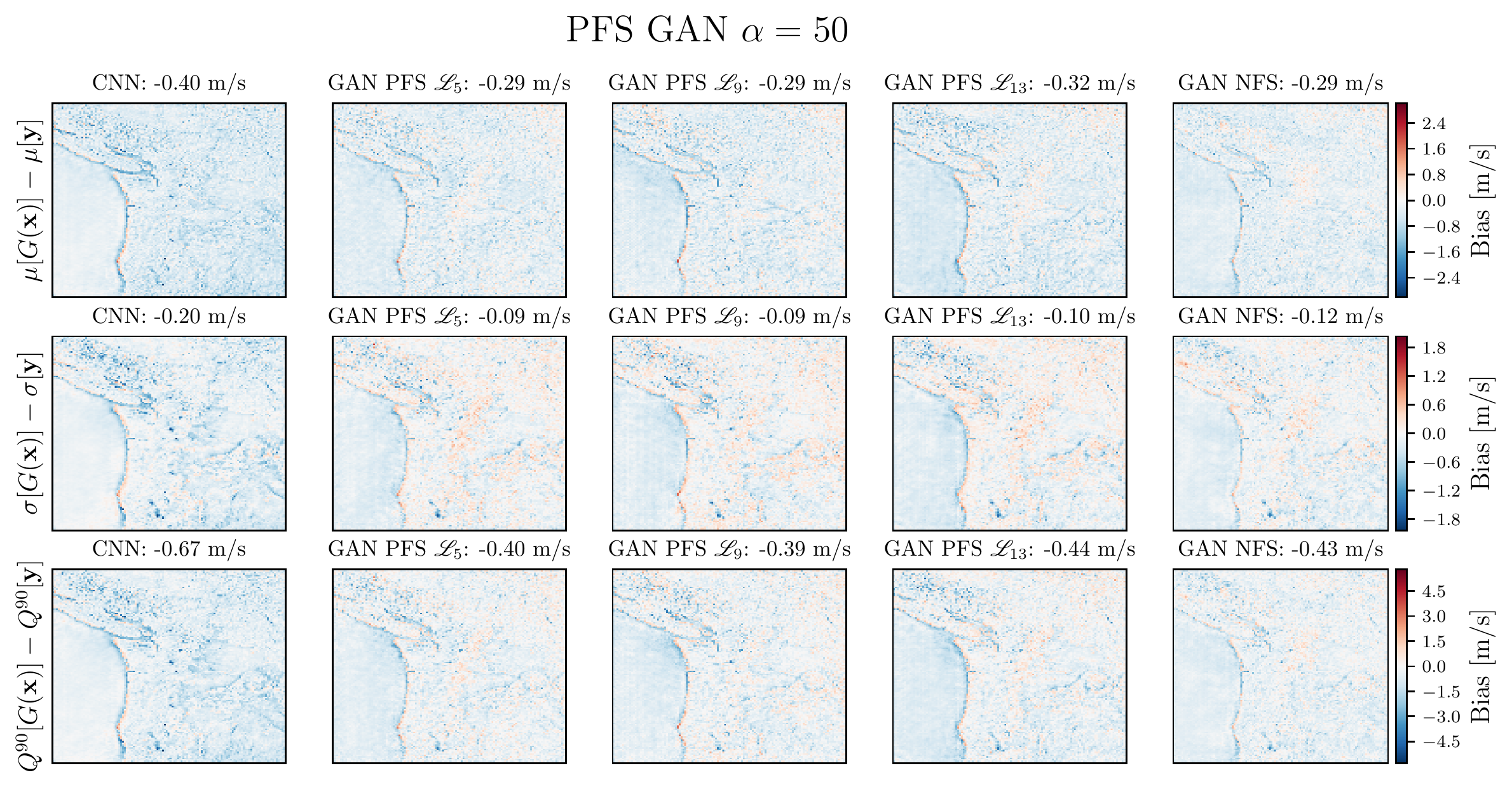}
    \caption{Map of the West region partial FS GAN ($\alpha = 50$) and other SR  wind speed bias in the mean (top row), standard deviation (middle row), and 90th percentile (bottom row) for each SR model. Statistics are calculated along the temporal dimension of the test set. The spatial mean of the bias is reported in the title of each panel.}
    \label{fig:S16}
\end{figure}

\clearpage

\newpage

\bibliographystyle{ametsocV6}
\bibliography{annau2023}

\newpage